\RequirePackage[l2tabu, orthodox]{nag}

\documentclass[conference, 10pt, letterpaper, oneside, draftcls, onecolumn]{IEEEtran} 
\usepackage{ifpdf}

\usepackage[noadjust]{cite}

\ifCLASSINFOpdf
   \usepackage[pdftex]{graphicx}
\else
  \usepackage[dvipdfmx]{graphicx}
\fi
\usepackage[cmex10]{amsmath}
\usepackage{array}

\ifCLASSOPTIONcompsoc
  \usepackage[caption=false,font=normalsize,labelfont=sf,textfont=sf]{subfig}
\else
  \usepackage[caption=false,font=footnotesize]{subfig}
\fi
\usepackage{fixltx2e}
\usepackage{url}

\usepackage{xcolor}
\definecolor{burgundy}{rgb}{0.545098,0,0}
\definecolor{navyblue}{rgb}{0.0, 0.0, 0.5}
\definecolor{leafgreen}{rgb}{0.290196, 0.470588, 0.0}
\definecolor{bluegreen}{rgb}{0, 0.470588, 0.415686}
\definecolor{zuhl}{rgb}{0.1875, 0.26171875, 0.46484375}
\definecolor{orange}{rgb}{1, 0.6470588235, 0}

\usepackage{amsthm}
\usepackage{mathtools}
\usepackage{amssymb}
\usepackage{bm, bbm}
\usepackage{array}

\theoremstyle{plain}
\newtheorem{definition}{Definition}
\newtheorem{lemma}{Lemma}

\newtheorem{theorem}{Theorem}
\newtheorem{corollary}{Corollary}
\newtheorem{remark}{Remark}

\newcommand{\bvec}[1]{\boldsymbol{#1}}
\newcommand{\sbvec}[1]{\boldsymbol{#1}}
\newcommand{\sgn}{\operatorname{sgn}}
\newcommand{\argmax}{\mathop{\mathrm{arg~max}}\limits}

\newcommand{\figref}[1]{Fig.~\ref{#1}}
\newcommand{\lemref}[1]{Lemma~\ref{#1}}
\newcommand{\thref}[1]{Theorem~\ref{#1}}
\newcommand{\defref}[1]{Definition~\ref{#1}}
\newcommand{\corref}[1]{Corollary~\ref{#1}}
\newcommand{\sectref}[1]{Section~\ref{#1}}
\newcommand{\remref}[1]{Remark~\ref{#1}}

\allowdisplaybreaks[3]

\usepackage[utf8]{inputenc}
\usepackage[font = footnotesize]{caption}
\usepackage{overpic}
\usepackage{pict2e}
\usepackage{lipsum}
\usepackage{docmute}

\ifCLASSINFOpdf
	\usepackage[colorlinks]{hyperref} 
	\usepackage{microtype} 
\else
\fi

\usepackage[all, warning]{onlyamsmath}

\begin{document}
%
\title{Sharp Bounds Between Two R\'{e}nyi Entropies of \\ Distinct Positive Orders}

\author{
\IEEEauthorblockN{Yuta Sakai and Ken-ichi Iwata}
\IEEEauthorblockA{Graduate School of Engineering, University of Fukui,\\
3-9-1 Bunkyo, Fukui, Fukui, 910-8507, Japan,\\
E-mail: \{y-sakai,~k-iwata\}@u-fukui.ac.jp}
}


%


\maketitle

\begin{abstract}
Many axiomatic definitions of entropy, such as the R\'{e}nyi entropy, of a random variable are closely related to the $\ell_{\alpha}$-norm of its probability distribution.
This study considers probability distributions on finite sets, and examines the sharp bounds of the $\ell_{\beta}$-norm with a fixed $\ell_{\alpha}$-norm, $\alpha \neq \beta$, for $n$-dimensional probability vectors with an integer $n \ge 2$.
From the results, we derive the sharp bounds of the R\'{e}nyi entropy of positive order $\beta$ with a fixed R\'{e}nyi entropy of another positive order $\alpha$.
As applications, we investigate sharp bounds of Ariomoto's mutual information of order $\alpha$ and Gallager's random coding exponents for uniformly focusing channels under the uniform input distribution.
\end{abstract}


%
\IEEEpeerreviewmaketitle

\section{Introduction}

Information measures for random variables play important roles in many fields of engineering, such as coding theorems.
Let the $n$-dimensional probability simplex be denoted by
\begin{align}
\Delta_{n}
\coloneqq
\Bigg\{ (p_{1}, p_{2}, \dots, p_{n}) \in \mathbb{R}^{n} \ \Bigg| \ p_{i} \ge 0 \ \mathrm{and} \ \sum_{i=1}^{n} p_{i} = 1 \Bigg\}
\end{align}
for an integer $n \ge 2$.
For a random variable $X \sim \bvec{p} \in \Delta_{n}$, the Shannon entropy \cite{shannon} is defined by
\begin{align}
H( X )
=
H( \bvec{p} )
\coloneqq
- \sum_{i=1}^{n} p_{i} \ln p_{i} ,
\end{align}
which is one of famous information measures of uncertainty of $X$.
Previously, the Shannon entropy was axiomatically generalized to several forms \cite{havrda, daroczy, renyi, tsallis, boekee, behara}.
One of famous extensions of the Shannon entropy is the R\'{e}nyi entropy \cite{renyi} of order $\alpha$, which is defined by
\begin{align}
H_{\alpha}( X )
=
H_{\alpha}( \bvec{p} )
\coloneqq
\frac{ \alpha }{ 1 - \alpha } \ln \| \bvec{p} \|_{\alpha}
\label{def:Renyi}
\end{align}
for $X \sim \bvec{p} \in \Delta_{n}$ and $\alpha \in (0, 1) \cup (1, \infty)$, where
\begin{align}
\| \bvec{p} \|_{\alpha}
\coloneqq
\Bigg( \sum_{i=1}^{n} p_{i}^{\alpha} \Bigg)^{1/\alpha}
\end{align}
denotes the $\ell_{\alpha}$-norm of $\bvec{p}$ for $\alpha \in (0, \infty)$.
Thus, we see from \eqref{def:Renyi} that the R\'{e}nyi entropy is closely related to the $\ell_{\alpha}$-norm.
Moreover, for $\alpha \in \{ 1, \infty \}$, the R\'{e}nyi entropy is also defined by
\begin{align}
H_{1}( \bvec{p} )
& \coloneqq
\lim_{\alpha \to 1} H_{\alpha}( \bvec{p} )
=
H( \bvec{p} ) ,
\label{def:Renyi_1} \\
H_{\infty}( \bvec{p} )
& \coloneqq
\lim_{\alpha \to \infty} H_{\alpha}( \bvec{p} )
=
- \ln \| \bvec{p} \|_{\infty} ,
\label{def:Renyi_infty}
\end{align}
where the last equality of \eqref{def:Renyi_1} follows by L'H\^{o}pital's rule and the $\ell_{\infty}$-norm used in \eqref{def:Renyi_infty} is given by
\begin{align}
\| \bvec{p} \|_{\infty}
\coloneqq
\lim_{\alpha \to \infty} \| \bvec{p} \|_{\alpha}
=
\max \big\{ p_{1}, p_{2}, \dots, p_{n} \big\} .
\end{align}
Note that the minimum error probability $P_{\mathrm{e}}(X)$ of guessing the value of $X \sim \bvec{p}$ is calculated by $P_{\mathrm{e}}( X ) = 1 - \| \bvec{p} \|_{\infty}$ (cf. \cite[Eq. (2)]{feder}).
The R\'{e}nyi entropy of order 0 is called the Hartley entropy or the max-entropy; however, it is omitted and we only consider the R\'{e}nyi entropy of \emph{positive} order in this study.

In the previous study, the sharp bounds of the Shannon entropy $H(X)$ with a fixed error probability $P_{\mathrm{e}}(X)$, i.e.,
\begin{align}
\min_{\bvec{p} \in \Delta_{n} : P_{\mathrm{e}}( X ) = p_{\mathrm{e}}} H( \bvec{p} )
\le
H( X )
\le
\max_{\bvec{p} \in \Delta_{n} : P_{\mathrm{e}}( X ) = p_{\mathrm{e}}} H( \bvec{p} )
\label{ineq:fano}
\end{align}
for $p_{\mathrm{e}} \in [0, (n-1)/n]$, were independently derived by Kovalevsky \cite{kovalevsky}, Tebbe and Dwyer \cite{tebbe}, and Feder and Merhav \cite{feder}.
Note that the upper bound of \eqref{ineq:fano} is the \emph{un}conditional version of Fano's inequality \cite{fano} (cf. the remark of \cite[p.~40]{cover}).
We now define the following two characteristic $n$-dimensional probability vectors:
(i) the distribution $\bvec{v}_{n}( \cdot )$ is defined by
\begin{align}
\bvec{v}_{n}( p )
& \coloneqq
( v_{1}, v_{2}, \dots, v_{n} )
\in \Delta_{n} ,
\\
v_{i}
& =
\begin{cases}
1 - (n-1) p
& \mathrm{if} \ i = 1 , \\
p
& \mathrm{otherwise}
\end{cases}
\end{align}
for $p \in [0, 1 / (n-1)]$, and
(ii) the distribution $\bvec{w}_{n}( \cdot )$ is defined by
\begin{align}
\bvec{w}_{n}( p )
& \coloneqq
( w_{1}, w_{2}, \dots, w_{n} )
\in \Delta_{n} ,
\\
w_{i}
& =
\begin{cases}
p
& \mathrm{if} \ i \le \lfloor 1/p \rfloor , \\
1 - \lfloor 1/p \rfloor \, p
& \mathrm{if} \ i = \lfloor 1/p \rfloor + 1 , \\
0
& \mathrm{otherwise}
\end{cases}
\end{align}
for $p \in [1/n, 1]$, where $\lfloor x \rfloor \coloneqq \max\{ z \in \mathbb{Z} \mid z \le x \}$ denotes the floor function of $x \in \mathbb{R}$.
Then, the results of \cite{kovalevsky, tebbe, feder} show that the upper and lower bounds of \eqref{ineq:fano} are attained by the distributions $\bvec{v}_{n}( \cdot )$ and $\bvec{w}_{n}( \cdot )$, respectively.
In addition, by using topological method, Harremo\"{e}s and Tops{\o}e \cite{topsoe} derived that the upper and lower bounds of the Shannon entropy with a fixed index of coincidence $IC( \bvec{p} ) \coloneqq \| \bvec{p} \|_{2}^{2}$ are also attained by the distributions $\bvec{v}_{n}( \cdot )$ and $\bvec{w}_{n}( \cdot )$, respectively.
In the above previous works, we note that the error probability $P_{\mathrm{e}}( X )$ and the index of coincidence $IC( \bvec{p} )$ are closely related to the $\ell_{\infty}$-norm and the $\ell_{2}$-norm, respectively.
Namely, these results are related to the min-entropy $H_{\infty}( X )$ and the collision entropy $H_{2}( X )$.
As a generalization of the above results, we derived in \cite{part1} that the extremal Shannon entropies with a fixed $\ell_{\alpha}$-norm, $\alpha \in (0, 1) \cup (1, \infty)$, are attained by the distributions $\bvec{v}_{n}( \cdot )$ and $\bvec{w}_{n}( \cdot )$.
Therefore, the sharp bounds of the R\'{e}nyi entropy of order $\alpha \in (0, 1) \cup (1, \infty)$ with a fixed Shannon entropy, and vice versa, were derived in \cite{part1}.
Furthermore, in \cite{part2}, we extended the result of \cite{part1} to the relations between the \emph{conditional} Shannon entropy and the \emph{expectation} of $\ell_{\alpha}$-norm.

In this study, we investigate the sharp bounds of $\ell_{\beta}$-norm with a fixed $\ell_{\alpha}$-norm for $n$-dimensional probability vectors, as shown in \thref{th:extremes} of \sectref{sect:unconditional}.
Note that the case $\alpha = \beta$ is omitted in the study since it is obvious.
Since the R\'{e}nyi entropy is closely related to the $\ell_{\alpha}$-norm, \thref{th:extremes} implies the sharp bounds between two R\'{e}nyi entropies of distinct orders, which is described in \thref{th:R_extremes} of \sectref{sect:unconditional}.
On the other hand, in \thref{th:convexhull} of \sectref{sect:conditional}, we show the exact feasible regions between \emph{expectations} of $\ell_{\alpha}$- and $\ell_{\beta}$-norms for probability distributions on a finite set.
Since Arimoto's conditional R\'{e}nyi entropy \cite{arimoto2} is closely related to the expectations of $\ell_{\alpha}$-norm, we can apply \thref{th:convexhull} to it as \figref{fig:convexhull} of \sectref{sect:focusing}.
As applications of the above results, \sectref{sect:focusing} investigates the sharp bounds on reliability functions, such as the mutual information of order $\alpha$ \cite{arimoto2} and the $E_{0}$ function \cite{red}, for uniformly focusing channels of \defref{def:focusing} of \sectref{sect:focusing} under the uniform input distribution.

\section{Extremality of $\ell_{\alpha}$-norm and R\'{e}nyi entropy}
\label{sect:unconditional}

In this section, the sharp bounds between the $\ell_{\alpha}$-norm and the $\ell_{\beta}$-norm, $\alpha \neq \beta$, for $n$-dimensional probability vectors are examined.
As the result, \thref{th:extremes} shows that the distributions $\bvec{v}_{n}( \cdot )$ and $\bvec{w}_{n}( \cdot )$ take extremal values of norms for $n$-dimensional probability vectors.
The following lemma shows the monotonicity of the $\ell_{\alpha}$-norms of distributions $\bvec{v}_{n}( p )$ and $\bvec{w}_{n}( p )$ with respect to $p$.

\begin{lemma}
\label{lem:Hv}
For a fixed $\alpha \in (0, 1)$, $\| \bvec{v}_{n}( p_{v} ) \|_{\alpha}$ (resp. $\| \bvec{w}_{n}( p_{w} ) \|_{\alpha}$) is strictly increasing (resp. strictly decreasing) for $p_{v} \in [0, 1/n]$ (resp. $p_{w} \in [1/n, 1]$).
Conversely, for a fixed $\alpha \in (1, \infty]$, $\| \bvec{v}_{n}( p_{v} ) \|_{\alpha}$ (resp. $\| \bvec{w}_{n}( p_{w} ) \|_{\alpha}$) is strictly decreasing (resp. strictly increasing) for $p_{v} \in [0, 1/n]$ (resp. $p_{w} \in [1/n, 1]$).
\end{lemma}


\begin{IEEEproof}[Proof of \lemref{lem:Hv}]
We first prove \lemref{lem:Hv} for $\| \bvec{v}_{n}( p ) \|_{\alpha}$.
If $\alpha = 1$, then \lemref{lem:Hv} is reduced to \cite[Lemma~1]{part1}.
In this proof, we omit the case of $\alpha = 1$ and we only consider $\| \bvec{v}_{n}( p ) \|_{\alpha}$ for $\alpha \in (0, 1) \cup (1, \infty]$.
A direct calculation shows
\begin{align}
\frac{ \partial \| \bvec{v}_{n}( p ) \|_{\alpha} }{ \partial p }
& =
\frac{ \partial }{ \partial p } \left( \vphantom{\sum} (1 - (n-1) p)^{\alpha} + (n-1) \, p^{\alpha} \right)^{1/\alpha}
\\
& =
\frac{1}{\alpha} \left( \vphantom{\sum} (1 - (n-1) p)^{\alpha} + (n-1) \, p^{\alpha} \right)^{(1/\alpha)-1} \left( \frac{ \partial }{ \partial p } \left( \vphantom{\sum} (1 - (n-1) p)^{\alpha} + (n-1) \, p^{\alpha} \right) \right)
\\
& =
\frac{1}{\alpha} \left( \vphantom{\sum} (1 - (n-1) p)^{\alpha} + (n-1) \, p^{\alpha} \right)^{(1/\alpha)-1} \left( \vphantom{\sum} - \alpha (n-1) (1 - (n-1) p)^{\alpha-1} + \alpha (n-1) \, p^{\alpha-1} \right)
\\
& =
(n-1) \left( \vphantom{\sum} (1 - (n-1) p)^{\alpha} + (n-1) \, p^{\alpha} \right)^{(1/\alpha)-1} \left( \vphantom{\sum} p^{\alpha-1} - (1 - (n-1) p)^{\alpha-1} \right)
\label{eq:diff1_norm_v}
\end{align}
for $p \in (0, 1/(n-1))$.
Let
\begin{align}
\sgn ( x )
\coloneqq
\begin{cases}
1
& \mathrm{if} \ x > 0 , \\
0
& \mathrm{if} \ x = 0 , \\
-1
& \mathrm{if} \ x < 0
\end{cases}
\end{align}
denote the sign function of $x \in \mathbb{R}$.
Since
\begin{align}
\sgn \left( \vphantom{\sum} p^{\alpha-1} - (1 - (n-1) p)^{\alpha-1} \right)
& =
\begin{cases}
1
& \mathrm{if} \ \alpha < 1 , \\
0
& \mathrm{if} \ \alpha = 1 , \\
-1
& \mathrm{if} \ \alpha > 1
\end{cases}
\label{eq:sgn_factor_diff_norm_v}
\end{align}
for $p \in (0, 1/n)$, we obtain
\begin{align}
\sgn \left( \frac{ \partial \| \bvec{v}_{n}( p ) \|_{\alpha} }{ \partial p } \right)
& \overset{\eqref{eq:diff1_norm_v}}{=}
\sgn \left( (n-1) \left( \vphantom{\sum} (1 - (n-1) p)^{\alpha} + (n-1) \, p^{\alpha} \right)^{(1/\alpha)-1} \left( \vphantom{\sum} p^{\alpha-1} - (1 - (n-1) p)^{\alpha-1} \right) \right)
\\
& =
\underbrace{ \sgn (n-1) }_{=1} \cdot \underbrace{ \sgn \left( \left( \vphantom{\sum} (1 - (n-1) p)^{\alpha} + (n-1) \, p^{\alpha} \right)^{(1/\alpha)-1} \right) }_{=1} \cdot \sgn \left( \vphantom{\sum} p^{\alpha-1} - (1 - (n-1) p)^{\alpha-1} \right)
\\
& \overset{\eqref{eq:sgn_factor_diff_norm_v}}{=}
\begin{cases}
1
& \mathrm{if} \ \alpha < 1 , \\
0
& \mathrm{if} \ \alpha = 1 , \\
-1
& \mathrm{if} \ \alpha > 1
\end{cases}
\end{align}
for $p \in (0, 1/n)$, which implies that
\begin{itemize}
\item
if $\alpha \in (0, 1)$, then $\| \bvec{v}_{n}( p ) \|_{\alpha}$ is strictly increasing for $p \in [0, 1/n]$, and
\item
if $\alpha \in (1, \infty)$, then $\| \bvec{v}_{n}( p ) \|_{\alpha}$ is strictly decreasing for $p \in [0, 1/n]$.
\end{itemize}
Moreover, since $\| \bvec{v}_{n}( p ) \|_{\infty} = 1 - (n-1) p$, the $\ell_{\infty}$-norm $\| \bvec{v}_{n}( p ) \|_{\infty}$ is also strictly decreasing for $p \in [0, 1/n]$.
%

We next prove \lemref{lem:Hv} for $\| \bvec{w}_{n}( p ) \|_{\alpha}$.
In this part, we also omit the case of $\alpha = 1$ and we only consider $\| \bvec{w}_{n}( p ) \|_{\alpha}$ for $\alpha \in (0, 1) \cup (1, \infty]$.
For an integer $m \in [2, n]$, we readily see that
\begin{align}
\| \bvec{w}_{n}( p ) \|_{\alpha}
=
\| \bvec{v}_{m}( p ) \|_{\alpha}
\end{align}
for $p \in [1/m, 1/(m-1)]$;
hence, we get
\begin{align}
\frac{ \partial \| \bvec{w}_{n}( p ) \|_{\alpha} }{ \partial p }
& \overset{\eqref{eq:diff1_norm_v}}{=}
(m-1) \left( \vphantom{\sum} (1 - (m-1) p)^{\alpha} + (m-1) \, p^{\alpha} \right)^{(1/\alpha)-1} \left( \vphantom{\sum} p^{\alpha-1} - (1 - (m-1) p)^{\alpha-1} \right)
\label{eq:diff1_norm_w}
\end{align}
for $p \in (1/m, 1/(m-1))$.
Since
\begin{align}
\sgn \left( \vphantom{\sum} p^{\alpha-1} - (1 - (m-1) p)^{\alpha-1} \right)
& =
\begin{cases}
1
& \mathrm{if} \ \alpha > 1 , \\
0
& \mathrm{if} \ \alpha = 1 , \\
-1
& \mathrm{if} \ \alpha < 1
\end{cases}
\label{eq:sgn_factor_diff_norm_w}
\end{align}
for $p \in (1/m, 1/(m-1))$, we obtain
\begin{align}
\sgn \left( \frac{ \partial \| \bvec{w}_{n}( p ) \|_{\alpha} }{ \partial p } \right)
& \overset{\eqref{eq:diff1_norm_w}}{=}
\sgn \left( (m-1) \left( \vphantom{\sum} (1 - (m-1) p)^{\alpha} + (m-1) \, p^{\alpha} \right)^{(1/\alpha)-1} \left( \vphantom{\sum} p^{\alpha-1} - (1 - (m-1) p)^{\alpha-1} \right) \right)
\\
& =
\underbrace{ \sgn (m-1) }_{ = 1 } \cdot \underbrace{ \sgn \left( \left( \vphantom{\sum} (1 - (m-1) p)^{\alpha} + (m-1) \, p^{\alpha} \right)^{(1/\alpha)-1} \right) }_{=1} \cdot \sgn \left( \vphantom{\sum} p^{\alpha-1} - (1 - (m-1) p)^{\alpha-1} \right)
\\
& \overset{\eqref{eq:sgn_factor_diff_norm_w}}{=}
\begin{cases}
1
& \mathrm{if} \ \alpha > 1 , \\
0
& \mathrm{if} \ \alpha = 1 , \\
-1
& \mathrm{if} \ \alpha < 1
\end{cases}
\label{eq:sgn_diff1_norm_w}
\end{align}
for $p \in (1/m, 1/(m-1))$.
Since
\begin{align}
\| \bvec{w}_{n}( 1/n ) \|_{\alpha}
& =
\bigg( \bigg\lfloor \frac{1}{p} \bigg\rfloor p^{\alpha} + \bigg( 1 - \bigg\lfloor \frac{1}{p} \bigg\rfloor p \bigg)^{\alpha} \bigg)^{1/\alpha} \Bigg|_{p = 1/n}
\\
& =
\left( n^{1-\alpha} + 0^{\alpha} \right)^{1/\alpha}
\\
& =
n^{(1/\alpha)-1} ,
\\
\lim_{p \to (1/m)^{-}} \| \bvec{w}_{n}( p ) \|_{\alpha}
& =
\lim_{p \to (1/m)^{-}} \bigg( \bigg\lfloor \frac{1}{p} \bigg\rfloor p^{\alpha} + \bigg( 1 - \bigg\lfloor \frac{1}{p} \bigg\rfloor p \bigg)^{\alpha} \bigg)^{1/\alpha}
\\
& =
\lim_{p \to (1/m)^{-}} \left( \vphantom{\sum} m \, p^{\alpha} + \left( 1 - m \, p \right)^{\alpha} \right)^{1/\alpha}
\\
& =
\left( \vphantom{\sum} m^{1-\alpha} + 0^{\alpha} \right)^{1/\alpha}
\\
& =
m^{(1/\alpha)-1} ,
\\
\lim_{p \to (1/m)^{+}} \| \bvec{w}_{n}( p ) \|_{\alpha}
& =
\lim_{p \to (1/m)^{+}} \bigg( \bigg\lfloor \frac{1}{p} \bigg\rfloor p^{\alpha} + \bigg( 1 - \bigg\lfloor \frac{1}{p} \bigg\rfloor p \bigg)^{\alpha} \bigg)^{1/\alpha}
\\
& =
\lim_{p \to (1/m)^{+}} \left( \vphantom{\sum} (m-1) \, p^{\alpha} + \left( 1 - (m-1) \, p \right)^{\alpha} \right)^{1/\alpha}
\\
& =
\left( \vphantom{\sum} (m-1) \, m^{-\alpha} + m^{-\alpha} \right)^{1/\alpha}
\\
& =
\left( \vphantom{\sum} m^{1-\alpha} \right)^{1/\alpha}
\\
& =
m^{(1/\alpha)-1} ,
\\
\| \bvec{w}_{n}( 1 ) \|_{\alpha}
& =
\bigg( \bigg\lfloor \frac{1}{p} \bigg\rfloor p^{\alpha} + \bigg( 1 - \bigg\lfloor \frac{1}{p} \bigg\rfloor p \bigg)^{\alpha} \bigg)^{1/\alpha} \Bigg|_{p = 1}
\\
& =
\left( \vphantom{\sum} 1^{\alpha} + 0^{\alpha} \right)^{1/\alpha}
\\
& =
1
\end{align}
for $\alpha \in (0, \infty)$ and an integer $m \in [2, n-1]$, the $\ell_{\alpha}$-norm $\| \bvec{w}_{n}( p ) \|_{\alpha}$ is continuous for $p \in [1/n, 1]$.
Thus, it follows from \eqref{eq:sgn_diff1_norm_w} that
\begin{itemize}
\item
if $\alpha \in (0, 1)$, then $\| \bvec{w}_{n}( p ) \|_{\alpha}$ is strictly decreasing for $p \in [1/n, 1]$, and
\item
if $\alpha \in (1, \infty)$, then $\| \bvec{w}_{n}( p ) \|_{\alpha}$ is strictly increasing for $p \in [1/n, 1]$.
\end{itemize}
Moreover, since $\| \bvec{w}_{n}( p ) \|_{\infty} = p$, the $\ell_{\infty}$-norm $\| \bvec{w}_{n}( p ) \|_{\infty}$ is also strictly increasing for $p \in [1/n, 1]$.
%
\end{IEEEproof}

Since the strictly monotonic function has an inverse function, \lemref{lem:Hv} implies that the distributions $\bvec{v}_{n}( \cdot )$ and $\bvec{w}_{n}( \cdot )$ are uniquely determined by a given $\ell_{\alpha}$-norm. 
By using distributions $\bvec{v}_{n}( \cdot )$ and $\bvec{w}_{n}( \cdot )$, we show extremal properties of $\ell_{\alpha}$-norm of $n$-dimensional probability vectors, as shown in Lemmas~\ref{lem:vector_v}~and~\ref{lem:vector_w}.

\begin{lemma}
\label{lem:vector_v}
For any $n \ge 2$, $\bvec{p} \in \Delta_{n}$, and $\alpha \in (0, 1) \cup (1, \infty]$, there exists $p \in [0, 1/n]$ such that 
\begin{align}
\| \bvec{v}_{n}( p ) \|_{\alpha}
& =
\| \bvec{p} \|_{\alpha} ,
\\
\| \bvec{v}_{n}( p ) \|_{\beta}
& \le
\| \bvec{p} \|_{\beta}
\quad \mathrm{for} \ \mathrm{all} \ \beta \in (\min\{ 1, \alpha \}, \max\{ 1, \alpha \}) ,
\\
\| \bvec{v}_{n}( p ) \|_{\beta}
& \ge
\| \bvec{p} \|_{\beta}
\quad \mathrm{for} \ \mathrm{all} \ \beta \in (0, \min\{ 1, \alpha \}) \cup (\max\{ 1, \alpha \}, \infty] .
\end{align}
\end{lemma}

\begin{IEEEproof}[Proof of \lemref{lem:vector_v}]
Following the notation used in \cite{marshall}, for a probability vector $\bvec{p} = (p_{1}, p_{2}, \dots, p_{n}) \in \Delta_{n}$, let
\begin{align}
p_{[1]} \ge p_{[2]} \ge \dots \ge p_{[n]}
\end{align}
denote the components of $\bvec{p}$ in decreasing order; and let
\begin{align}
\bvec{p}_{\downarrow}
\coloneqq
(p_{[1]}, p_{[2]}, \dots, p_{[n]})
\label{def:rearrangement}
\end{align}
denote the decreasing rearrangement of $\bvec{p}$.
Note that
\begin{align}
1 & \le \| \bvec{p} \|_{\alpha} \le n^{(1/\alpha) - 1}
&& \mathrm{for} \ 0 < \alpha \le 1 ,
\\
n^{(1/\alpha) - 1} & \le \| \bvec{p} \|_{\alpha} \le 1
&& \mathrm{for} \ 1 \le \alpha \le \infty
\label{ineq:norm_alpha_ge_1}
\end{align}
for $\bvec{p} \in \Delta_{n}$, where it is assumed in \eqref{ineq:norm_alpha_ge_1} that $n^{(1/\alpha) - 1} = 1/n$ when $\alpha = \infty$.

We first consider the obvious case of \lemref{lem:vector_v} such that $n = 2$, $\| \bvec{p} \|_{\alpha} = 1$, or $\| \bvec{p} \|_{\alpha} = n^{(1/\alpha) - 1}$.
It can be easily seen that, for any $\bvec{p} = (p_{1}, p_{2}) \in \Delta_{2}$, there exists $p \in [0, 1/2]$ such that $\bvec{p}_{\downarrow} = \bvec{v}_{2}( p )$;
therefore, the lemma obviously holds when $n = 2$.

It follows from Minkowski's inequality (see, e.g., the proof of \cite[Theorem~3]{boekee}) that
\begin{itemize}
\item
$\| \bvec{p} \|_{\alpha}$ is strictly concave in $\bvec{p} \in \Delta_{n}$ for a fixed $\alpha \in (0, 1)$ and
\item
$\| \bvec{p} \|_{\alpha}$ is strictly convex in $\bvec{p} \in \Delta_{n}$ for a fixed $\alpha \in (1, \infty)$.
\end{itemize}
In addition, since
\begin{align}
\| \lambda \bvec{p} + (1-\lambda) \bvec{q} \|_{\infty}
& =
\max_{1 \le i \le n} \{ \lambda p_{i} + (1 - \lambda) q_{i} \}
\\
& \le
\lambda \Big( \max_{1 \le i \le n} p_{i} \Big) + (1-\lambda) \Big( \max_{1 \le i \le n} q_{i} \Big)
\label{convexity:infty_norm} \\
& =
\lambda \| \bvec{p} \|_{\infty} + (1-\lambda) \| \bvec{q} \|_{\infty}
\end{align}
for $\bvec{p}, \bvec{q} \in \Delta_{n}$ and $\lambda \in [0, 1]$, we also get that $\| \bvec{p} \|_{\infty}$ is convex in $\bvec{p} \in \Delta_{n}$.
Note that the convexity of $\| \bvec{p} \|_{\infty}$ is not strict since the inequality of \eqref{convexity:infty_norm} holds with equality if
\begin{align}
\argmax_{1 \le i \le n} p_{i}
=
\argmax_{1 \le i \le n} q_{i} .
\end{align}
The above convexity and concavity imply that
\begin{align}
\forall \alpha \in (0, \infty], \quad \| \bvec{p} \|_{\alpha} & = n^{(1/\alpha)-1}
& \iff &&
\bvec{p} & = \bigg( \frac{1}{n}, \frac{1}{n}, \dots, \frac{1}{n} \bigg)  = \bvec{v}_{n} \bigg( \frac{1}{n} \bigg) ,
\label{equiv:unif}
\\
\forall \alpha \in (0, \infty], \quad \| \bvec{p} \|_{\alpha} & = 1
& \iff &&
\bvec{p}_{\downarrow} & = (1, 0, \dots, 0) = \bvec{v}_{n}( 0 ) .
\label{equiv:det}
\end{align}
Therefore, the lemma obviously holds when either $\| \bvec{p} \|_{\alpha} = 1$ or $\| \bvec{p} \|_{\alpha} = n^{(1/\alpha)-1}$.

Thus, \lemref{lem:vector_v} is proved for the cases $n = 2$ and $\| \bvec{p} \|_{\alpha} \in \{ 1, n^{(1/\alpha)-1} \}$ for $\alpha \in (0, 1) \cup (1, \infty]$.
Hence, it is enough to prove the lemma for $\bvec{p} \in \Delta_{n}$ such that $n \ge 3$ and $\| \bvec{p} \|_{\alpha} \in ( \min\{ 1, n^{(1/\alpha)-1} \}, \max\{ 1, n^{(1/\alpha)-1} \} )$ for $\alpha \in (0, 1) \cup (1, \infty]$ in the later analyses.

We now begin to prove \lemref{lem:vector_v} for $\alpha \neq \infty$ and $\beta \neq \infty$.
For fixed $n \ge 3$, $\alpha \in (0, 1) \cup (1, \infty)$, and $A \in (\min\{ 1, n^{(1/\alpha)-1} \}, \max\{ 1, n^{(1/\alpha)-1} \})$, we assume for $\bvec{p} \in \Delta_{n}$ that 
\begin{align}
\| \bvec{p} \|_{\alpha} = A .
\label{eq:fixed_beta}
\end{align}
For that $\bvec{p}$ satisfying \eqref{eq:fixed_beta}, let $k \in \{ 2, 3, \dots, n-1 \}$ be an index such that $p_{[k-1]} > p_{[k+1]} = p_{[n]}$;
namely, the index $k$ is chosen to satisfy the following inequalities:
\begin{align}
p_{[1]} \ge p_{[2]} \ge \dots \ge p_{[k-1]} \ge p_{[k]} \ge p_{[k+1]} = p_{[k+2]} = \dots = p_{[n]}
\qquad (p_{[k-1]} > p_{[k+1]}) .
\label{eq:equal_k+1_to_n}
\end{align}
Since
$
p_{1} + p_{2} + \dots + p_{n} = 1
$,
we observe that
\begin{align}
&&
\sum_{i=1}^{n} p_{i}
& =
1
\\
& \ \Longrightarrow \ &
\frac{ \mathrm{d} }{ \mathrm{d} p_{[k]} } \left( \sum_{i=1}^{n} p_{i} \right)
& =
\frac{ \mathrm{d} }{ \mathrm{d} p_{[k]} } (1)
\\
& \iff &
\frac{ \mathrm{d} }{ \mathrm{d} p_{[k]} } \left( \sum_{i=1}^{n} p_{[i]} \right)
& =
0
\\
& \iff &
\frac{ \mathrm{d} p_{[k]} }{ \mathrm{d} p_{[k]} }
+
\sum_{i=1 : i \neq k}^{n} \frac{ \mathrm{d} p_{[i]} }{ \mathrm{d} p_{[k]} }
& =
0
\\
& \iff &
\sum_{i=1 : i \neq k}^{n} \frac{ \mathrm{d} p_{[i]} }{ \mathrm{d} p_{[k]} }
& =
-1 .
\label{eq:total_diff_prob}
\end{align}
In this proof, we assume that
\begin{align}
\frac{ \mathrm{d} p_{[i]} }{ \mathrm{d} p_{[k]} }
& =
0
\label{eq:hypo1}
\end{align}
for $i \in \{ 2, 3, \dots, k-1 \}$ and
\begin{align}
\frac{ \mathrm{d} p_{[j]} }{ \mathrm{d} p_{[k]} }
& =
\frac{ \mathrm{d} p_{[n]} }{ \mathrm{d} p_{[k]} }
\label{eq:hypo2}
\end{align}
for $j \in \{ k+1, k+2, \dots, n \}$.
By constraints \eqref{eq:hypo1} and \eqref{eq:hypo2}, we get
\begin{align}
&&
\sum_{i=1}^{n} p_{i}
& =
1
\\
& \ \overset{\eqref{eq:total_diff_prob}}{\Longrightarrow} \ &
\sum_{i = 1 : i \neq k}^{n} \frac{ \mathrm{d} p_{[i]} }{ \mathrm{d} p_{[k]} }
& =
- 1
\\
& \iff &
\sum_{i = 1}^{k-1} \frac{ \mathrm{d} p_{[i]} }{ \mathrm{d} p_{[k]} } + \sum_{j = k+1}^{n} \frac{ \mathrm{d} p_{[j]} }{ \mathrm{d} p_{[k]} }
& =
- 1
\\
& \overset{\eqref{eq:hypo1}}{\iff} &
\frac{ \mathrm{d} p_{[1]} }{ \mathrm{d} p_{[k]} } + \sum_{j = k+1}^{n} \frac{ \mathrm{d} p_{[j]} }{ \mathrm{d} p_{[k]} }
& =
-1
\\
& \overset{\eqref{eq:hypo2}}{\iff} &
\frac{ \mathrm{d} p_{[1]} }{ \mathrm{d} p_{[k]} } + (n-k) \frac{ \mathrm{d} p_{[n]} }{ \mathrm{d} p_{[k]} }
& =
-1
\\
& \iff &
\frac{ \mathrm{d} p_{[1]} }{ \mathrm{d} p_{[k]} }
& =
- 1 - (n-k) \frac{ \mathrm{d} p_{[n]} }{ \mathrm{d} p_{[k]} } .
\label{eq:total_prob_hypo}
\end{align}
Defining
the $\alpha$-logarithm function \cite{tsallis2} as
\begin{align}
\ln_{\alpha} x
\coloneqq
\begin{cases}
\ln x
& \mathrm{if} \ \alpha = 1 , \\
\dfrac{ x^{1-\alpha} - 1 }{ 1 - \alpha }
& \mathrm{if} \ \alpha \neq 1
\end{cases}
\label{def:qlog}
\end{align}
for $x > 0$, where note that L'H\^{o}pital's rule shows
\begin{align}
\lim_{\alpha \to 1} \frac{ x^{1-\alpha} - 1 }{ 1 - \alpha }
=
\ln x ,
\end{align}
we observe from the constraint \eqref{eq:fixed_beta} that
\begin{align}
&&
\| \bvec{p} \|_{\alpha}
& =
A
\\
& \iff &
\sum_{i=1}^{n} p_{i}^{\alpha}
& =
A^{\alpha}
\\
& \ \Longrightarrow \ &
\frac{ \mathrm{d} }{ \mathrm{d} p_{[k]} } \left( \sum_{i=1}^{n} p_{i}^{\alpha} \right)
& =
\frac{ \mathrm{d} }{ \mathrm{d} p_{[k]} } (A^{\alpha})
\\
& \iff &
\frac{ \mathrm{d} }{ \mathrm{d} p_{[k]} } \left( \sum_{i=1}^{n} p_{[i]}^{\alpha} \right)
& =
0
\\
& \iff &
\sum_{i=1}^{n} \frac{ \mathrm{d} }{ \mathrm{d} p_{[k]} } (p_{[i]}^{\alpha})
& =
0
\\
& \iff &
\frac{ \mathrm{d} }{ \mathrm{d} p_{[k]} } (p_{[k]}^{\alpha}) + \sum_{i = 1 : i \neq k}^{n} \frac{ \mathrm{d} }{ \mathrm{d} p_{[k]} } (p_{[i]}^{\alpha})
& =
0
\\
& \iff &
\alpha \, p_{[k]}^{\alpha-1} + \sum_{i = 1 : i \neq k}^{n} \frac{ \mathrm{d} }{ \mathrm{d} p_{[k]} } (p_{[i]}^{\alpha})
& =
0
\\
& \iff &
\sum_{i = 1 : i \neq k}^{n} \frac{ \mathrm{d} }{ \mathrm{d} p_{[k]} } (p_{[i]}^{\alpha})
& =
- \alpha \, p_{[k]}^{\alpha-1}
\\
& \overset{\text{(a)}}{\iff} &
\sum_{i = 1 : i \neq k}^{n} \left( \frac{ \mathrm{d} p_{[i]} }{ \mathrm{d} p_{[k]} } \right) \left( \frac{ \mathrm{d} }{ \mathrm{d} p_{[i]} } (p_{[i]}^{\alpha}) \right)
& =
- \alpha \, p_{[k]}^{\alpha-1}
\\
& \iff &
\sum_{i = 1 : i \neq k}^{n} \left( \frac{ \mathrm{d} p_{[i]} }{ \mathrm{d} p_{[k]} } \right) (\alpha \, p_{[i]}^{\alpha-1})
& =
- \alpha \, p_{[k]}^{\alpha-1}
\label{eq:diff1_beta_halfway} \\
& \iff &
\sum_{i = 1 : i \neq k}^{n} \left( \frac{ \mathrm{d} p_{[i]} }{ \mathrm{d} p_{[k]} } \right) p_{[i]}^{\alpha-1}
& =
- p_{[k]}^{\alpha-1}
\label{eq:diff1_beta_halfway_2} \\
& \iff &
\sum_{i = 1}^{k-1} \left( \frac{ \mathrm{d} p_{[i]} }{ \mathrm{d} p_{[k]} } \right) p_{[i]}^{\alpha-1} + \sum_{j = k+1}^{n} \left( \frac{ \mathrm{d} p_{[j]} }{ \mathrm{d} p_{[k]} } \right) p_{[j]}^{\alpha-1}
& =
- p_{[k]}^{\alpha-1}
\\
& \overset{\eqref{eq:equal_k+1_to_n}}{\iff} &
\sum_{i = 1}^{k-1} \left( \frac{ \mathrm{d} p_{[i]} }{ \mathrm{d} p_{[k]} } \right) p_{[i]}^{\alpha-1} + p_{[n]}^{\alpha-1} \sum_{j = k+1}^{n} \left( \frac{ \mathrm{d} p_{[j]} }{ \mathrm{d} p_{[k]} } \right)
& =
- p_{[k]}^{\alpha-1}
\\
& \overset{\eqref{eq:hypo1}}{\iff} &
\left( \frac{ \mathrm{d} p_{[1]} }{ \mathrm{d} p_{[k]} } \right) p_{[1]}^{\alpha-1} + p_{[n]}^{\alpha-1} \sum_{j = k+1}^{n} \left( \frac{ \mathrm{d} p_{[j]} }{ \mathrm{d} p_{[k]} } \right)
& =
- p_{[k]}^{\alpha-1}
\\
& \overset{\eqref{eq:hypo2}}{\iff} &
\left( \frac{ \mathrm{d} p_{[1]} }{ \mathrm{d} p_{[k]} } \right) p_{[1]}^{\alpha-1} + p_{[n]}^{\alpha-1} (n-k) \left( \frac{ \mathrm{d} p_{[n]} }{ \mathrm{d} p_{[k]} } \right)
& =
- p_{[k]}^{\alpha-1}
\\
& \overset{\eqref{eq:total_prob_hypo}}{\iff} &
\left( - 1 - (n-k) \frac{ \mathrm{d} p_{[n]} }{ \mathrm{d} p_{[k]} } \right) p_{[1]}^{\alpha-1} + (n-k) \left( \frac{ \mathrm{d} p_{[n]} }{ \mathrm{d} p_{[k]} } \right) p_{[n]}^{\alpha-1}
& =
- p_{[k]}^{\alpha-1}
\\
& \iff &
- p_{[1]}^{\alpha-1} - (n-k) \left( \frac{ \mathrm{d} p_{[n]} }{ \mathrm{d} p_{[k]} } \right) p_{[1]}^{\alpha-1} + (n-k) \left( \frac{ \mathrm{d} p_{[n]} }{ \mathrm{d} p_{[k]} } \right) p_{[n]}^{\alpha-1}
& =
- p_{[k]}^{\alpha-1}
\\
& \iff &
(n-k) \left( \frac{ \mathrm{d} p_{[n]} }{ \mathrm{d} p_{[k]} } \right) \left( p_{[n]}^{\alpha-1} - p_{[1]}^{\alpha-1} \right)
& =
- \left( p_{[k]}^{\alpha-1} - p_{[1]}^{\alpha-1} \right)
\\
& \iff &
(n-k) \left( \frac{ \mathrm{d} p_{[n]} }{ \mathrm{d} p_{[k]} } \right)
& =
- \frac{ p_{[k]}^{\alpha-1} - p_{[1]}^{\alpha-1} }{ p_{[n]}^{\alpha-1} - p_{[1]}^{\alpha-1} }
\\
& \iff &
(n-k) \left( \frac{ \mathrm{d} p_{[n]} }{ \mathrm{d} p_{[k]} } \right)
& =
- \frac{ \left( \frac{ p_{[k]} }{ p_{[1]} } \right)^{\alpha-1} - 1 }{ \left( \frac{ p_{[n]} }{ p_{[1]} } \right)^{\alpha-1} - 1 }
\\
& \iff &
(n-k) \left( \frac{ \mathrm{d} p_{[n]} }{ \mathrm{d} p_{[k]} } \right)
& =
- \frac{ \left( \frac{ p_{[1]} }{ p_{[k]} } \right)^{1-\alpha} - 1 }{ \left( \frac{ p_{[1]} }{ p_{[n]} } \right)^{1-\alpha} - 1 }
\\
& \overset{\eqref{def:qlog}}{\iff} &
(n-k) \left( \frac{ \mathrm{d} p_{[n]} }{ \mathrm{d} p_{[k]} } \right)
& =
- \frac{ \ln_{\alpha} \left( \frac{ p_{[1]} }{ p_{[k]} } \right) }{ \ln_{\alpha} \left( \frac{ p_{[1]} }{ p_{[n]} } \right) }
\\
& \iff &
\frac{ \mathrm{d} p_{[n]} }{ \mathrm{d} p_{[k]} }
& =
- \frac{ 1 }{ n-k } \left( \frac{ \ln_{\alpha} \left( \frac{ p_{[1]} }{ p_{[k]} } \right) }{ \ln_{\alpha} \left( \frac{ p_{[1]} }{ p_{[n]} } \right) } \right)
\label{eq:total_norm_hypo}
\end{align}
where (a) follows by the chain rule of the derivative.

We now check the sign of the right-hand side of \eqref{eq:total_norm_hypo}.
Since $\ln_{\alpha} x$ is a strictly increasing function of $x > 0$ for every $\alpha \in \mathbb{R}$, we get
\begin{align}
0
\le
\ln_{\alpha} \left( \frac{ p_{[1]} }{ p_{[k]} } \right)
<
\ln_{\alpha} \left( \frac{ p_{[1]} }{ p_{[n]} } \right)
\label{ineq:qlog_frac_p1_pk_pn}
\end{align}
for $1 > p_{[1]} \ge p_{[k]} \ge p_{[n]} > 0 \ (p_{[1]} > p_{[n]})$, where the left-hand inequality holds with equality if and only if $p_{[1]} = p_{[k]}$.
Note that the monotonicity of $\alpha$-logarithm function can be verified as follows:
For $\alpha \in \mathbb{R} \setminus \{ 1 \}$, a simple calculation yields
\begin{align}
\frac{ \partial \ln_{\alpha} x }{ \partial x }
& =
\frac{ \partial }{ \partial x } \left( \frac{ x^{1-\alpha} - 1 }{ 1 - \alpha } \right)
\\
& =
\frac{ 1 }{ 1 - \alpha } \left( \frac{ \partial }{ \partial x } ( x^{1-\alpha} ) \right)
\\
& =
\frac{ 1 }{ 1 - \alpha } \left( \vphantom{\sum} (1-\alpha) x^{-\alpha} \right)
\\
& =
x^{-\alpha}
\\
& >
0
\label{eq:diff_qlog}
\end{align}
for $x > 0$, which implies that $\ln_{\alpha} x$ is a strictly increasing function of $x > 0$ for every $\alpha \in \mathbb{R} \setminus \{ 1 \}$;
on the other hand, if $\alpha = 1$, then $\ln_{1} x = \ln x$ is clearly a strictly increasing function of $x > 0$.
It follows from \eqref{ineq:qlog_frac_p1_pk_pn} that
\begin{align}
0 \le \frac{ \ln_{\alpha} \left( \frac{ p_{[1]} }{ p_{[k]} } \right) }{ \ln_{\alpha} \left( \frac{ p_{[1]} }{ p_{[n]} } \right) } < 1 ;
\label{ineq:qlog_beta_frac}
\end{align}
and therefore, we obtain
\begin{align}
\sgn \left( \frac{ \mathrm{d} p_{[n]} }{ \mathrm{d} p_{[k]} } \right)
=
\begin{cases}
0
& \mathrm{if} \ p_{[1]} = p_{[k]} , \\
-1
& \mathrm{otherwise} ,
\end{cases}
\label{eq:sign_d1dk}
\end{align}
which implies that $p_{[n]}$ is strictly decreasing for $p_{[k]}$ under the constraints \eqref{eq:fixed_beta}, \eqref{eq:equal_k+1_to_n}, \eqref{eq:hypo1}, and \eqref{eq:hypo2}.

Similarly, we check the sign of the right-hand side of \eqref{eq:total_prob_hypo}:
\begin{align}
\frac{ \mathrm{d} p_{[1]} }{ \mathrm{d} p_{[k]} }
& =
- 1 - (n-k) \frac{ \mathrm{d} p_{[n]} }{ \mathrm{d} p_{[k]} }
\\
& \overset{\eqref{eq:total_norm_hypo}}{=}
- 1 + \frac{ \ln_{\alpha} \left( \frac{ p_{[1]} }{ p_{[k]} } \right) }{ \ln_{\alpha} \left( \frac{ p_{[1]} }{ p_{[n]} } \right) } .
\end{align}
It follows from \eqref{ineq:qlog_beta_frac} that
\begin{align}
-1 \le \frac{ \mathrm{d} p_{[1]} }{ \mathrm{d} p_{[k]} } < 0 ,
\label{eq:sign_dndk}
\end{align}
which implies that $p_{[1]}$ is also strictly decreasing for $p_{[k]}$ under the constraints \eqref{eq:fixed_beta}, \eqref{eq:equal_k+1_to_n}, \eqref{eq:hypo1}, and \eqref{eq:hypo2}.

We next consider $\| \bvec{p} \|_{\beta}$ for a fixed $\beta \in (0, 1) \cup (1, \infty)$.
A direct calculation yields
\begin{align}
\frac{ \mathrm{d} \| \bvec{p} \|_{\beta}^{\beta} }{ \mathrm{d} p_{[k]} }
& =
\frac{ \mathrm{d} }{ \mathrm{d} p_{[k]} } \left( \sum_{i=1}^{n} p_{i}^{\beta} \right)
\\
& \overset{\eqref{eq:diff1_beta_halfway}}{=}
\beta \left( p_{[k]}^{\beta-1} + \sum_{i=1 : i \neq k}^{n} \left( \frac{ \mathrm{d} p_{[i]} }{ \mathrm{d} p_{[k]} } \right) p_{[i]}^{\beta-1} \right)
\\
& =
\beta \left( p_{[k]}^{\beta-1} + \sum_{i=1}^{k-1} \left( \frac{ \mathrm{d} p_{[i]} }{ \mathrm{d} p_{[k]} } \right) p_{[i]}^{\beta-1} + \sum_{j=k+1}^{n} \left( \frac{ \mathrm{d} p_{[j]} }{ \mathrm{d} p_{[k]} } \right) p_{[j]}^{\beta-1} \right)
\\
& \overset{\eqref{eq:equal_k+1_to_n}}{=}
\beta \left( p_{[k]}^{\beta-1} + \sum_{i=1}^{k-1} \left( \frac{ \mathrm{d} p_{[i]} }{ \mathrm{d} p_{[k]} } \right) p_{[i]}^{\beta-1} + p_{[n]}^{\beta-1} \sum_{j=k+1}^{n} \left( \frac{ \mathrm{d} p_{[j]} }{ \mathrm{d} p_{[k]} } \right) \right)
\\
& \overset{\eqref{eq:hypo1}}{=}
\beta \left( p_{[k]}^{\beta-1} + \left( \frac{ \mathrm{d} p_{[1]} }{ \mathrm{d} p_{[k]} } \right) p_{[1]}^{\beta-1} + p_{[n]}^{\beta-1} \sum_{j=k+1}^{n} \left( \frac{ \mathrm{d} p_{[j]} }{ \mathrm{d} p_{[k]} } \right) \right)
\\
& \overset{\eqref{eq:hypo2}}{=}
\beta \left( p_{[k]}^{\beta-1} + \left( \frac{ \mathrm{d} p_{[1]} }{ \mathrm{d} p_{[k]} } \right) p_{[1]}^{\beta-1} + p_{[n]}^{\beta-1} (n-k) \left( \frac{ \mathrm{d} p_{[n]} }{ \mathrm{d} p_{[k]} } \right) \right)
\\
& \overset{\eqref{eq:total_prob_hypo}}{=}
\beta \left( p_{[k]}^{\beta-1} + \left( - 1 - (n-k) \frac{ \mathrm{d} p_{[n]} }{ \mathrm{d} p_{[k]} } \right) p_{[1]}^{\beta-1} + (n-k) \left( \frac{ \mathrm{d} p_{[n]} }{ \mathrm{d} p_{[k]} } \right) p_{[n]}^{\beta-1} \right)
\\
& =
\beta \left( p_{[k]}^{\beta-1} - p_{[1]}^{\beta-1} - (n-k) \left( \frac{ \mathrm{d} p_{[n]} }{ \mathrm{d} p_{[k]} } \right) p_{[1]}^{\beta-1} + (n-k) \left( \frac{ \mathrm{d} p_{[n]} }{ \mathrm{d} p_{[k]} } \right) p_{[n]}^{\beta-1} \right)
\\
& =
\beta \left( (p_{[k]}^{\beta-1} - p_{[1]}^{\beta-1}) + (n-k) \left( \frac{ \mathrm{d} p_{[n]} }{ \mathrm{d} p_{[k]} } \right) (p_{[n]}^{\beta-1} - p_{[1]}^{\beta-1}) \right)
\\
& \overset{\eqref{eq:total_norm_hypo}}{=}
\beta \left( (p_{[k]}^{\beta-1} - p_{[1]}^{\beta-1}) + (n-k) \left( - \frac{ 1 }{ n-k } \left( \frac{ \ln_{\alpha} \left( \frac{ p_{[1]} }{ p_{[k]} } \right) }{ \ln_{\alpha} \left( \frac{ p_{[1]} }{ p_{[n]} } \right) } \right) \right) (p_{[n]}^{\beta-1} - p_{[1]}^{\beta-1}) \right)
\\
& =
\beta \left( (p_{[k]}^{\beta-1} - p_{[1]}^{\beta-1}) - \left( \frac{ \ln_{\alpha} \left( \frac{ p_{[1]} }{ p_{[k]} } \right) }{ \ln_{\alpha} \left( \frac{ p_{[1]} }{ p_{[n]} } \right) } \right) (p_{[n]}^{\beta-1} - p_{[1]}^{\beta-1}) \right)
\\
& =
\beta \left( \vphantom{\sum} p_{[n]}^{\beta-1} - p_{[1]}^{\beta-1} \right) \left( \frac{ p_{[k]}^{\beta-1} - p_{[1]}^{\beta-1} }{ p_{[n]}^{\beta-1} - p_{[1]}^{\beta-1} } - \frac{ \ln_{\alpha} \left( \frac{ p_{[1]} }{ p_{[k]} } \right) }{ \ln_{\alpha} \left( \frac{ p_{[1]} }{ p_{[n]} } \right) } \right)
\\
& \overset{\eqref{def:qlog}}{=}
\beta \left( \vphantom{\sum} p_{[n]}^{\beta-1} - p_{[1]}^{\beta-1} \right) \left( \frac{ \ln_{\beta} \left( \frac{ p_{[1]} }{ p_{[k]} } \right) }{ \ln_{\beta} \left( \frac{ p_{[1]} }{ p_{[n]} } \right) } - \frac{ \ln_{\alpha} \left( \frac{ p_{[1]} }{ p_{[k]} } \right) }{ \ln_{\alpha} \left( \frac{ p_{[1]} }{ p_{[n]} } \right) } \right) .
\label{eq:diff1_norm_alpha_v}
\end{align}
Since $\| \bvec{p} \|_{\beta} = (\| \bvec{p} \|_{\beta}^{\beta})^{1/\beta}$, it follows by the chain rule that
\begin{align}
\frac{ \mathrm{d} \| \bvec{p} \|_{\beta} }{ \mathrm{d} p_{[k]} }
& =
\left( \frac{1}{\beta} \frac{ \| \bvec{p} \|_{\beta} }{ \| \bvec{p} \|_{\beta}^{\beta} } \right) \cdot \left( \frac{ \mathrm{d} \| \bvec{p} \|_{\beta}^{\beta} }{ \mathrm{d} p_{[k]} } \right)
\\
& \overset{\eqref{eq:diff1_norm_alpha_v}}{=}
\left( \sum_{i=1}^{n} p_{i}^{\beta} \right)^{(1/\beta)-1} \left( \vphantom{\sum} p_{[n]}^{\beta-1} - p_{[1]}^{\beta-1} \right) \left( \frac{ \ln_{\beta} \left( \frac{ p_{[1]} }{ p_{[k]} } \right) }{ \ln_{\beta} \left( \frac{ p_{[1]} }{ p_{[n]} } \right) } - \frac{ \ln_{\alpha} \left( \frac{ p_{[1]} }{ p_{[k]} } \right) }{ \ln_{\alpha} \left( \frac{ p_{[1]} }{ p_{[n]} } \right) } \right) .
\end{align}
Thus, we get
\begin{align}
\sgn \left( \frac{ \mathrm{d} \| \bvec{p} \|_{\beta} }{ \mathrm{d} p_{[k]} } \right)
& =
\sgn \left( \left( \sum_{i=1}^{n} p_{i}^{\beta} \right)^{(1/\beta)-1} \left( \vphantom{\sum} p_{[n]}^{\beta-1} - p_{[1]}^{\beta-1} \right) \left( \frac{ \ln_{\beta} \left( \frac{ p_{[1]} }{ p_{[k]} } \right) }{ \ln_{\beta} \left( \frac{ p_{[1]} }{ p_{[n]} } \right) } - \frac{ \ln_{\alpha} \left( \frac{ p_{[1]} }{ p_{[k]} } \right) }{ \ln_{\alpha} \left( \frac{ p_{[1]} }{ p_{[n]} } \right) } \right) \right)
\\
& =
\underbrace{ \sgn \left( \left( \sum_{i=1}^{n} p_{i}^{\beta} \right)^{(1/\beta)-1} \right) }_{ = 1 } \cdot \sgn \left( \vphantom{\sum} p_{[n]}^{\beta-1} - p_{[1]}^{\beta-1} \right) \cdot \sgn \left( \frac{ \ln_{\beta} \left( \frac{ p_{[1]} }{ p_{[k]} } \right) }{ \ln_{\beta} \left( \frac{ p_{[1]} }{ p_{[n]} } \right) } - \frac{ \ln_{\alpha} \left( \frac{ p_{[1]} }{ p_{[k]} } \right) }{ \ln_{\alpha} \left( \frac{ p_{[1]} }{ p_{[n]} } \right) } \right)
\\
& =
\sgn \left( \vphantom{\sum} p_{[n]}^{\beta-1} - p_{[1]}^{\beta-1} \right) \cdot \sgn \left( \frac{ \ln_{\beta} \left( \frac{ p_{[1]} }{ p_{[k]} } \right) }{ \ln_{\beta} \left( \frac{ p_{[1]} }{ p_{[n]} } \right) } - \frac{ \ln_{\alpha} \left( \frac{ p_{[1]} }{ p_{[k]} } \right) }{ \ln_{\alpha} \left( \frac{ p_{[1]} }{ p_{[n]} } \right) } \right) .
\label{eq:sgn_diff_Norm_v_factor12}
\end{align}
For $p_{[1]} > p_{[n]} > 0$, we readily see that
\begin{align}
\sgn \left( \vphantom{\sum} p_{[n]}^{\beta-1} - p_{[1]}^{\beta-1} \right)
& =
\begin{cases}
1
& \mathrm{if} \ \beta < 1 , \\
0
& \mathrm{if} \ \beta = 1 , \\
-1
& \mathrm{if} \ \beta > 1 .
\end{cases}
\label{eq:sgn_diff_Norm_v_factor1}
\end{align}
It was derived in \cite[Lemma~5]{part1} that, for $\alpha < \beta$ and $1 \le x \le y \ (y \neq 1)$,
\begin{align}
\frac{ \ln_{\alpha} x }{ \ln_{\alpha} y }
\le
\frac{ \ln_{\beta} x }{ \ln_{\beta} y }
\label{eq:frac_qlog}
\end{align}
with equality if and only if $x \in \{ 1, y \}$.
Therefore, since
\begin{align}
1 \le \frac{ p_{[1]} }{ p_{[k]} } \le \frac{ p_{[1]} }{ p_{[n]} }
\quad
\left( 1 \neq \frac{ p_{[1]} }{ p_{[n]} } \right)
\end{align}
for $1 > p_{[1]} \ge p_{[k]} \ge p_{[n]} > 0 \ (p_{[1]} > p_{[n]})$, it follows from \eqref{eq:frac_qlog} that
\begin{align}
\sgn \left( \frac{ \ln_{\beta} \left( \frac{ p_{[1]} }{ p_{[k]} } \right) }{ \ln_{\beta} \left( \frac{ p_{[1]} }{ p_{[n]} } \right) } - \frac{ \ln_{\alpha} \left( \frac{ p_{[1]} }{ p_{[k]} } \right) }{ \ln_{\alpha} \left( \frac{ p_{[1]} }{ p_{[n]} } \right) } \right)
& =
\begin{cases}
1
& \mathrm{if} \ \beta > \alpha \ \mathrm{and} \ p_{[1]} > p_{[k]} > p_{[n]} , \\
0
& \mathrm{if} \ \beta = \alpha \ \mathrm{or} \ p_{[1]} = p_{[k]} \ \mathrm{or} \ p_{[k]} = p_{[n]} , \\
-1
& \mathrm{if} \ \beta < \alpha \ \mathrm{and} \ p_{[1]} > p_{[k]} > p_{[n]} .
\end{cases}
\label{eq:sgn_diff_Norm_v_factor2}
\end{align}
Combining \eqref{eq:sgn_diff_Norm_v_factor1} and \eqref{eq:sgn_diff_Norm_v_factor2}, if $p_{[1]} > p_{[k]} > p_{[n]}$, we obtain
\begin{align}
\sgn \left( \frac{ \mathrm{d} \| \bvec{p} \|_{\beta} }{ \mathrm{d} p_{[k]} } \right)
& \overset{\eqref{eq:sgn_diff_Norm_v_factor12}}{=}
\sgn \left( \vphantom{\sum} p_{[n]}^{\beta-1} - p_{[1]}^{\beta-1} \right) \cdot \sgn \left( \frac{ \ln_{\beta} \left( \frac{ p_{[1]} }{ p_{[k]} } \right) }{ \ln_{\beta} \left( \frac{ p_{[1]} }{ p_{[n]} } \right) } - \frac{ \ln_{\alpha} \left( \frac{ p_{[1]} }{ p_{[k]} } \right) }{ \ln_{\alpha} \left( \frac{ p_{[1]} }{ p_{[n]} } \right) } \right)
\\
& =
\begin{cases}
1
& \mathrm{if} \ \beta \in (\min\{ 1, \alpha \}, \max\{ 1, \alpha \}) , \\
0
& \mathrm{if} \ \beta \in \{ 1, \alpha \} , \\
-1
& \mathrm{if} \ \beta \in (0, \min\{ 1, \alpha \}) \cup (\max\{ 1, \alpha \}, \infty) ,
\end{cases}
\label{eq:sgn_diff_Norm_v}
\end{align}
which implies that
\begin{itemize}
\item
if $\alpha \in (0, 1)$, then
\begin{itemize}
\item
$\| \bvec{p} \|_{\beta}$ with a fixed $\beta \in (\alpha, 1)$ is strictly increasing for $p_{[k]}$, and
\item
$\| \bvec{p} \|_{\beta}$ with a fixed $\beta \in (0, \alpha) \cup (1, \infty)$ is strictly decreasing for $p_{[k]}$,
\end{itemize}
\item
if $\alpha \in (1, \infty)$, then
\begin{itemize}
\item
$\| \bvec{p} \|_{\beta}$ with a fixed $\beta \in (1, \alpha)$ is strictly increasing for $p_{[k]}$, and
\item
$\| \bvec{p} \|_{\beta}$ with a fixed $\beta \in (0, 1) \cup (\alpha, \infty)$ is strictly decreasing for $p_{[k]}$.
\end{itemize}
\end{itemize}
Note that the above monotonicity hold under the constraints \eqref{eq:fixed_beta}, \eqref{eq:equal_k+1_to_n}, \eqref{eq:hypo1}, and \eqref{eq:hypo2}.

To accomplish the proof of \lemref{lem:vector_v} for $\alpha \neq \infty$ and $\beta \neq \infty$ by using the above relations, we repeat the following operation until the vector $\bvec{p} = (p_{1}, p_{2}, \dots, p_{n})$ satisfies $p_{[2]} = p_{[n]}$, i.e.,
\begin{align}
p_{[1]} > p_{[2]} = p_{[3]} = \dots = p_{[n]} > 0 ,
\label{eq:proof_vector_v}
\end{align}
which is equivalent to the vector $\bvec{v}_{n}( \cdot )$.
If the index $k \in \{ 3, 4, \dots, n-1 \}$ of constraint \eqref{eq:equal_k+1_to_n} satisfies $p_{[k-1]} > p_{[k]} = p_{[k+1]}$, then we reset the index $k$ to $k - 1$;
namely, we choose the index $k \in \{ 2, 3, \dots, n-1 \}$ to satisfy the following inequalities:
\begin{align}
p_{[1]} \ge p_{[2]} \ge \dots \ge p_{[k-1]} \ge p_{[k]} > p_{[k+1]} = p_{[k+2]} = \dots = p_{[n]} \ge 0 .
\label{eq:choose_k}
\end{align}
For that index $k$, we consider to decrease the value $p_{[k]}$ under the constraints \eqref{eq:fixed_beta}, \eqref{eq:equal_k+1_to_n}, \eqref{eq:hypo1}, and \eqref{eq:hypo2}.
It follows from \eqref{eq:sign_d1dk} that the value $p_{[1]}$ is strictly increased by according to decreasing $p_{[k]}$.
Hence, if $p_{[k]}$ is decreased, then the strict inequality $p_{[1]} > p_{[2]}$ must be held.
Similarly, it follows from \eqref{eq:hypo2} and \eqref{eq:sign_dndk} that, for all indices $j \in \{ k+1, k+2, \dots, n \}$, the value $p_{[j]}$ is also strictly increased by according to decreasing $p_{[k]}$.
Hence, if $p_{[k]}$ is decreased, then $p_{[k+1]} = p_{[k+2]} = \dots = p_{[n]} > 0$ also must be held.
Let $\bvec{q} = (q_{1}, q_{2}, \dots, q_{n})$ denote the probability vector such that made from $\bvec{p}$ by decreasing $p_{[k]}$ until the equality $p_{[k]} = p_{[k+1]}$ holds under the conditions of \eqref{eq:fixed_beta}, \eqref{eq:equal_k+1_to_n}, \eqref{eq:hypo1}, \eqref{eq:hypo2}, and \eqref{eq:choose_k}.
Namely, the vector $\bvec{q}$ satisfies the following inequalities:
\begin{align}
q_{[1]} > q_{[2]} \ge q_{[3]} \ge \dots \ge q_{[k-1]} > q_{[k]} = q_{[k+1]} = \dots = q_{[n]} > 0 .
\end{align}
Since $\bvec{q}$ is made from $\bvec{p}$ under the constraint \eqref{eq:fixed_beta}, note that
\begin{align}
\| \bvec{q} \|_{\alpha} = \| \bvec{p} \|_{\alpha}
\end{align}
for a fixed $\alpha \in (0, 1) \cup (1, \infty)$.
Moreover, it follows from \eqref{eq:sgn_diff_Norm_v} that
\begin{align}
\| \bvec{q} \|_{\beta} \le \| \bvec{p} \|_{\beta}
\quad & \mathrm{for} \ \beta \in (\min\{ 1, \alpha \}, \max\{ 1, \alpha \}) ,
\\
\| \bvec{q} \|_{\beta} \ge \| \bvec{p} \|_{\beta}
\quad & \mathrm{for} \ \beta \in (0, \min\{ 1, \alpha \}) \cup (\max\{ 1, \alpha \}, \infty) .
\end{align}
Repeating these operation until \eqref{eq:proof_vector_v} holds, we have
\begin{align}
\| \bvec{v}_{n}( p ) \|_{\alpha}
& =
\| \bvec{p} \|_{\alpha} ,
\label{eq:vector_v_proof_halfway_1} \\
\| \bvec{v}_{n}( p ) \|_{\beta}
& \le
\| \bvec{p} \|_{\beta}
\quad \mathrm{for} \ \beta \in (\min\{ 1, \alpha \}, \max\{ 1, \alpha \}) ,
\label{eq:vector_v_proof_halfway_2} \\
\| \bvec{v}_{n}( p ) \|_{\beta}
& \ge
\| \bvec{p} \|_{\beta}
\quad \mathrm{for} \ \beta \in (0, \min\{ 1, \alpha \}) \cup (\max\{ 1, \alpha \}, \infty)
\label{eq:vector_v_proof_halfway_3}
\end{align}
for any $\bvec{p} \in \Delta_{n}$, any fixed $\alpha \in (0, 1) \cup (1, \infty)$, and some $p \in [0, 1/n]$.

Finally, we consider the $\ell_{\beta}$-norm with a fixed $\ell_{\infty}$-norm for $\beta \in (0, 1) \cup (1, \infty)$, i.e., $\alpha = \infty$.
In a similar way to \cite[p.~1214]{he}, we now prove the Schur-convexity of the $\ell_{\beta}$-norm for $\beta \in (0, 1) \cup (1, \infty)$ by introducing the method of majorization \cite{marshall}.
A probability vector $\bvec{p} = (p_{1}, p_{2}, \dots, p_{n}) \in \Delta_{n}$ is said to be majorized by $\bvec{q} = (q_{1}, q_{2}, \dots, q_{n}) \in \Delta_{n}$ if
\begin{align}
\forall k = 1, 2, \dots, n-1,
\quad
\sum_{i = 1}^{k} p_{[i]}
\le
\sum_{i = 1}^{k} q_{[i]} ,
\end{align}
and we write it as $\bvec{p} \prec \bvec{q}$.
Then, a function $\phi : \Delta_{n} \to \mathbb{R}$ is said to be \emph{Schur-convex} if
\begin{align}
\bvec{p} \prec \bvec{q}
\ \Longrightarrow \
\phi( \bvec{p} ) \le \phi( \bvec{q} )
\label{def:Schur_convex}
\end{align}
for $\bvec{p}, \bvec{q} \in \Delta_{n}$, where $\phi( \cdot )$ is also said to be \emph{strictly} Schur-convex if the inequality of \eqref{def:Schur_convex} is strict.
Similarly, a function $\phi : \Delta_{n} \to \mathbb{R}$ is said to be \emph{Schur-concave} if
\begin{align}
\bvec{p} \prec \bvec{q}
\ \Longrightarrow \
\phi( \bvec{p} ) \ge \phi( \bvec{q} )
\label{def:Schur_concave}
\end{align}
for $\bvec{p}, \bvec{q} \in \Delta_{n}$, where $\phi( \cdot )$ is also said to be \emph{strictly} Schur-concave if the inequality of \eqref{def:Schur_concave} is strict.
It is known that if a function $g : [0, 1] \to \mathbb{R}$ is convex, then the function
\begin{align}
\phi( \bvec{p} )
=
\sum_{i=1}^{n} g( p_{i} )
\end{align}
is Schur-convex in $\bvec{p} \in \Delta_{n}$ (cf. \cite[p.~64]{marshall}).
Since $- \phi( \cdot )$ is Schur-concave when $\phi( \cdot )$ is Schur-convex, it also follows that if a function $g : [0, 1] \to \mathbb{R}$ is concave, then the function
\begin{align}
\phi( \bvec{p} )
=
\sum_{i=1}^{n} g( p_{i} )
\end{align}
is Schur-concave in $\bvec{p} \in \Delta_{n}$.
Thus, since $x^{\beta}$ is strictly concave in $x \ge 0$ for every $\beta \in (0, 1)$, we see that
\begin{align}
\| \bvec{p} \|_{\beta}^{\beta}
& =
\sum_{i=1}^{n} p_{i}^{\beta}
\end{align}
is Schur-concave in $\bvec{p} \in \Delta_{n}$ for every $\alpha \in (0, 1)$;
similarly, since $x^{\beta}$ is strictly convex in $x \ge 0$ for every $\beta \in (1, \infty)$, we also see that $\| \bvec{p} \|_{\beta}^{\beta}$ is strictly Schur-convex in $\bvec{p} \in \Delta_{n}$ for every $\alpha \in (1, \infty)$.
Therefore, since $x \mapsto x^{1/\beta}$ is a strictly increasing function of $x \ge 0$ for every $\beta \in (0, \infty)$, we have that $\| \bvec{p} \|_{\beta} = (\| \bvec{p} \|_{\beta}^{\beta})^{1/\beta}$ is strictly Schur-concave in $\bvec{p} \in \Delta_{n}$ for every $\beta \in (0, 1)$ and $\| \bvec{p} \|_{\beta}$ is strictly Schur-convex in $\bvec{p} \in \Delta_{n}$ for every $\beta \in (1, \infty)$.

To consider probability vectors $\bvec{p} \in \Delta_{n}$ with a fixed $\ell_{\infty}$-norm, we assume in \eqref{eq:fixed_beta} that $\alpha = \infty$, i.e,
\begin{align}
\| \bvec{p} \|_{\infty}
=
p_{[1]}
=
A
\label{eq:hypo_fixed_infty}
\end{align}
for a constant $A \in (1/n, 1)$.
Then, since
\begin{align}
\bvec{v}_{n} \bigg( \frac{ 1 - A }{ (n-1) } \bigg)
=
\bigg( A, \underbrace{ \frac{ 1 - A }{ (n-1) }, \frac{ 1 - A }{ (n-1) }, \dots, \frac{ 1 - A }{ (n-1) } }_{\text{$(n-1)$ times}} \bigg)
\prec
\bvec{p}
\end{align}
for all $\bvec{p} \in \Delta_{n}$ under the constrain \eqref{eq:hypo_fixed_infty}, it follows from the Schur-convexity of the $\ell_{\beta}$-norm that
\begin{align}
\bigg\| \, \bvec{v}_{n} \bigg( \frac{ 1 - A }{ (n-1) } \bigg) \bigg\|_{\beta}
& \le
\| \bvec{p} \|_{\beta}
\quad
\mathrm{for} \ \beta \in (1, \infty) ,
\label{eq:vector_v_proof_halfway_4} \\
\bigg\| \, \bvec{v}_{n} \bigg( \frac{ 1 - A }{ (n-1) } \bigg) \bigg\|_{\beta}
& \ge
\| \bvec{p} \|_{\beta}
\quad
\mathrm{for} \ \beta \in (0, 1)
\label{eq:vector_v_proof_halfway_5}
\end{align}
for all $\bvec{p} \in \Delta_{n}$ under the constraint \eqref{eq:hypo_fixed_infty}, where note that
\begin{align}
0 < \frac{ 1 - A }{ (n-1) } < \frac{1}{n}
\end{align}
for $A \in (1/n, 1)$.
Combining \eqref{eq:vector_v_proof_halfway_1}--\eqref{eq:vector_v_proof_halfway_3}, \eqref{eq:vector_v_proof_halfway_4}, and \eqref{eq:vector_v_proof_halfway_5}, for any $n \ge 2$, $\bvec{p} \in \Delta_{n}$, and $\alpha \in (0, 1) \cup (1, \infty]$, we have that there exists $p \in [0, 1/n]$ such that
\begin{align}
\| \bvec{v}_{n}( p ) \|_{\alpha}
& =
\| \bvec{p} \|_{\alpha} ,
\label{eq:vector_v_proof_1} \\
\| \bvec{v}_{n}( p ) \|_{\beta}
& \le
\| \bvec{p} \|_{\beta}
\quad \mathrm{for} \ \beta \in (\min\{ 1, \alpha \}, \max\{ 1, \alpha \}) ,
\label{eq:vector_v_proof_2} \\
\| \bvec{v}_{n}( p ) \|_{\beta}
& \ge
\| \bvec{p} \|_{\beta}
\quad \mathrm{for} \ \beta \in (0, \min\{ 1, \alpha \}) \cup (\max\{ 1, \alpha \}, \infty] ,
\label{eq:vector_v_proof_3}
\end{align}
where the inequality \eqref{eq:vector_v_proof_3} for $\beta = \infty$ follows from \eqref{eq:vector_v_proof_halfway_4}, \eqref{eq:vector_v_proof_halfway_5}, and the monotonicity of $\| \bvec{v}_{n}( p ) \|_{\infty}$ for $p \in [0, 1/n]$ (cf. \lemref{lem:Hv}).
This completes the proof of \lemref{lem:vector_v}.
\end{IEEEproof}

\begin{lemma}
\label{lem:vector_w}
For any $n \ge 2$, $\bvec{p} \in \Delta_{n}$, and $\alpha \in (0, 1) \cup (1, \infty]$, there exists $p \in [1/n, 1]$ such that 
\begin{align}
\| \bvec{w}_{n}( p ) \|_{\alpha}
& =
\| \bvec{p} \|_{\alpha} ,
\\
\| \bvec{w}_{n}( p ) \|_{\beta}
& \ge
\| \bvec{p} \|_{\beta}
\quad \mathrm{for} \ \mathrm{all} \ \beta \in (\min\{ 1, \alpha \}, \max\{ 1, \alpha \}) ,
\\
\| \bvec{w}_{n}( p ) \|_{\beta}
& \le
\| \bvec{p} \|_{\beta}
\quad \mathrm{for} \ \mathrm{all} \ \beta \in (0, \min\{ 1, \alpha \}) \cup (\max\{ 1, \alpha \}, \infty] .
\end{align}
\end{lemma}

\begin{IEEEproof}[Proof of \lemref{lem:vector_w}]
It can be easily seen that, for any $\bvec{p} \in \Delta_{2}$, there exists $p \in [1/2, 1]$ such that $\bvec{p}_{\downarrow} = \bvec{w}_{2}( p )$;
therefore, the lemma obviously holds when $n = 2$.
As with \eqref{equiv:unif} and \eqref{equiv:det}, we readily see that
\begin{align}
\forall \alpha \in (0, \infty],
\quad
\| \bvec{p} \|_{\alpha} & = n^{(1/\alpha)-1}
& \iff &&
\bvec{p} & = \bigg( \frac{1}{n}, \frac{1}{n}, \dots, \frac{1}{n} \bigg) = \bvec{w}_{n} \bigg( \frac{1}{n} \bigg) ,
\\
\forall \alpha \in (0, \infty],
\quad
\| \bvec{p} \|_{\alpha} & = 1
& \iff &&
\bvec{p}_{\downarrow} & = (1, 0, \dots, 0) = \bvec{w}_{n}( 1 )
\end{align}
Therefore, the lemma obviously holds when $\| \bvec{p} \|_{\alpha} \in \{ 1, n^{(1/\alpha)-1} \}$;
thus, \lemref{lem:vector_w} is proved for the case $n = 2$ and {\linebreak} $\| \bvec{p} \|_{\alpha} \in \{ 1, n^{(1/\alpha)-1} \}$ for $\alpha \in (0, 1) \cup (1, \infty)$.
Hence, it is enough to prove the lemma for $\bvec{p} \in \Delta_{n}$ such that $n \ge 3$ and $\| \bvec{p} \|_{\alpha} \in (\min\{ 1, n^{(1/\alpha)-1} \}, \max\{ 1, n^{(1/\alpha)-1} \})$ for $\alpha \in (0, 1) \cup (1, \infty]$ in the late analyses.

We now begin to prove \lemref{lem:vector_w} for $\alpha \neq \infty$ and $\beta \neq \infty$.
For fixed $n \ge 3$, $\alpha \in (0, 1) \cup (1, \infty)$, and $A \in (\min\{ 1, n^{(1/\alpha) - 1} \}, \max\{ 1, n^{(1/\alpha)-1} \})$, we assume for $\bvec{p} \in \Delta_{n}$ that 
\begin{align}
\| \bvec{p} \|_{\alpha} = A .
\label{eq:fixed_beta_w}
\end{align}
For that $\bvec{p}$ satisfying \eqref{eq:fixed_beta_w}, let $k, l \in \{ 2, 3, \dots, n \} \ (k < l)$ be indices such that $p_{[1]} = p_{[k-1]} > p_{[k+1]}$ and $p_{[l]} > p_{[l+1]} = 0$;
namely, the indices $k, l$ are chosen to satisfy the following inequalities:
\begin{align}
p_{[1]} = \dots = p_{[k-1]} \ge p_{[k]} \ge p_{[k+1]} \ge \dots \ge p_{[l-1]} \ge p_{[l]} > p_{[l+1]} = \dots = p_{[n]} = 0
\quad (p_{[k-1]} > p_{[k+1]}) .
\label{eq:equal_1_to_k-1}
\end{align}
Since
$
p_{1} + p_{2} + \dots + p_{n} = 1
$,
we observe as with \eqref{eq:total_diff_prob} that
\begin{align}
\sum_{i=1}^{n} p_{i}
=
1
\qquad \Longrightarrow \qquad
\sum_{i = 1 : i \neq k}^{n} \frac{ \mathrm{d} p_{[i]} }{ \mathrm{d} p_{[k]} }
=
- 1 .
\label{eq:total_diff_prob_w}
\end{align}
In this proof, we assume that
\begin{align}
\frac{ \mathrm{d} p_{[i]} }{ \mathrm{d} p_{[k]} }
=
\frac{ \mathrm{d} p_{[1]} }{ \mathrm{d} p_{[k]} }
\label{eq:hypo1_w}
\end{align}
for $i \in \{ 2, 3, \dots, k-1 \}$, 
\begin{align}
\frac{ \mathrm{d} p_{[j]} }{ \mathrm{d} p_{[k]} }
=
1
\label{eq:hypo2_w}
\end{align}
for $j \in \{ k+1, k+2, \dots, l-1 \}$, and
\begin{align}
\frac{ \mathrm{d} p_{[m]} }{ \mathrm{d} p_{[k]} }
=
0
\label{eq:hypo3_w}
\end{align}
for $m \in \{ l+1, l+2, \dots, n \}$.
By constraints \eqref{eq:hypo1_w}, \eqref{eq:hypo2_w}, and \eqref{eq:hypo3_w}, we get
\begin{align}
&&
\sum_{i=1}^{n} p_{i}
& =
1
\\
& \ \overset{\eqref{eq:total_diff_prob_w}}{\Longrightarrow} \ &
\sum_{i = 1 : i \neq k}^{n} \frac{ \mathrm{d} p_{[i]} }{ \mathrm{d} p_{[k]} }
& =
- 1
\\
& \iff &
\sum_{i = 1}^{k-1} \frac{ \mathrm{d} p_{[i]} }{ \mathrm{d} p_{[k]} } + \sum_{j = k+1}^{l-1} \frac{ \mathrm{d} p_{[j]} }{ \mathrm{d} p_{[k]} } + \frac{ \mathrm{d} p_{[l]} }{ \mathrm{d} p_{[k]} } + \sum_{m = l+1}^{n} \frac{ \mathrm{d} p_{[m]} }{ \mathrm{d} p_{[l]} }
& =
- 1
\\
& \overset{\eqref{eq:hypo1_w}}{\iff} &
(k-1) \frac{ \mathrm{d} p_{[1]} }{ \mathrm{d} p_{[k]} } + \sum_{j = k+1}^{l-1} \frac{ \mathrm{d} p_{[j]} }{ \mathrm{d} p_{[k]} } + \frac{ \mathrm{d} p_{[l]} }{ \mathrm{d} p_{[k]} } + \sum_{m = l+1}^{n} \frac{ \mathrm{d} p_{[m]} }{ \mathrm{d} p_{[l]} }
& =
- 1
\\
& \overset{\eqref{eq:hypo2_w}}{\iff} &
(k-1) \frac{ \mathrm{d} p_{[1]} }{ \mathrm{d} p_{[k]} } + (l - k - 1) + \frac{ \mathrm{d} p_{[l]} }{ \mathrm{d} p_{[k]} } + \sum_{m = l+1}^{n} \frac{ \mathrm{d} p_{[m]} }{ \mathrm{d} p_{[l]} }
& =
- 1
\\
& \overset{\eqref{eq:hypo3_w}}{\iff} &
(k-1) \frac{ \mathrm{d} p_{[1]} }{ \mathrm{d} p_{[k]} } + (l - k - 1) + \frac{ \mathrm{d} p_{[l]} }{ \mathrm{d} p_{[k]} }
& =
- 1
\\
& \iff &
(k-1) \frac{ \mathrm{d} p_{[1]} }{ \mathrm{d} p_{[k]} } + \frac{ \mathrm{d} p_{[l]} }{ \mathrm{d} p_{[k]} }
& =
- (l-k)
\\
& \iff &
(k-1) \frac{ \mathrm{d} p_{[1]} }{ \mathrm{d} p_{[k]} } 
& =
- (l-k) - \frac{ \mathrm{d} p_{[l]} }{ \mathrm{d} p_{[k]} }
\\
& \iff &
\frac{ \mathrm{d} p_{[1]} }{ \mathrm{d} p_{[k]} }
& =
- \frac{1}{k-1} \left( (l - k) +  \frac{ \mathrm{d} p_{[l]} }{ \mathrm{d} p_{[k]} } \right) ,
\label{eq:total_prob_hypo_w}
\end{align}
where note in \eqref{eq:total_prob_hypo_w} that $k \ge 2$.
Moreover, we observe from the constraint \eqref{eq:fixed_beta_w} that
\begin{align}
&&
\| \bvec{p} \|_{\alpha}
& =
A
\\
& \iff &
\sum_{i = 1}^{n} p_{i}^{\alpha}
& =
A^{\alpha}
\\
& \overset{\eqref{eq:equal_1_to_k-1}}{\iff} &
\sum_{i = 1}^{l} p_{i}^{\alpha}
& =
A^{\alpha}
\\
& \ \overset{\eqref{eq:diff1_beta_halfway_2}}{\Longrightarrow} \ &
\sum_{i = 1 : i \neq k}^{l} \left( \frac{ \mathrm{d} p_{[i]} }{ \mathrm{d} p_{[k]} } \right) p_{[i]}^{\alpha-1}
& =
- p_{[k]}^{\alpha-1}
\\
& \iff &
\sum_{i = 1}^{k-1} \left( \frac{ \mathrm{d} p_{[i]} }{ \mathrm{d} p_{[k]} } \right) p_{[i]}^{\alpha-1} + \sum_{j = k+1}^{l-1} \left( \frac{ \mathrm{d} p_{[j]} }{ \mathrm{d} p_{[k]} } \right) p_{[j]}^{\alpha-1} + \left( \frac{ \mathrm{d} p_{[l]} }{ \mathrm{d} p_{[k]} } \right) p_{[l]}^{\alpha-1}
& =
- p_{[k]}^{\alpha-1}
\\
& \overset{\eqref{eq:equal_1_to_k-1}}{\iff} &
p_{[1]}^{\alpha-1} \sum_{i = 1}^{k-1} \left( \frac{ \mathrm{d} p_{[i]} }{ \mathrm{d} p_{[k]} } \right) + \sum_{j = k+1}^{l-1} \left( \frac{ \mathrm{d} p_{[j]} }{ \mathrm{d} p_{[k]} } \right) p_{[j]}^{\alpha-1} + \left( \frac{ \mathrm{d} p_{[l]} }{ \mathrm{d} p_{[k]} } \right) p_{[l]}^{\alpha-1}
& =
- p_{[k]}^{\alpha-1}
\\
& \overset{\eqref{eq:hypo1_w}}{\iff} &
(k-1) \, p_{[1]}^{\alpha-1} \left( \frac{ \mathrm{d} p_{[1]} }{ \mathrm{d} p_{[k]} } \right) + \sum_{j = k+1}^{l-1} \left( \frac{ \mathrm{d} p_{[j]} }{ \mathrm{d} p_{[k]} } \right) p_{[j]}^{\alpha-1} + \left( \frac{ \mathrm{d} p_{[l]} }{ \mathrm{d} p_{[k]} } \right) p_{[l]}^{\alpha-1}
& =
- p_{[k]}^{\alpha-1}
\\
& \overset{\eqref{eq:hypo2_w}}{\iff} &
(k-1) \, p_{[1]}^{\alpha-1} \left( \frac{ \mathrm{d} p_{[1]} }{ \mathrm{d} p_{[k]} } \right) + \sum_{j = k+1}^{l-1} p_{[j]}^{\alpha-1} + \left( \frac{ \mathrm{d} p_{[l]} }{ \mathrm{d} p_{[k]} } \right) p_{[l]}^{\alpha-1}
& =
- p_{[k]}^{\alpha-1}
\\
& \iff &
(k-1) \, p_{[1]}^{\alpha-1} \left( \frac{ \mathrm{d} p_{[1]} }{ \mathrm{d} p_{[k]} } \right) + \left( \frac{ \mathrm{d} p_{[l]} }{ \mathrm{d} p_{[k]} } \right) p_{[l]}^{\alpha-1}
& =
- \sum_{j = k}^{l-1} p_{[j]}^{\alpha-1}
\\
& \overset{\eqref{eq:total_prob_hypo_w}}{\iff} &
(k-1) \, p_{[1]}^{\alpha-1} \left( - \frac{1}{k-1} \left( (l - k) + \frac{ \mathrm{d} p_{[l]} }{ \mathrm{d} p_{[k]} } \right) \right) + \left( \frac{ \mathrm{d} p_{[l]} }{ \mathrm{d} p_{[k]} } \right) p_{[l]}^{\alpha-1}
& =
- \sum_{j = k}^{l-1} p_{[j]}^{\alpha-1}
\\
& \iff &
- (l - k) \, p_{[1]}^{\alpha-1} - \left( \frac{ \mathrm{d} p_{[l]} }{ \mathrm{d} p_{[k]} } \right) p_{[1]}^{\alpha-1} + \left( \frac{ \mathrm{d} p_{[l]} }{ \mathrm{d} p_{[k]} } \right) p_{[l]}^{\alpha-1}
& =
- \sum_{j = k}^{l-1} p_{[j]}^{\alpha-1}
\\
& \iff &
- (l - k) \, p_{[1]}^{\alpha-1} + \left( \frac{ \mathrm{d} p_{[l]} }{ \mathrm{d} p_{[k]} } \right) \left( p_{[l]}^{\alpha-1} - p_{[1]}^{\alpha-1} \right)
& =
- \sum_{j = k}^{l-1} p_{[j]}^{\alpha-1}
\\
& \iff &
\left( \frac{ \mathrm{d} p_{[l]} }{ \mathrm{d} p_{[k]} } \right) \left( p_{[1]}^{\alpha-1} - p_{[l]}^{\alpha-1} \right)
& =
- \sum_{j = k}^{l-1} \left( p_{[j]}^{\alpha-1} - p_{[1]}^{\alpha-1} \right)
\\
& \iff &
\frac{ \mathrm{d} p_{[l]} }{ \mathrm{d} p_{[k]} }
& =
- \frac{ \sum_{j = k}^{l-1} \left( p_{[j]}^{\alpha-1} - p_{[1]}^{\alpha-1} \right) }{ p_{[l]}^{\alpha-1} - p_{[1]}^{\alpha-1} }
\\
& \iff &
\frac{ \mathrm{d} p_{[l]} }{ \mathrm{d} p_{[k]} }
& =
- \frac{ \sum_{j = k}^{l-1} \left( \left( \frac{ p_{[j]} }{ p_{[1]} } \right)^{\alpha-1} - 1 \right) }{ \left( \frac{ p_{[l]} }{ p_{[1]} } \right)^{\alpha-1} - 1 }
\\
& \iff &
\frac{ \mathrm{d} p_{[l]} }{ \mathrm{d} p_{[k]} }
& =
- \frac{ \sum_{j = k}^{l-1} \left( \left( \frac{ p_{[1]} }{ p_{[j]} } \right)^{1-\alpha} - 1 \right) }{ \left( \frac{ p_{[1]} }{ p_{[l]} } \right)^{1-\alpha} - 1 }
\\
& \overset{\eqref{def:qlog}}{\iff} &
\frac{ \mathrm{d} p_{[l]} }{ \mathrm{d} p_{[k]} }
& =
- \sum_{j = k}^{l-1} \left( \frac{ \ln_{\alpha} \left( \frac{ p_{[1]} }{ p_{[j]} } \right) }{ \ln_{\alpha} \left( \frac{ p_{[1]} }{ p_{[l]} } \right) } \right) .
\label{eq:total_norm_hypo_w}
\end{align}
We now check the sign of the right-hand side of \eqref{eq:total_norm_hypo_w}.
Since $p_{[1]} \ge p_{[k]} \ge p_{[j]}$ for $j \ge k \ge 2$, we get
\begin{align}
\frac{ \mathrm{d} p_{[l]} }{ \mathrm{d} p_{[k]} }
& \overset{\eqref{eq:total_norm_hypo_w}}{=}
- \sum_{j = k}^{l-1} \left( \frac{ \ln_{\alpha} \left( \frac{ p_{[1]} }{ p_{[j]} } \right) }{ \ln_{\alpha} \left( \frac{ p_{[1]} }{ p_{[l]} } \right) } \right)
\\
& \overset{\text{(a)}}{\le}
- \sum_{j = k}^{l-1} \left( \frac{ \ln_{\alpha} \left( \frac{ p_{[1]} }{ p_{[k]} } \right) }{ \ln_{\alpha} \left( \frac{ p_{[1]} }{ p_{[l]} } \right) } \right)
\\
& =
- (l - k) \left( \frac{ \ln_{\alpha} \left( \frac{ p_{[1]} }{ p_{[k]} } \right) }{ \ln_{\alpha} \left( \frac{ p_{[1]} }{ p_{[l]} } \right) } \right)
\\
& \overset{\text{(b)}}{\le}
0 ,
\label{ineq:dldk_w}
\end{align}
where (a) follows by the monotonicity of the $\alpha$-logarithm function (cf. \eqref{eq:diff_qlog}), the inequality (a) holds with equality if and only if $p_{[k]} = p_{[l-1]}$, and the inequality (b) holds with equality if and only if $p_{[1]} = p_{[k]}$ for $k \le l - 1$.
Thus, it follows from \eqref{ineq:dldk_w} that $p_{[l]}$ is strictly decreasing for $p_{[k]}$ under the constraints \eqref{eq:fixed_beta_w}, \eqref{eq:equal_1_to_k-1}, \eqref{eq:hypo1_w}, \eqref{eq:hypo2_w}, and \eqref{eq:hypo3_w}.

Similarly, we check the sign of the right-hand side of \eqref{eq:total_prob_hypo_w} as follows:
\begin{align}
\frac{ \mathrm{d} p_{[1]} }{ \mathrm{d} p_{[k]} }
& =
- \frac{1}{k-1} \left( (l - k) +  \frac{ \mathrm{d} p_{[l]} }{ \mathrm{d} p_{[k]} } \right)
\\
& \overset{\eqref{eq:total_norm_hypo_w}}{=}
- \frac{l - k}{k-1} + \frac{1}{k-1} \sum_{j = k}^{l-1} \left( \frac{ \ln_{\alpha} \left( \frac{ p_{[1]} }{ p_{[j]} } \right) }{ \ln_{\alpha} \left( \frac{ p_{[1]} }{ p_{[l]} } \right) } \right)
\\
& \overset{\text{(a)}}{<}
- \frac{l - k}{k-1} + \frac{1}{k-1} \sum_{j = k}^{l-1} \left( \frac{ \ln_{\alpha} \left( \frac{ p_{[1]} }{ p_{[l]} } \right) }{ \ln_{\alpha} \left( \frac{ p_{[1]} }{ p_{[l]} } \right) } \right)
\\
& =
- \frac{l - k}{k-1} + \frac{1}{k-1} \sum_{j = k}^{l-1} 1
\\
& =
- \frac{l - k}{k-1} + \frac{l-k}{k-1}
\\
& =
0 ,
\label{ineq:d1dk_w}
\end{align}
where (a) follows by $p_{[1]} > p_{[l]}$ (cf. the constraint \eqref{eq:equal_1_to_k-1}) and the monotonicity of the $\alpha$-logarithm function (cf. \eqref{eq:diff_qlog}).
Hence, it follows from \eqref{ineq:d1dk_w} that $p_{[1]}$ is also strictly decreasing for $p_{[k]}$ under the constraints \eqref{eq:fixed_beta_w}, \eqref{eq:equal_1_to_k-1}, \eqref{eq:hypo1_w}, \eqref{eq:hypo2_w}, and \eqref{eq:hypo3_w}.

We next consider $\| \bvec{p} \|_{\beta}$ for a fixed $\beta \in (0, 1) \cup (1, \infty)$.
A direct calculation yields
\begin{align}
\frac{ \mathrm{d} \| \bvec{p} \|_{\beta}^{\beta} }{ \mathrm{d} p_{[k]} }
& =
\frac{ \mathrm{d} }{ \mathrm{d} p_{[k]} } \left( \sum_{i=1}^{n} p_{i}^{\beta} \right)
\\
& \overset{\eqref{eq:equal_1_to_k-1}}{=}
\frac{ \mathrm{d} }{ \mathrm{d} p_{[k]} } \left( \sum_{i=1}^{l} p_{i}^{\beta} \right)
\\
& \overset{\eqref{eq:diff1_beta_halfway}}{=}
\beta \left( p_{[k]}^{\beta-1} + \sum_{i=1 : i \neq k}^{l} \left( \frac{ \mathrm{d} p_{[i]} }{ \mathrm{d} p_{[k]} } \right) p_{[i]}^{\beta-1} \right)
\\
& =
\beta \left( p_{[k]}^{\beta-1} + \sum_{i=1}^{k-1} \left( \frac{ \mathrm{d} p_{[i]} }{ \mathrm{d} p_{[k]} } \right) p_{[i]}^{\beta-1} + \sum_{j=k+1}^{l-1} \left( \frac{ \mathrm{d} p_{[j]} }{ \mathrm{d} p_{[k]} } \right) p_{[j]}^{\beta-1} + \left( \frac{ \mathrm{d} p_{[l]} }{ \mathrm{d} p_{[k]} } \right) p_{[l]}^{\beta-1} \right)
\\
& \overset{\eqref{eq:equal_1_to_k-1}}{=}
\beta \left( p_{[k]}^{\beta-1} + p_{[1]}^{\beta-1} \sum_{i=1}^{k-1} \left( \frac{ \mathrm{d} p_{[i]} }{ \mathrm{d} p_{[k]} } \right) + \sum_{j=k+1}^{l-1} \left( \frac{ \mathrm{d} p_{[j]} }{ \mathrm{d} p_{[k]} } \right) p_{[j]}^{\beta-1} + \left( \frac{ \mathrm{d} p_{[l]} }{ \mathrm{d} p_{[k]} } \right) p_{[l]}^{\beta-1} \right)
\\
& \overset{\eqref{eq:hypo1_w}}{=}
\beta \left( p_{[k]}^{\beta-1} + (k-1) \, p_{[1]}^{\beta-1} \left( \frac{ \mathrm{d} p_{[1]} }{ \mathrm{d} p_{[k]} } \right) + \sum_{j=k+1}^{l-1} \left( \frac{ \mathrm{d} p_{[j]} }{ \mathrm{d} p_{[k]} } \right) p_{[j]}^{\beta-1} + \left( \frac{ \mathrm{d} p_{[l]} }{ \mathrm{d} p_{[k]} } \right) p_{[l]}^{\beta-1} \right)
\\
& \overset{\eqref{eq:hypo2_w}}{=}
\beta \left( p_{[k]}^{\beta-1} + (k-1) \, p_{[1]}^{\beta-1} \left( \frac{ \mathrm{d} p_{[1]} }{ \mathrm{d} p_{[k]} } \right) + \sum_{j=k+1}^{l-1} p_{[j]}^{\beta-1} + \left( \frac{ \mathrm{d} p_{[l]} }{ \mathrm{d} p_{[k]} } \right) p_{[l]}^{\beta-1} \right)
\\
& =
\beta \left( (k-1) \, p_{[1]}^{\beta-1} \left( \frac{ \mathrm{d} p_{[1]} }{ \mathrm{d} p_{[k]} } \right) + \sum_{j=k}^{l-1} p_{[j]}^{\beta-1} + \left( \frac{ \mathrm{d} p_{[l]} }{ \mathrm{d} p_{[k]} } \right) p_{[l]}^{\beta-1} \right)
\\
& \overset{\eqref{eq:total_prob_hypo_w}}{=}
\beta \left( (k-1) \, p_{[1]}^{\beta-1} \left( - \frac{1}{k-1} \left( (l - k) +  \frac{ \mathrm{d} p_{[l]} }{ \mathrm{d} p_{[k]} } \right) \right) + \sum_{j=k}^{l-1} p_{[j]}^{\beta-1} + \left( \frac{ \mathrm{d} p_{[l]} }{ \mathrm{d} p_{[k]} } \right) p_{[l]}^{\beta-1} \right)
\\
& =
\beta \left( - (l - k) \, p_{[1]}^{\beta-1} - \left( \frac{ \mathrm{d} p_{[l]} }{ \mathrm{d} p_{[k]} } \right) p_{[1]}^{\beta-1} + \sum_{j=k}^{l-1} p_{[j]}^{\beta-1} + \left( \frac{ \mathrm{d} p_{[l]} }{ \mathrm{d} p_{[k]} } \right) p_{[l]}^{\beta-1} \right)
\\
& =
\beta \left( \sum_{j=k}^{l-1} \left( p_{[j]}^{\beta-1} - p_{[1]}^{\beta-1} \right) + \left( \frac{ \mathrm{d} p_{[l]} }{ \mathrm{d} p_{[k]} } \right) \left( p_{[l]}^{\beta-1} - p_{[1]}^{\beta-1} \right) \right)
\\
& \overset{\eqref{eq:total_norm_hypo_w}}{=}
\beta \left( \sum_{j=k}^{l-1} \left( p_{[j]}^{\beta-1} - p_{[1]}^{\beta-1} \right) - \sum_{j = k}^{l-1} \left( \frac{ \ln_{\alpha} \left( \frac{ p_{[1]} }{ p_{[j]} } \right) }{ \ln_{\alpha} \left( \frac{ p_{[1]} }{ p_{[l]} } \right) } \right) \left( p_{[l]}^{\beta-1} - p_{[1]}^{\beta-1} \right) \right)
\\
& =
\beta \left( p_{[l]}^{\beta-1} - p_{[1]}^{\beta-1} \right) \left( \sum_{j=k}^{l-1} \left( \frac{ p_{[j]}^{\beta-1} - p_{[1]}^{\beta-1} }{ p_{[l]}^{\beta-1} - p_{[1]}^{\beta-1} } \right) - \sum_{j = k}^{l-1} \left( \frac{ \ln_{\alpha} \left( \frac{ p_{[1]} }{ p_{[j]} } \right) }{ \ln_{\alpha} \left( \frac{ p_{[1]} }{ p_{[l]} } \right) } \right) \right)
\\
& \overset{\eqref{def:qlog}}{=}
\beta \left( p_{[l]}^{\beta-1} - p_{[1]}^{\beta-1} \right) \sum_{j=k}^{l-1} \left( \frac{ \ln_{\beta} \left( \frac{ p_{[1]} }{ p_{[j]} } \right) }{ \ln_{\beta} \left( \frac{ p_{[1]} }{ p_{[l]} } \right) } - \frac{ \ln_{\alpha} \left( \frac{ p_{[1]} }{ p_{[j]} } \right) }{ \ln_{\alpha} \left( \frac{ p_{[1]} }{ p_{[l]} } \right) } \right) .
\label{eq:diff1_norm_alpha_w}
\end{align}
Since $\| \bvec{p} \|_{\beta} = (\| \bvec{p} \|_{\beta}^{\beta})^{1/\beta}$, it follows by the chain rule that
\begin{align}
\frac{ \mathrm{d} \| \bvec{p} \|_{\beta} }{ \mathrm{d} p_{[k]} }
& =
\left( \frac{1}{\beta} \frac{ \| \bvec{p} \|_{\beta} }{ \| \bvec{p} \|_{\beta}^{\beta} } \right) \cdot \left( \frac{ \mathrm{d} \| \bvec{p} \|_{\beta}^{\beta} }{ \mathrm{d} p_{[k]} } \right)
\\
& \overset{\eqref{eq:diff1_norm_alpha_w}}{=}
\left( \sum_{i=1}^{n} p_{i}^{\beta} \right)^{(1/\beta)-1} \left( p_{[l]}^{\beta-1} - p_{[1]}^{\beta-1} \right) \sum_{j=k}^{l-1} \left( \frac{ \ln_{\beta} \left( \frac{ p_{[1]} }{ p_{[j]} } \right) }{ \ln_{\beta} \left( \frac{ p_{[1]} }{ p_{[l]} } \right) } - \frac{ \ln_{\alpha} \left( \frac{ p_{[1]} }{ p_{[j]} } \right) }{ \ln_{\alpha} \left( \frac{ p_{[1]} }{ p_{[l]} } \right) } \right) .
\end{align}
Thus, we get
\begin{align}
\sgn \left( \frac{ \mathrm{d} \| \bvec{p} \|_{\beta} }{ \mathrm{d} p_{[k]} } \right)
& =
\sgn \left( \left( \sum_{i=1}^{n} p_{i}^{\beta} \right)^{(1/\beta)-1} \left( \vphantom{\sum} p_{[l]}^{\beta-1} - p_{[1]}^{\beta-1} \right) \sum_{j=k}^{l-1} \left( \frac{ \ln_{\beta} \left( \frac{ p_{[1]} }{ p_{[j]} } \right) }{ \ln_{\beta} \left( \frac{ p_{[1]} }{ p_{[l]} } \right) } - \frac{ \ln_{\alpha} \left( \frac{ p_{[1]} }{ p_{[j]} } \right) }{ \ln_{\alpha} \left( \frac{ p_{[1]} }{ p_{[l]} } \right) } \right) \right)
\\
& =
\underbrace{ \sgn \left( \left( \sum_{i=1}^{n} p_{i}^{\beta} \right)^{(1/\beta)-1} \right) }_{ = 1 } \cdot \sgn \left( \vphantom{\sum} p_{[l]}^{\beta-1} - p_{[1]}^{\beta-1} \right) \cdot \sgn \left( \sum_{j=k}^{l-1} \left( \frac{ \ln_{\beta} \left( \frac{ p_{[1]} }{ p_{[j]} } \right) }{ \ln_{\beta} \left( \frac{ p_{[1]} }{ p_{[l]} } \right) } - \frac{ \ln_{\alpha} \left( \frac{ p_{[1]} }{ p_{[j]} } \right) }{ \ln_{\alpha} \left( \frac{ p_{[1]} }{ p_{[l]} } \right) } \right) \right)
\\
& =
\sgn \left( \vphantom{\sum} p_{[l]}^{\beta-1} - p_{[1]}^{\beta-1} \right) \cdot \sgn \left( \sum_{j=k}^{l-1} \left( \frac{ \ln_{\beta} \left( \frac{ p_{[1]} }{ p_{[j]} } \right) }{ \ln_{\beta} \left( \frac{ p_{[1]} }{ p_{[l]} } \right) } - \frac{ \ln_{\alpha} \left( \frac{ p_{[1]} }{ p_{[j]} } \right) }{ \ln_{\alpha} \left( \frac{ p_{[1]} }{ p_{[l]} } \right) } \right) \right) .
\label{eq:sgn_diff_Norm_w_factor12}
\end{align}
Since $p_{[1]} > p_{[l]} > 0$ by the constraint \eqref{eq:equal_1_to_k-1}, we readily see that
\begin{align}
\sgn \left( \vphantom{\sum} p_{[l]}^{\beta-1} - p_{[1]}^{\beta-1} \right)
& =
\begin{cases}
1
& \mathrm{if} \ \beta < 1 , \\
0
& \mathrm{if} \ \beta = 1 , \\
-1
& \mathrm{if} \ \beta > 1 .
\end{cases}
\label{eq:sgn_diff_Norm_w_factor1}
\end{align}
Moreover, since
\begin{align}
1 \le \frac{ p_{[1]} }{ p_{[j]} } \le \frac{ p_{[1]} }{ p_{[l]} }
\quad
\left( 1 \neq \frac{ p_{[1]} }{ p_{[l]} } \right)
\end{align}
for $j \in \{ k, k+1, \dots, l-1 \}$, it follows from \eqref{eq:frac_qlog} (cf. \cite[Eq. (20)]{part1}) that
\begin{align}
\sgn \left( \frac{ \ln_{\beta} \left( \frac{ p_{[1]} }{ p_{[j]} } \right) }{ \ln_{\beta} \left( \frac{ p_{[1]} }{ p_{[l]} } \right) } - \frac{ \ln_{\alpha} \left( \frac{ p_{[1]} }{ p_{[j]} } \right) }{ \ln_{\alpha} \left( \frac{ p_{[1]} }{ p_{[l]} } \right) } \right)
& =
\begin{cases}
1
& \mathrm{if} \ \beta > \alpha \ \mathrm{and} \ p_{[1]} > p_{[j]} > p_{[l]} , \\
0
& \mathrm{if} \ \beta = \alpha \ \mathrm{or} \ p_{[1]} = p_{[j]} \ \mathrm{or} \ p_{[j]} = p_{[l]} , \\
-1
& \mathrm{if} \ \beta < \alpha \ \mathrm{and} \ p_{[1]} > p_{[j]} > p_{[l]}
\end{cases}
\end{align}
for $j \in \{ k, k+1, \dots, l-1 \}$;
and therefore, we have
\begin{align}
\sgn \left( \sum_{j=k}^{l-1} \left( \frac{ \ln_{\beta} \left( \frac{ p_{[1]} }{ p_{[j]} } \right) }{ \ln_{\beta} \left( \frac{ p_{[1]} }{ p_{[l]} } \right) } - \frac{ \ln_{\alpha} \left( \frac{ p_{[1]} }{ p_{[j]} } \right) }{ \ln_{\alpha} \left( \frac{ p_{[1]} }{ p_{[l]} } \right) } \right) \right)
& =
\begin{cases}
1
& \mathrm{if} \ \beta > \alpha \ \mathrm{and} \ (p_{[1]} > p_{[k]} \ \mathrm{or} \ p_{[k]} > p_{[l]}) , \\
0
& \mathrm{if} \ \beta = \alpha \ \mathrm{or} \ (p_{[1]} = p_{[k]} \ \mathrm{and} \ p_{[k+1]} = p_{[l]})  \ \mathrm{or} \ p_{[k]} = p_{[l]} , \\
-1
& \mathrm{if} \ \beta < \alpha \ \mathrm{and} \ (p_{[1]} > p_{[k]} \ \mathrm{or} \ p_{[k]} > p_{[l]}) .
\end{cases}
\label{eq:sgn_diff_Norm_w_factor2}
\end{align}
Combining \eqref{eq:sgn_diff_Norm_w_factor1} and \eqref{eq:sgn_diff_Norm_w_factor2}, if $p_{[1]} > p_{[k]}$ or $p_{[k]} > p_{[l]}$, we obtain
\begin{align}
\sgn \left( \frac{ \mathrm{d} \| \bvec{p} \|_{\beta} }{ \mathrm{d} p_{[k]} } \right)
& \overset{\eqref{eq:sgn_diff_Norm_w_factor12}}{=}
\sgn \left( \vphantom{\sum} p_{[l]}^{\beta-1} - p_{[1]}^{\beta-1} \right) \cdot \sgn \left( \sum_{j=k}^{l-1} \left( \frac{ \ln_{\beta} \left( \frac{ p_{[1]} }{ p_{[j]} } \right) }{ \ln_{\beta} \left( \frac{ p_{[1]} }{ p_{[l]} } \right) } - \frac{ \ln_{\alpha} \left( \frac{ p_{[1]} }{ p_{[j]} } \right) }{ \ln_{\alpha} \left( \frac{ p_{[1]} }{ p_{[l]} } \right) } \right) \right) 
\\
& =
\begin{cases}
1
& \mathrm{if} \ \beta \in (\min\{ 1, \alpha \}, \max\{ 1, \alpha \}) , \\
0
& \mathrm{if} \ \beta \in \{ 1, \alpha \} , \\
-1
& \mathrm{if} \ \beta \in (0, \min\{ 1, \alpha \}) \cup (\max\{ 1, \alpha \}, \infty) ,
\end{cases}
\label{eq:sgn_diff_Norm_w}
\end{align}
which implies that
\begin{itemize}
\item
if $\alpha \in (0, 1)$, then
\begin{itemize}
\item
$\| \bvec{p} \|_{\beta}$ with a fixed $\beta \in (\alpha, 1)$ is strictly increasing for $p_{[k]}$, and
\item
$\| \bvec{p} \|_{\beta}$ with a fixed $\beta \in (0, \alpha) \cup (1, \infty)$ is strictly decreasing for $p_{[k]}$,
\end{itemize}
\item
if $\alpha \in (1, \infty)$, then
\begin{itemize}
\item
$\| \bvec{p} \|_{\beta}$ with a fixed $\beta \in (1, \alpha)$ is strictly increasing for $p_{[k]}$, and
\item
$\| \bvec{p} \|_{\beta}$ with a fixed $\beta \in (0, 1) \cup (\alpha, \infty)$ is strictly decreasing for $p_{[k]}$.
\end{itemize}
\end{itemize}
Note that the above monotonicity hold under the constraints \eqref{eq:fixed_beta_w}, \eqref{eq:equal_1_to_k-1}, \eqref{eq:hypo1_w}, \eqref{eq:hypo2_w}, and \eqref{eq:hypo3_w}.

To accomplish the proof of \lemref{lem:vector_w} for $\alpha \neq \infty$ and $\beta \neq \infty$ by using the above relations, we repeat the following operation until the vector $\bvec{p} = (p_{1}, p_{2}, \dots, p_{n})$ satisfies $p_{[k-1]} = p_{[k]}$ and $l = k+1$, i.e.,
\begin{align}
p_{[1]} = p_{[2]} = \dots = p_{[k-1]} = p_{[k]} \ge p_{[k+1]} > p_{[k+2]} = p_{[k+3]} \dots = p_{[n]} = 0 ,
\label{eq:proof_vector_w}
\end{align}
which is equivalent to the vector $\bvec{w}_{n}( \cdot )$.
If $p_{[k-1]} = p_{[k]}$ and $k < l-1$, then we reset the index $k \in \{ 2, 3, \dots, n-2 \}$ to $k + 1$;
namely,
we now choose the indices $k, l \in \{ 2, 3, \dots, n \} \ (k < l)$ to satisfy the following inequalities:
\begin{align}
p_{[1]} = p_{[2]} = \dots = p_{[k-1]} > p_{[k]} \ge p_{[k+1]} \ge \dots \ge p_{[l-1]} \ge p_{[l]} > p_{[l+1]} = p_{[l+2]} = \dots = p_{[n]} = 0 .
\label{eq:choose_k2}
\end{align}
For that index $k$, we consider to increase $p_{[k]}$ under the constraints \eqref{eq:fixed_beta_w}, \eqref{eq:equal_1_to_k-1}, \eqref{eq:hypo1_w}, \eqref{eq:hypo2_w}, and \eqref{eq:hypo3_w}.
Note that the constraint \eqref{eq:hypo2_w} implies that, for $j \in \{ k+1, k+2, \dots, l-1 \}$, the value $p_{[j]}$ is increased at the same speed as $p_{[k]}$.
It follows from \eqref{eq:hypo1_w} and \eqref{ineq:d1dk_w} that, for $i \in \{ 1, 2, \dots, k-1 \}$, the value $p_{[i]}$ is strictly decreased by according to increasing $p_{[k]}$.
Similarly, it follows from \eqref{ineq:dldk_w} that $p_{[l]}$ is also strictly decreased by according to increasing $p_{[k]}$.
Let $\bvec{q} = (q_{1}, q_{2}, \dots, q_{n})$ denote the probability vector such that made from $\bvec{p}$ by increasing $p_{[k]}$ until the equality $p_{[k-1]} = p_{[k]}$ or $p_{[l]} = 0$ holds under the conditions of \eqref{eq:fixed_beta_w}, \eqref{eq:hypo1_w}, \eqref{eq:hypo2_w}, \eqref{eq:hypo3_w}, and \eqref{eq:choose_k2}.
Namely, the vector $\bvec{q}$ satisfies either
\begin{align}
q_{[1]} = q_{[2]} = \dots = q_{[k-1]} = q_{[k]} \ge q_{[k+1]} \ge \dots \ge q_{[l-1]} > q_{[l]} \ge q_{[l+1]} = q_{[l+2]} = \dots = q_{[n]} = 0
\label{ineq:q_w_1}
\end{align}
or
\begin{align}
q_{[1]} = q_{[2]} = \dots = q_{[k-1]} \ge q_{[k]} \ge q_{[k+1]} \ge \dots \ge q_{[l-1]} > q_{[l]} = q_{[l+1]} = q_{[l+2]} = \dots = q_{[n]} = 0 .
\label{ineq:q_w_2}
\end{align}
Note that there is a possibility that both of \eqref{ineq:q_w_1} and \eqref{ineq:q_w_2} hold as follows:
\begin{align}
q_{[1]} = q_{[2]} = \dots = q_{[k-1]} = q_{[k]} \ge q_{[k+1]} \ge \dots \ge q_{[l-1]} > q_{[l]} = q_{[l+1]} = q_{[l+2]} = \dots = q_{[n]} = 0 .
\end{align}
Since $\bvec{q}$ is made from $\bvec{p}$ under the constraint \eqref{eq:fixed_beta_w}, note that
\begin{align}
\| \bvec{q} \|_{\alpha} = \| \bvec{p} \|_{\alpha}
\end{align}
for a fixed $\alpha \in (0, 1) \cup (1, \infty)$.
Moreover, it follows from \eqref{eq:sgn_diff_Norm_w} that
\begin{align}
\| \bvec{q} \|_{\beta} \ge \| \bvec{p} \|_{\beta}
\quad & \mathrm{for} \ \beta \in (\min\{ 1, \alpha \}, \max\{ 1, \alpha \}) ,
\\
\| \bvec{q} \|_{\beta} \le \| \bvec{p} \|_{\beta}
\quad & \mathrm{for} \ \beta \in (0, \min\{ 1, \alpha \}) \cup (\max\{ 1, \alpha \}, \infty) .
\end{align}
Repeating these operation until \eqref{eq:proof_vector_w} holds, we have
\begin{align}
\| \bvec{w}_{n}( p ) \|_{\alpha}
& =
\| \bvec{p} \|_{\alpha} ,
\label{eq:vector_w_proof_halfway_1} \\
\| \bvec{w}_{n}( p ) \|_{\beta}
& \ge
\| \bvec{p} \|_{\beta}
\quad \mathrm{for} \ \beta \in (\min\{ 1, \alpha \}, \max\{ 1, \alpha \}) ,
\label{eq:vector_w_proof_halfway_2} \\
\| \bvec{w}_{n}( p ) \|_{\beta}
& \le
\| \bvec{p} \|_{\beta}
\quad \mathrm{for} \ \beta \in (0, \min\{ 1, \alpha \}) \cup (\max\{ 1, \alpha \}, \infty)
\label{eq:vector_w_proof_halfway_3}
\end{align}
for any $\bvec{p} \in \Delta_{n}$, any fixed $\alpha \in (0, 1) \cup (1, \infty)$, and some $p \in [1/n, 1]$.

Finally, we consider the $\ell_{\beta}$-norm with a fixed $\ell_{\infty}$-norm for $\beta \in (0, 1) \cup (1, \infty)$.
To consider probability vectors $\bvec{p} \in \Delta_{n}$ with a fixed $\ell_{\infty}$-norm, we assume in \eqref{eq:fixed_beta_w} that $\alpha = \infty$, i.e,
\begin{align}
\| \bvec{p} \|_{\infty}
=
p_{[1]}
=
A
\label{eq:hypo_fixed_infty_w}
\end{align}
for a constant $A \in (1/n, 1)$.
Recall from the proof of \lemref{lem:vector_v} that
\begin{itemize}
\item
$\| \bvec{p} \|_{\beta}$ is strictly Schur-concave in $\bvec{p} \in \Delta_{n}$ for every $\beta \in (0, 1)$ and
\item
$\| \bvec{p} \|_{\beta}$ is strictly Schur-convex in $\bvec{p} \in \Delta_{n}$ for every $\beta \in (1, \infty)$.
\end{itemize}
Thus, since
\begin{align}
\bvec{w}_{n} ( A )
=
\bigg( \underbrace{ A, A, \dots, A }_{\text{$\lfloor 1/A \rfloor$ times}}, 1 - \bigg\lfloor \frac{ 1 }{ A } \bigg\rfloor A, 0, 0, \dots, 0 \bigg)
\succ
\bvec{p}
\end{align}
for all $\bvec{p} \in \Delta_{n}$ under the constrain \eqref{eq:hypo_fixed_infty_w}, it follows from the Schur-convexity of the $\ell_{\beta}$-norm that
\begin{align}
\| \bvec{w}_{n} ( A ) \|_{\beta}
& \ge
\| \bvec{p} \|_{\beta}
\quad
\mathrm{for} \ \beta \in (1, \infty) ,
\label{eq:vector_w_proof_halfway_4} \\
\| \bvec{w}_{n} ( A ) \|_{\beta}
& \le
\| \bvec{p} \|_{\beta}
\quad
\mathrm{for} \ \beta \in (0, 1)
\label{eq:vector_w_proof_halfway_5}
\end{align}
for all $\bvec{p} \in \Delta_{n}$ under the constraint \eqref{eq:hypo_fixed_infty}.
Combining \eqref{eq:vector_w_proof_halfway_1}--\eqref{eq:vector_w_proof_halfway_3}, \eqref{eq:vector_w_proof_halfway_4}, and \eqref{eq:vector_w_proof_halfway_5}, for any $n \ge 2$, $\bvec{p} \in \Delta_{n}$, and $\alpha \in (0, 1) \cup (1, \infty]$, we have that there exists $p \in [1/n, 1]$ such that
\begin{align}
\| \bvec{w}_{n}( p ) \|_{\alpha}
& =
\| \bvec{p} \|_{\alpha} ,
\label{eq:vector_w_proof_1} \\
\| \bvec{w}_{n}( p ) \|_{\beta}
& \le
\| \bvec{p} \|_{\beta}
\quad \mathrm{for} \ \beta \in (\min\{ 1, \alpha \}, \max\{ 1, \alpha \}) ,
\label{eq:vector_w_proof_2} \\
\| \bvec{w}_{n}( p ) \|_{\beta}
& \ge
\| \bvec{p} \|_{\beta}
\quad \mathrm{for} \ \beta \in (0, \min\{ 1, \alpha \}) \cup (\max\{ 1, \alpha \}, \infty] ,
\label{eq:vector_w_proof_3}
\end{align}
where the inequality \eqref{eq:vector_w_proof_3} for $\beta = \infty$ follows from \eqref{eq:vector_w_proof_halfway_4}, \eqref{eq:vector_w_proof_halfway_5}, and the monotonicity of $\| \bvec{w}_{n}( p ) \|_{\infty}$ for $p \in [1/n, 1]$ (cf. \lemref{lem:Hv}).
This completes the proof of \lemref{lem:vector_w}.
\end{IEEEproof}

Lemmas~\ref{lem:vector_v}~and~\ref{lem:vector_w} show that, among all $n$-dimensional probability vectors with a fixed $\ell_{\alpha}$-norm for a given $\alpha \in (0, 1) \cup (1, \infty]$, the distributions $\bvec{v}_{n}( \cdot )$ and $\bvec{w}_{n}( \cdot )$ take extremal $\ell_{\beta}$-norm for all positive $\beta \ (\neq \alpha)$.
Combining Lemmas~\ref{lem:vector_v}~and~\ref{lem:vector_w}, we have the following theorem.

\begin{theorem}
\label{th:extremes}
For any $n \ge 2$, $\bvec{p} \in \Delta_{n}$, and $\alpha \in (0, 1) \cup (1, \infty]$, there exist unique numbers $p_{v} \in [0, 1/n]$ and $p_{w} \in [1/n, 1]$ such that 
\begin{align}
\| \bvec{p} \|_{\alpha}
=
\| \bvec{v}_{n}( p_{v} ) \|_{\alpha}
=
\| \bvec{w}_{n}( p_{w} ) \|_{\alpha} & ,
\label{eq:extremes_alpha} \\
\| \bvec{v}_{n}( p_{v} ) \|_{\beta} \le \| \bvec{p} \|_{\beta} \le \| \bvec{w}_{n}( p_{w} ) \|_{\beta}
& \qquad \mathrm{for} \ \mathrm{all} \ \beta \in (\min\{ 1, \alpha \}, \max\{ 1, \alpha \}) ,
\label{ineq:extremes_beta1} \\
\| \bvec{w}_{n}( p_{w} ) \|_{\beta} \le \| \bvec{p} \|_{\beta} \le \| \bvec{v}_{n}( p_{v} ) \|_{\beta}
& \qquad \mathrm{for} \ \mathrm{all} \ \beta \in (0, \min\{ 1, \alpha \}) \cup (\max\{ 1, \alpha \}, \infty] .
\label{ineq:extremes_beta2}
\end{align}
\end{theorem}

\begin{IEEEproof}[Proof of Theorem \ref{th:extremes}]
For $n \ge 2$, $\bvec{p} \in \Delta_{n}$, and $\alpha \in (0, 1) \cup (1, \infty)$, it follows from Lemmas~\ref{lem:vector_v}~and~\ref{lem:vector_w} that there exist $p_{v} \in [0, 1/n]$ and $p_{w} \in [1/n, 1]$ such that satisfy the following:
\begin{align}
\| \bvec{p} \|_{\alpha}
=
\| \bvec{v}_{n}( p_{v} ) \|_{\alpha}
=
\| \bvec{w}_{n}( p_{w} ) \|_{\alpha} & ,
\label{eq:same_norm_q} \\
\| \bvec{w}_{n}( p_{w} ) \|_{\beta} \le \| \bvec{p} \|_{\beta} \le \| \bvec{v}_{n}( p_{v} ) \|_{\beta}
& \qquad \mathrm{for} \ \mathrm{all} \ \beta \in (\min\{ 1, \alpha \}, \max\{ 1, \alpha \}) ,
\\
\| \bvec{v}_{n}( p_{v} ) \|_{\beta} \le \| \bvec{p} \|_{\beta} \le \| \bvec{w}_{n}( p_{w} ) \|_{\beta}
& \qquad \mathrm{for} \ \mathrm{all} \ \beta \in (0, \min\{ 1, \alpha \}) \cup (\max\{ 1, \alpha \}, \infty) ,
\end{align}
which are equivalent to \eqref{eq:extremes_alpha}--\eqref{ineq:extremes_beta2}.
The uniqueness of the values $p_{v}$ and $p_{w}$ follow by \lemref{lem:Hv}. 
\end{IEEEproof}

\begin{figure*}[!t]
\centering
\subfloat[Case: $\alpha = 3$ and $\beta = 1/2$.]{
\begin{overpic}[width = 0.3\hsize, clip]{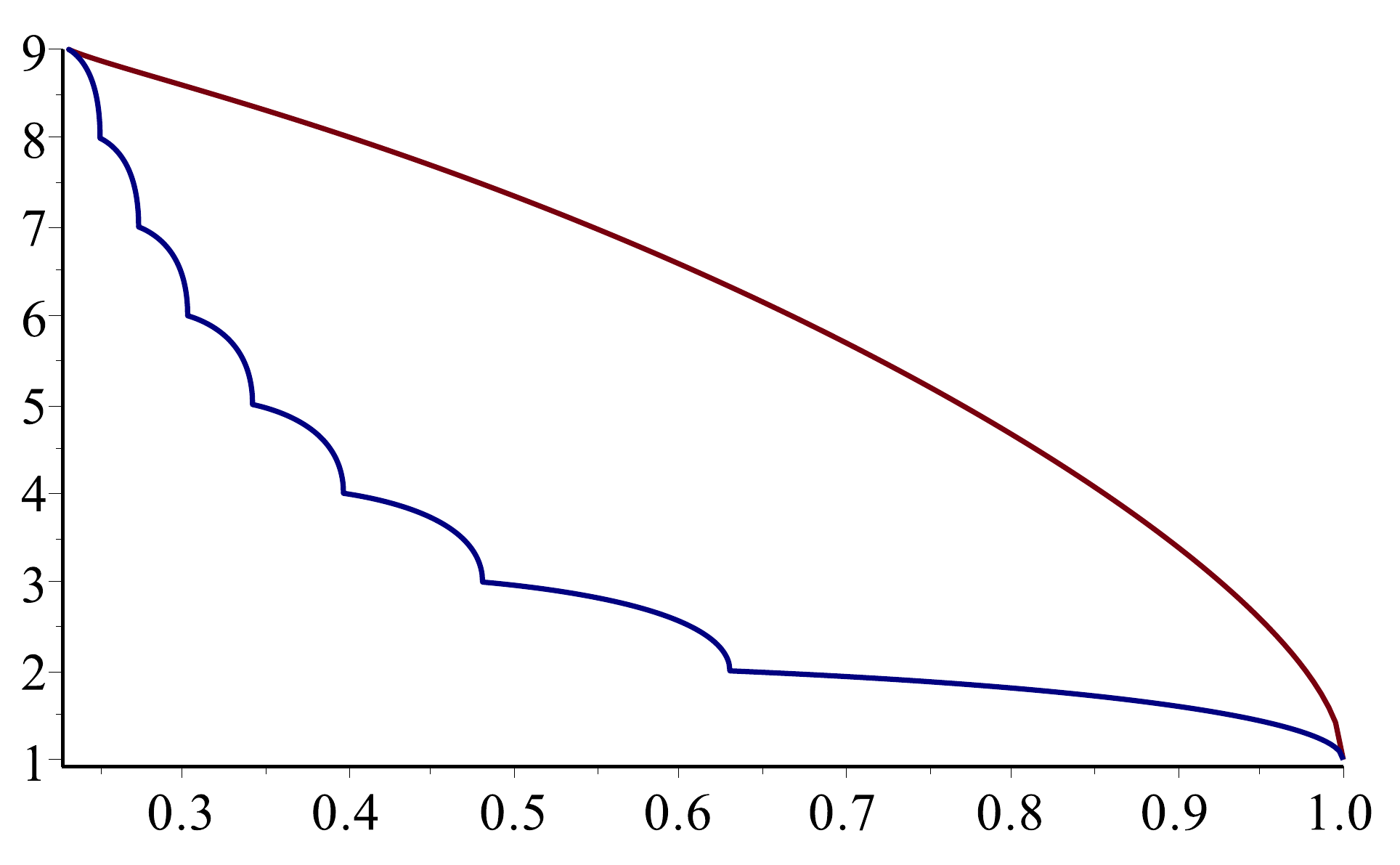}
\put(88, -3){$\| \bvec{p} \|_{\alpha}$}
\put(-6, 28){\rotatebox{90}{$\| \bvec{p} \|_{\beta}$}}
\put(60, 42){\color{burgundy} $\bvec{v}_{n}( \cdot )$}
\put(15, 18){\color{navyblue} $\bvec{w}_{n}( \cdot )$}
\end{overpic}
}\hspace{0.01\hsize}
\subfloat[Case: $\alpha = 3$ and $\beta = 3/2$.]{
\begin{overpic}[width = 0.3\hsize, clip]{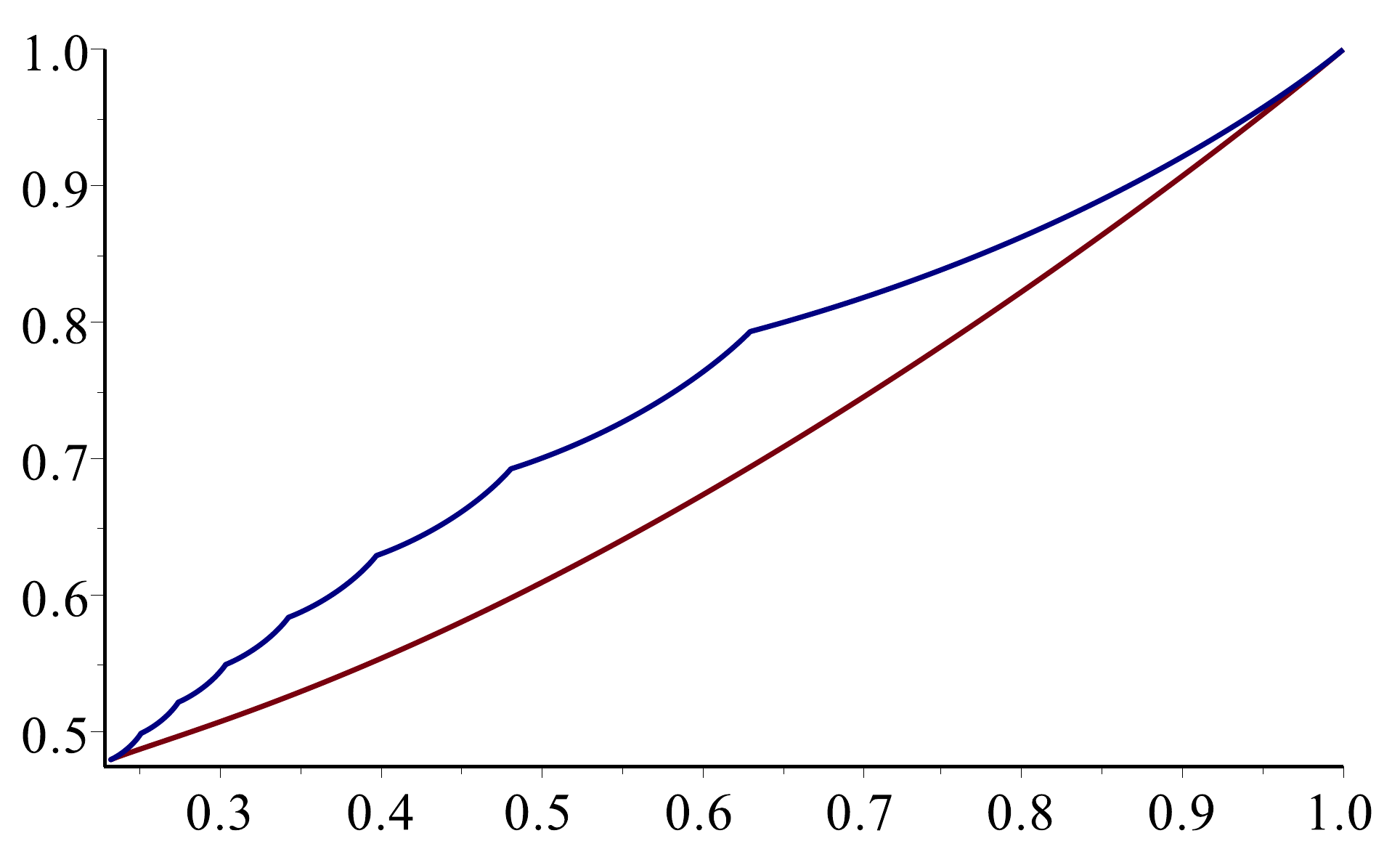}
\put(88, -3){$\| \bvec{p} \|_{\alpha}$}
\put(-6, 28){\rotatebox{90}{$\| \bvec{p} \|_{\beta}$}}
\put(70, 30){\color{burgundy} $\bvec{v}_{n}( \cdot )$}
\put(30, 35){\color{navyblue} $\bvec{w}_{n}( \cdot )$}
\end{overpic}
}\hspace{0.01\hsize}
\subfloat[Case: $\alpha = 3$ and $\beta = 100$.]{
\begin{overpic}[width = 0.3\hsize, clip]{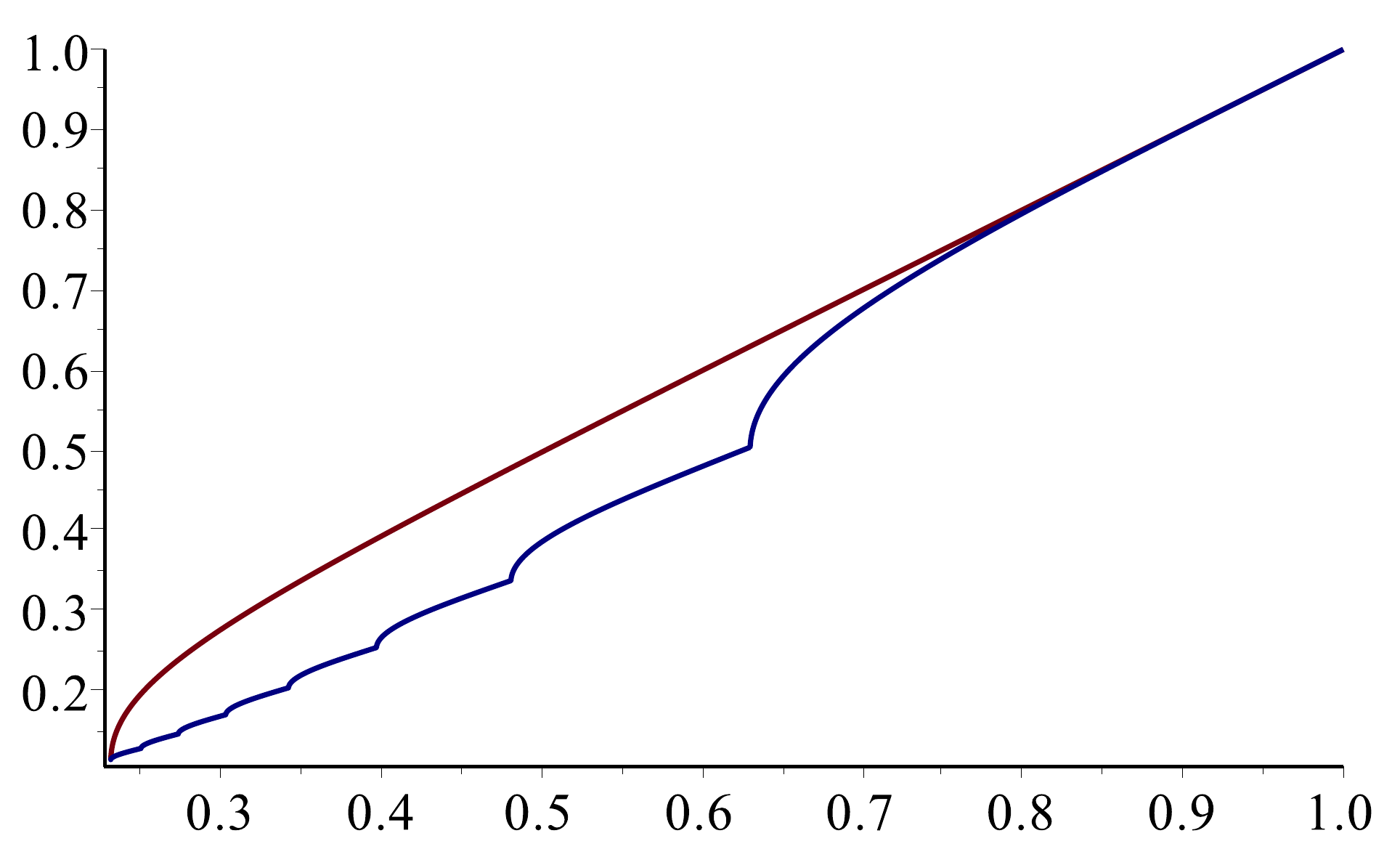}
\put(88, -3){$\| \bvec{p} \|_{\alpha}$}
\put(-6, 28){\rotatebox{90}{$\| \bvec{p} \|_{\beta}$}}
\put(28, 34){\color{burgundy} $\bvec{v}_{n}( \cdot )$}
\put(60, 27){\color{navyblue} $\bvec{w}_{n}( \cdot )$}
\end{overpic}
}
\caption{Sharp bounds between $\| \bvec{p} \|_{\alpha}$ and $\| \bvec{p} \|_{\beta}$ for $\bvec{p} \in \Delta_{n}$ with $n = 9$.
The boundaries are correspond to either the distribution $\bvec{v}_{n}( \cdot )$ or $\bvec{w}_{n}( \cdot )$.}
\label{fig:norm}
\end{figure*}
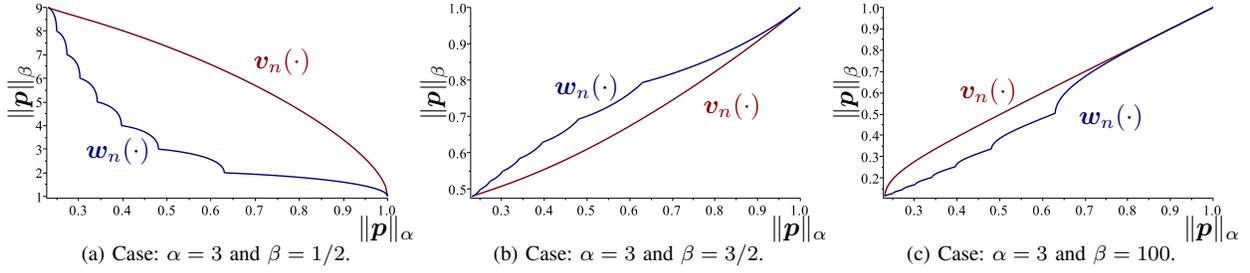

\thref{th:extremes} shows that, among all $n$-dimensional probability vectors with a fixed $\ell_{\alpha}$-norm for a given $\alpha \in (0, 1) \cup (1, \infty]$, the distributions $\bvec{v}_{n}( \cdot )$ and $\bvec{w}_{n}( \cdot )$ take the extremal $\ell_{\beta}$-norm for $\beta \in (0, 1) \cup (1, \infty]$, $\beta \neq \alpha$.
Hence, \thref{th:extremes} identifies the boundaries of the region
\begin{align}
\mathcal{R}_{n}( \alpha, \beta )
\coloneqq
\big\{ \big( \| \bvec{p} \|_{\alpha}, \| \bvec{p} \|_{\beta} \big) \: \big| \ \bvec{p} \in \Delta_{n} \big\}
\label{eq:region_unconditional}
\end{align}
for $n \ge 2$ and $\alpha, \beta \in (0, 1) \cup (1, \infty]$, $\alpha \neq \beta$.
We illustrate boundaries of $\mathcal{R}_{n}( \alpha, \beta )$ in \figref{fig:norm}.

By using \thref{th:extremes}, we derive the sharp bounds of the R\'{e}nyi entropy of order $\beta$ with a fixed R\'{e}nyi entropy of another order $\alpha$, as shown in \thref{th:R_extremes}.

\begin{theorem}
\label{th:R_extremes}
For any $n \ge 2$, $\bvec{p} \in \Delta_{n}$, and $\alpha \in (0, \infty]$, there exist unique numbers $p_{v} \in [0, 1/n]$ and $p_{w} \in [1/n, 1]$ such that 
\begin{align}
H_{\alpha}( \bvec{p} )
=
H_{\alpha}( \bvec{v}_{n}( p_{v} ) )
=
H_{\alpha}( \bvec{w}_{n}( p_{w} ) ) & ,
\label{eq:R_extremes_alpha} \\
H_{\beta}( \bvec{w}_{n}( p_{w} ) ) \le H_{\beta}( \bvec{p} ) \le H_{\beta}( \bvec{v}_{n}( p_{v} ) )
& \qquad \mathrm{for} \ \mathrm{all} \ \beta \in (0, \alpha) ,
\label{ineq:R_extremes_beta1} \\
H_{\beta}( \bvec{v}_{n}( p_{v} ) ) \le H_{\beta}( \bvec{p} ) \le H_{\beta}( \bvec{w}_{n}( p_{w} ) )
& \qquad \mathrm{for} \ \mathrm{all} \ \beta \in (\alpha, \infty] .
\label{ineq:R_extremes_beta2}
\end{align}
\end{theorem}

\begin{IEEEproof}[Proof of \thref{th:R_extremes}]
If $\alpha = 1$, then \thref{th:R_extremes} is reduced to \cite[Corollary~1]{part1}.
Similarly, if $\beta = 1$, then \thref{th:R_extremes} is also reduced to \cite[Theorem~2]{part1}.
Therefore, in this proof, it is enough to prove \thref{th:R_extremes} for $\alpha \in (0, 1) \cup (1, \infty]$ and $\beta \in (0, 1) \cup (1, \infty]$.

Consider the function
\begin{align}
f_{\alpha}( x )
=
\begin{cases}
\dfrac{ \alpha }{ 1 - \alpha } \ln x
& \mathrm{if} \ \alpha \in (0, 1) \cup (1, \infty) , \\
- \ln x
& \mathrm{if} \ \alpha = \infty
\end{cases}
\label{def:f_alpha}
\end{align}
for $x > 0$.
Then, we readily see that
\begin{align}
H_{\alpha}( \bvec{p} )
& =
f_{\alpha}( \| \bvec{p} \|_{\alpha} )
\label{eq:R_f_alpha}
\end{align}
for $\alpha \in (0, 1) \cup (1, \infty]$ and $\bvec{p} \in \Delta_{n}$.
It is clear that $f_{\alpha}( x )$ is a strictly monotonic function of $x > 0$ for every $\alpha \in (0, 1) \cup (1, \infty]$;
hence, it follows from \eqref{eq:R_f_alpha} that \eqref{ineq:extremes_beta1} of \thref{th:extremes} implies the equalities:
\begin{align}
H_{\alpha}( \bvec{p} )
=
H_{\alpha}( \bvec{v}_{n}( p_{v} ) )
=
H_{\alpha}( \bvec{w}_{n}( p_{w} ) ) ,
\end{align}
which is equivalent to \eqref{eq:R_extremes_alpha} of \thref{th:R_extremes}.

We now suppose that $\alpha \in (0, 1)$.
Then, Eq.~\eqref{ineq:extremes_beta1} of \thref{th:extremes} can be written as
\begin{align}
\| \bvec{v}_{n}( p_{v} ) \|_{\beta} \le \| \bvec{p} \|_{\beta} \le \| \bvec{w}_{n}( p_{w} ) \|_{\beta}
& \qquad \mathrm{for} \ \mathrm{all} \ \beta \in (\alpha, 1) .
\label{ineq:extremes_beta1_1}
\end{align}
Since $f_{\beta}( x )$ of \eqref{def:f_alpha} is a strictly increasing function of $x > 0$ for every $\beta \in (0, 1)$, it follows from \eqref{eq:R_f_alpha} that \eqref{ineq:extremes_beta1_1} implies the inequalities:
\begin{align}
H_{\beta}( \bvec{v}_{n}( p_{v} ) ) \le H_{\beta}( \bvec{p} ) \le H_{\beta}( \bvec{w}_{n}( p_{w} ) )
& \qquad \mathrm{for} \ \mathrm{all} \ \beta \in (\alpha, 1) .
\label{ineq:extremes_beta_1_1}
\end{align}
In addition, Eq.~\eqref{ineq:extremes_beta2} of \thref{th:extremes} can be written as
\begin{align}
\| \bvec{w}_{n}( p_{w} ) \|_{\beta} \le \| \bvec{p} \|_{\beta} \le \| \bvec{v}_{n}( p_{v} ) \|_{\beta}
& \qquad \mathrm{for} \ \mathrm{all} \ \beta \in (0, \alpha) \cup (1, \infty) .
\label{ineq:extremes_beta2_1}
\end{align}
Since $f_{\beta}( x )$ is a strictly increasing function of $x > 0$ for every $\beta \in (0, 1)$, it follows from \eqref{eq:R_f_alpha} that \eqref{ineq:extremes_beta2_1} implies the inequalities:
\begin{align}
H_{\beta}( \bvec{w}_{n}( p_{w} ) ) \le H_{\beta}( \bvec{p} ) \le H_{\beta}( \bvec{v}_{n}( p_{v} ) )
& \qquad \mathrm{for} \ \mathrm{all} \ \beta \in (0, \alpha) ;
\label{ineq:extremes_beta_1_2}
\end{align}
similarly, since $f_{\beta}( x )$ is a strictly decreasing function of $x > 0$ for every $\beta \in (1, \infty]$, it follows from \eqref{eq:R_f_alpha} that \eqref{ineq:extremes_beta2_1} implies the inequalities:
\begin{align}
H_{\beta}( \bvec{v}_{n}( p_{v} ) ) \le H_{\beta}( \bvec{p} ) \le H_{\beta}( \bvec{w}_{n}( p_{w} ) )
& \qquad \mathrm{for} \ \mathrm{all} \ \beta \in (1, \infty] .
\label{ineq:extremes_beta_1_3}
\end{align}
Combining \eqref{ineq:extremes_beta_1_1}, \eqref{ineq:extremes_beta_1_2}, and \eqref{ineq:extremes_beta_1_3}, we have
\begin{align}
H_{\beta}( \bvec{w}_{n}( p_{w} ) ) \le H_{\beta}( \bvec{p} ) \le H_{\beta}( \bvec{v}_{n}( p_{v} ) )
& \qquad \mathrm{for} \ \mathrm{all} \ \beta \in (0, \alpha) ,
\\
H_{\beta}( \bvec{v}_{n}( p_{v} ) ) \le H_{\beta}( \bvec{p} ) \le H_{\beta}( \bvec{w}_{n}( p_{w} ) )
& \qquad \mathrm{for} \ \mathrm{all} \ \beta \in (\alpha, 1) \cup (1, \infty] ,
\end{align}
which are \eqref{ineq:R_extremes_beta1} and \eqref{ineq:R_extremes_beta2} of \thref{th:R_extremes}, respectively, when $\alpha \in (0, 1)$.

On the other hand, we now suppose that $\alpha \in (1, \infty]$.
Then, Eq.~\eqref{ineq:extremes_beta1} of \thref{th:extremes} can be written as
\begin{align}
\| \bvec{v}_{n}( p_{v} ) \|_{\beta} \le \| \bvec{p} \|_{\beta} \le \| \bvec{w}_{n}( p_{w} ) \|_{\beta}
& \qquad \mathrm{for} \ \mathrm{all} \ \beta \in (1, \alpha) .
\label{ineq:extremes_beta1_2}
\end{align}
Since $f_{\beta}( x )$ of \eqref{def:f_alpha} is a strictly decreasing function of $x > 0$ for every $\beta \in (1, \infty]$, it follows from \eqref{eq:R_f_alpha} that \eqref{ineq:extremes_beta1_2} implies the inequalities:
\begin{align}
H_{\beta}( \bvec{w}_{n}( p_{w} ) ) \le H_{\beta}( \bvec{p} ) \le H_{\beta}( \bvec{v}_{n}( p_{v} ) )
& \qquad \mathrm{for} \ \mathrm{all} \ \beta \in (1, \alpha) .
\label{ineq:extremes_beta_2_1}
\end{align}
In addition, Eq.~\eqref{ineq:extremes_beta2} of \thref{th:extremes} can be written as
\begin{align}
\| \bvec{w}_{n}( p_{w} ) \|_{\beta} \le \| \bvec{p} \|_{\beta} \le \| \bvec{v}_{n}( p_{v} ) \|_{\beta}
& \qquad \mathrm{for} \ \mathrm{all} \ \beta \in (0, 1) \cup (\alpha, \infty) .
\label{ineq:extremes_beta2_2}
\end{align}
Since $f_{\beta}( x )$ is a strictly increasing function of $x > 0$ for every $\beta \in (0, 1)$, it follows from \eqref{eq:R_f_alpha} that \eqref{ineq:extremes_beta2_2} implies the inequalities:
\begin{align}
H_{\beta}( \bvec{w}_{n}( p_{w} ) ) \le H_{\beta}( \bvec{p} ) \le H_{\beta}( \bvec{v}_{n}( p_{v} ) )
& \qquad \mathrm{for} \ \mathrm{all} \ \beta \in (0, 1) ;
\label{ineq:extremes_beta_2_2}
\end{align}
similarly, since $f_{\beta}( x )$ is a strictly decreasing function of $x > 0$ for every $\beta \in (1, \infty]$, it follows from \eqref{eq:R_f_alpha} that \eqref{ineq:extremes_beta2_2} implies the inequalities:
\begin{align}
H_{\beta}( \bvec{v}_{n}( p_{v} ) ) \le H_{\beta}( \bvec{p} ) \le H_{\beta}( \bvec{w}_{n}( p_{w} ) )
& \qquad \mathrm{for} \ \mathrm{all} \ \beta \in (\alpha, \infty] .
\label{ineq:extremes_beta_2_3}
\end{align}
Combining \eqref{ineq:extremes_beta_2_1}, \eqref{ineq:extremes_beta_2_2}, and \eqref{ineq:extremes_beta_2_3}, we have
\begin{align}
H_{\beta}( \bvec{w}_{n}( p_{w} ) ) \le H_{\beta}( \bvec{p} ) \le H_{\beta}( \bvec{v}_{n}( p_{v} ) )
& \qquad \mathrm{for} \ \mathrm{all} \ \beta \in (0, 1) ,
\\
H_{\beta}( \bvec{v}_{n}( p_{v} ) ) \le H_{\beta}( \bvec{p} ) \le H_{\beta}( \bvec{w}_{n}( p_{w} ) )
& \qquad \mathrm{for} \ \mathrm{all} \ \beta \in (0, 1) \cup (\alpha, \infty] ,
\end{align}
which are \eqref{ineq:R_extremes_beta1} and \eqref{ineq:R_extremes_beta2} of \thref{th:R_extremes}, respectively, when $\alpha \in (1, \infty]$.
\end{IEEEproof}

\begin{figure}[!t]
\centering
\begin{overpic}[width = 0.8\hsize, clip]{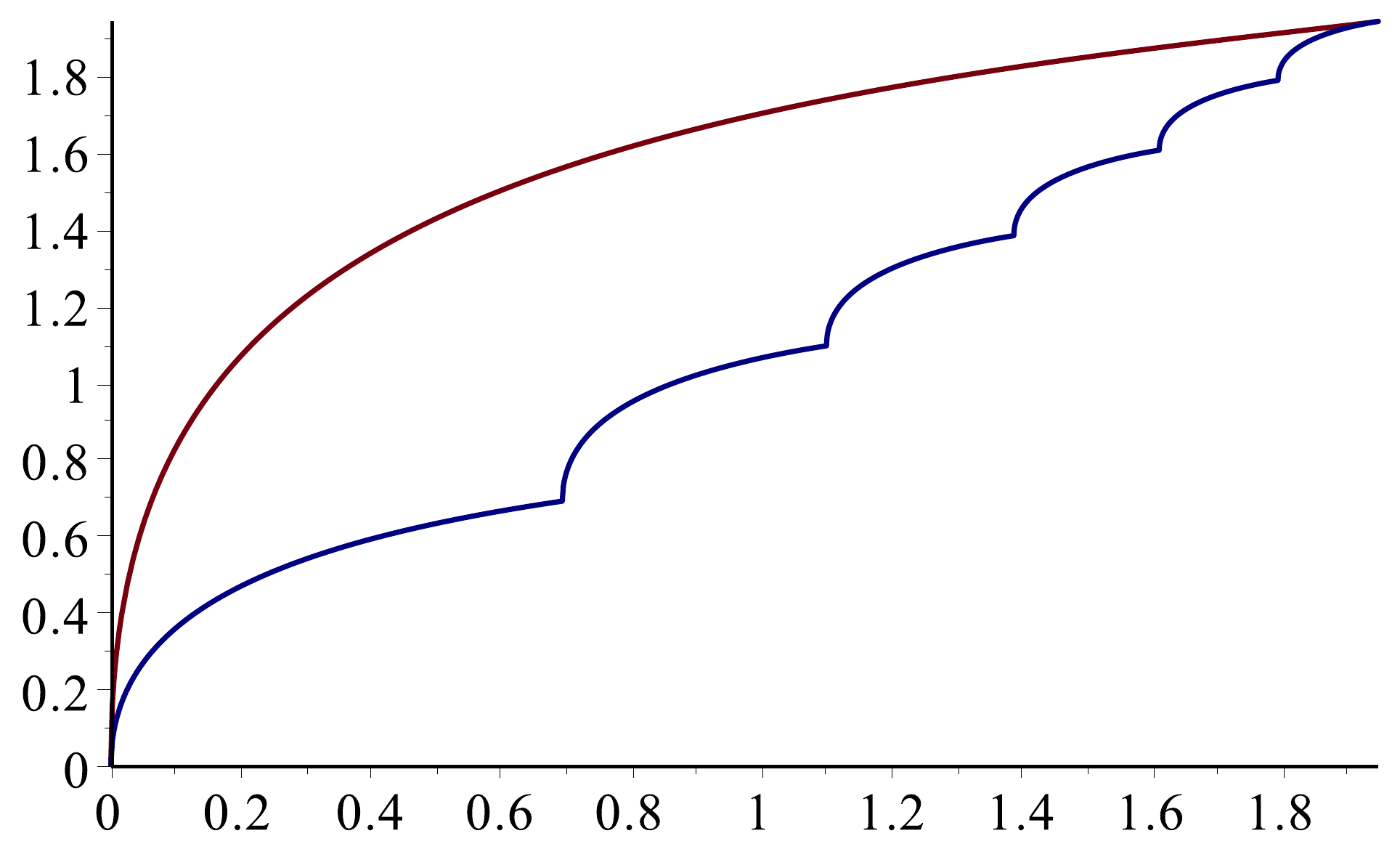}
\put(80, 10){$H_{\alpha}( \bvec{p} )$}
\put(98, 2){\scriptsize [nats]}
\put(-5, 30){\rotatebox{90}{$H_{\beta}( \bvec{p} )$}}
\put(-3, 60){\scriptsize [nats]}
\put(25, 51){\color{burgundy} $\bvec{v}_{n}( \cdot )$}
\put(70, 38){\color{navyblue} $\bvec{w}_{n}( \cdot )$}
\end{overpic}
\caption{Boundaries of the region $\big\{ \big( H_{\alpha}( \bvec{p} ), H_{\beta}( \bvec{p} ) \big) \ \big| \ \bvec{p} \in \Delta_{n} \big\}$ with $n = 7$, $\alpha = 2$, and $\beta = 1/2$.
The upper and lower bounds are established by the distributions $\bvec{v}_{n}( \cdot )$ and $\bvec{w}_{n}( \cdot )$, respectively.}
\label{fig:Renyi}
\end{figure}
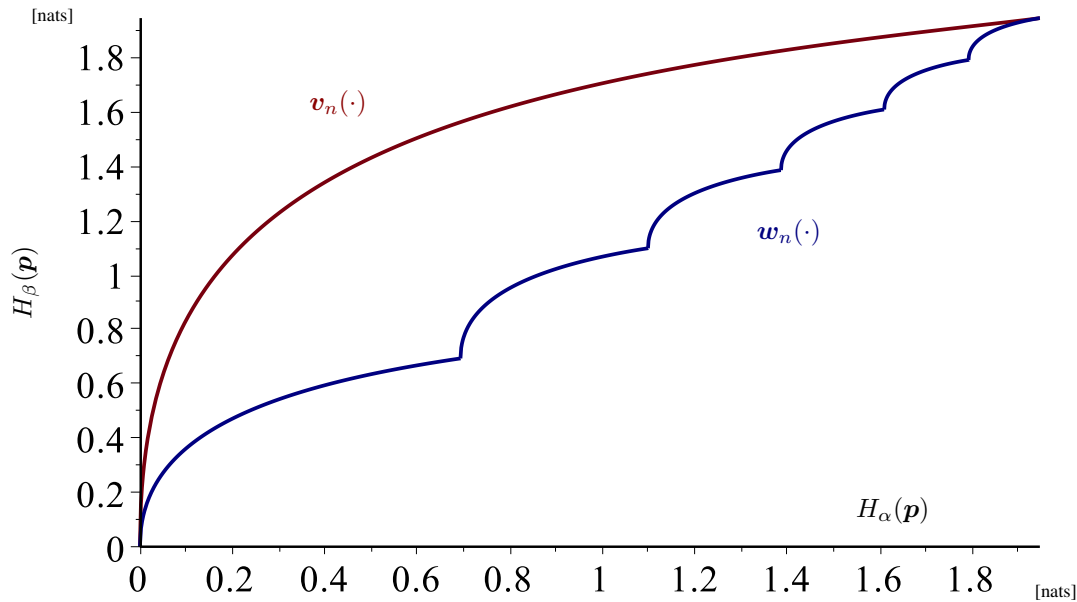

\thref{th:R_extremes} shows that, for a fixed R\'{e}nyi entropy of order $\alpha \in (0, \infty]$, the distributions $\bvec{v}_{n}( \cdot )$ and $\bvec{w}_{n}( \cdot )$ take the maximum and minimum R\'{e}nyi entropy of order $\beta \in (0, \alpha)$, respectively, and the distributions $\bvec{v}_{n}( \cdot )$ and $\bvec{w}_{n}( \cdot )$ take the minimum and maximum R\'{e}nyi entropy of order $\beta \in (\alpha, \infty]$, respectively.
Therefore, the distributions $\bvec{v}_{n}( \cdot )$ and $\bvec{w}_{n}( \cdot )$ have extremal properties in the sense between two R\'{e}nyi entropies of distinct positive orders.
We plot the bounds of \thref{th:R_extremes} in \figref{fig:Renyi}.

\begin{remark}
\label{remark:note}
We remark that \thref{th:R_extremes} is a generalization of \emph{\cite[Corollary~1]{part1}} since the R\'{e}nyi entropy of order $1$ is the Shannon entropy.
Many axiomatic definitions of entropy are closely related to the $\ell_{\alpha}$-norm of the probability distribution (cf. \emph{\cite[Table~1]{part1}}).
Therefore, in a similar way to \thref{th:R_extremes}, we can obtain the sharp bounds on several entropies \emph{\cite{tsallis, havrda, behara, daroczy, boekee}} with two distinct orders.
\end{remark}

We now consider to extend \thref{th:R_extremes} from the R\'{e}nyi entropy to the R\'{e}nyi divergence \cite{renyi}.
For $\bvec{p}, \bvec{q} \in \Delta_{n}$ such that $\bvec{p} \ll \bvec{q}$, i.e., $\bvec{p}$ is absolutely continuous with respect to $\bvec{q}$, the R\'{e}nyi divergence of order $\alpha$ is defined by
\begin{align}
D_{\alpha}( \bvec{p} \, \| \, \bvec{q} )
\coloneqq
\frac{ 1 }{ \alpha - 1 } \ln \sum_{i=1 : q_{i} > 0}^{n} p_{i}^{\alpha} q_{i}^{1-\alpha}
\end{align}
for $\alpha \in (0, 1) \cup (1, \infty)$.
Moreover, it is also defined that
\begin{align}
D_{1}( \bvec{p} \, \| \, \bvec{q} )
& \coloneqq
\lim_{\alpha \to 1} D_{\alpha}( \bvec{p} \, \| \, \bvec{q} )
=
\sum_{i=1 : p_{i} > 0}^{n} p_{i} \ln \frac{ p_{i} }{ q_{i} } ,
\label{def:KL} \\
D_{\infty}( \bvec{p} \, \| \, \bvec{q} )
& \coloneqq
\lim_{\alpha \to \infty} D_{\alpha}( \bvec{p} \, \| \, \bvec{q} )
=
\ln \max_{1 \le i \le n : q_{i} > 0} \bigg( \frac{ p_{i} }{ q_{i} } \bigg) ,
\end{align}
where the most right-hand side of \eqref{def:KL} is called the relative entropy or the Kullback-Leibler divergence.
For the $n$-ary uniform distribution
$
\bvec{u}_{n}
\coloneqq
( 1/n, 1/n, \dots, 1/n )
\in \Delta_{n}
$,
since
\begin{align}
D_{\alpha}( \bvec{p} \, \| \, \bvec{u}_{n} )
=
\ln n - H_{\alpha}( \bvec{p} )
\label{eq:Div_unif}
\end{align}
for $\alpha \in (0, \infty]$, the R\'{e}nyi divergence from the uniform distribution $\bvec{u}_{n}$ is a strictly decreasing function of the R\'{e}nyi entropy;
thus, the following corollary holds from \thref{th:R_extremes}.

\begin{corollary}
\label{cor:Div_extremes}
For any $n \ge 2$, $\bvec{p} \in \Delta_{n}$, and $\alpha \in (0, \infty]$, there exist unique values $p_{v} \in [0, 1/n]$ and $p_{w} \in [1/n, 1]$ such that
\begin{align}
D_{\alpha}( \bvec{p} \, \| \, \bvec{u}_{n} )
=
D_{\alpha}( \bvec{v}_{n}( p_{v} ) \, \| \, \bvec{u}_{n} )
& =
D_{\alpha}( \bvec{w}_{n}( p_{w} ) \, \| \, \bvec{u}_{n} ) ,
\\
D_{\beta}( \bvec{v}_{n}( p_{v} ) \, \| \, \bvec{u}_{n} )
\le
D_{\beta}( \bvec{p} \, \| \, \bvec{u}_{n} )
& \le
D_{\beta}( \bvec{w}_{n}( p_{w} ) \, \| \, \bvec{u}_{n} )
&& \mathrm{for} \ \mathrm{all} \ \beta \in (0, \alpha) ,
\\
D_{\beta}( \bvec{w}_{n}( p_{w} ) \, \| \, \bvec{u}_{n} )
\le
D_{\beta}( \bvec{p} \, \| \, \bvec{u}_{n} )
& \le
D_{\beta}( \bvec{v}_{n}( p_{v} ) \, \| \, \bvec{u}_{n} )
&& \mathrm{for} \ \mathrm{all} \ \beta \in (\alpha, \infty] .
\end{align}
\end{corollary}

As with \thref{th:R_extremes}, \corref{cor:Div_extremes} also shows the sharp bounds of R\'{e}nyi divergence from the uniform distribution of order $\beta$ with a fixed R\'{e}nyi divergence from the uniform distribution of another order $\alpha$.

\section{Feasible regions of Arimoto's conditional R\'{e}nyi entropies for two distinct orders}
\label{sect:conditional}

In this section, we extend the results of \sectref{sect:unconditional} from \emph{un}conditional settings to \emph{conditional} settings.
Consider a pair of discrete random variables $(X, Y) \sim P_{X|Y} P_{Y}$ such that $P_{Y}( y ) > 0$ for all $y \in \mathcal{Y}$.
The conditional Shannon entropy \cite{shannon} of $X$ given $Y$ is defined by
\begin{align}
H(X \mid Y)
& \coloneqq
\sum_{y \in \mathcal{Y}} P_{Y}( y ) H(X \mid Y = y)
\\
& =
\sum_{y \in \mathcal{Y}} P_{Y}( y ) \Bigg( - \sum_{x \in \mathcal{X}} P_{X|Y}(x \mid y) \ln P_{X|Y}(x \mid y) \Bigg) ,
\end{align}
which is also called the equivocation.
When we denote by
\begin{align}
N_{\alpha}(X \mid Y)
& \coloneqq
\sum_{y \in \mathcal{Y}} P_{Y}( y ) \Bigg( \sum_{x \in \mathcal{X}} P_{X|Y}(x \mid y)^{\alpha} \Bigg)^{1/\alpha}
\label{def:N}
\end{align}
the expectation of $\ell_{\alpha}$-norm for $\alpha \in (0, \infty]$ and $(X, Y) \sim P_{X|Y} P_{Y}$, Arimoto \cite{arimoto2} defined the conditional R\'{e}nyi entropy of order $\alpha$ as
\begin{align}
H_{\alpha}(X \mid Y)
\coloneqq
\frac{ \alpha }{ 1 - \alpha } \ln N_{\alpha}(X \mid Y)
\label{def:cond_Renyi}
\end{align}
for $\alpha \in (0, 1) \cup (1, \infty)$.
As with \eqref{def:Renyi_1} and \eqref{def:Renyi_infty}, it is defined that
\begin{align}
H_{1}(X \mid Y)
& \coloneqq
\lim_{\alpha \to 1} H_{\alpha}(X \mid Y)
=
H(X \mid Y) ,
\label{eq:lim_Hto1} \\
H_{\infty}(X \mid Y)
& \coloneqq
\lim_{\alpha \to \infty} H_{\alpha}(X \mid Y)
=
- \ln N_{\infty}(X \mid Y) ,
\label{eq:lim_Htoinfty}
\end{align}
where the last equality of \eqref{eq:lim_Hto1} is shown in \cite[Theorem~2]{arimoto2} and \cite[Proposition~1]{fehr}, and the last equality of \eqref{eq:lim_Htoinfty} is shown in \cite[Proposition~1]{fehr}.
%
If the cardinality of the finite set is denoted by $| \cdot |$, for the region
\begin{align}
\mathcal{R}_{n}^{\mathrm{cond}}( \alpha, \beta )
\coloneqq
\Big\{ \big( N_{\alpha}(X \mid Y), N_{\beta}(X \mid Y) \big) \ \Big| \
(X, Y) \in \mathcal{X} \times \mathcal{Y} , \, |\mathcal{X}| = n, \ \mathrm{and} \ | \mathcal{Y} | \ge 2
\Big\} ,
\end{align}
we present the following theorem.

\begin{theorem}
\label{th:convexhull}
$
\mathcal{R}_{n}^{\mathrm{cond}}( \alpha, \beta )
=
\mathrm{Conv} \big( \mathcal{R}_{n}( \alpha, \beta ) \big)
$,
where $\mathrm{Conv} \big( \mathcal{R} \big)$ denotes the convex hull of the set $\mathcal{R} \subset \mathbb{R}^{k} \ (k \in \mathbb{N})$.
\end{theorem}

\begin{IEEEproof}[Proof of \thref{th:convexhull}]
We prove \thref{th:convexhull} in a similar way to \cite[Theorem~3]{part2}.
It is clear that if $| \mathcal{X} | = n$, then the point $\big( N_{\alpha}(X \mid Y), N_{\beta}(X \mid Y) \big)$ is a convex combination of the points $\big( \| \bvec{p}_{i} \|_{\alpha}, \| \bvec{p}_{i} \|_{\beta} \big)$ for $\bvec{p}_{1}, \bvec{p}_{2}, \dots \in \Delta_{n}$ (cf. \eqref{def:N}).
Therefore, \thref{th:convexhull} holds.
\end{IEEEproof}

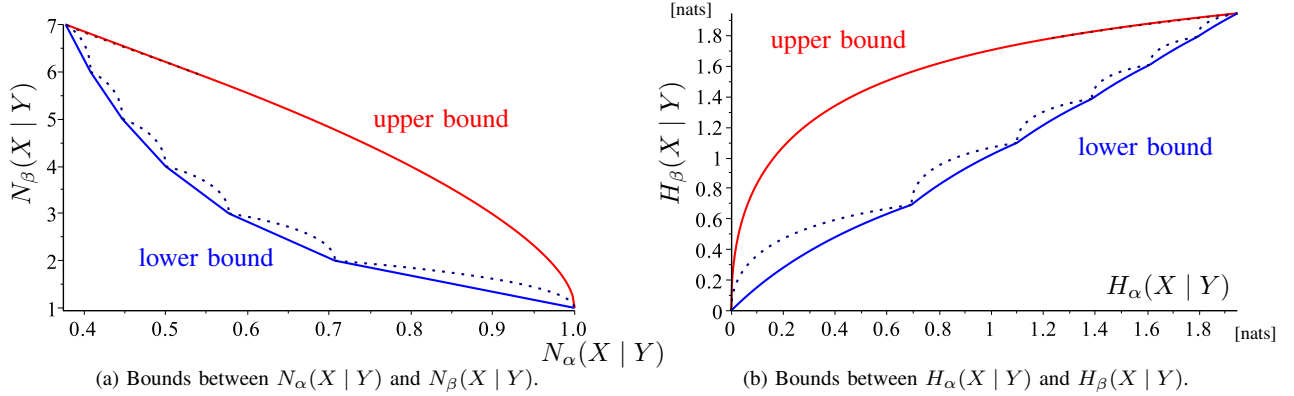
\begin{figure}[!t]
\centering
\subfloat[Bounds between $N_{\alpha}(X \mid Y)$ and $N_{\beta}(X \mid Y)$.]{
\begin{overpic}[width = 0.45\hsize, clip]{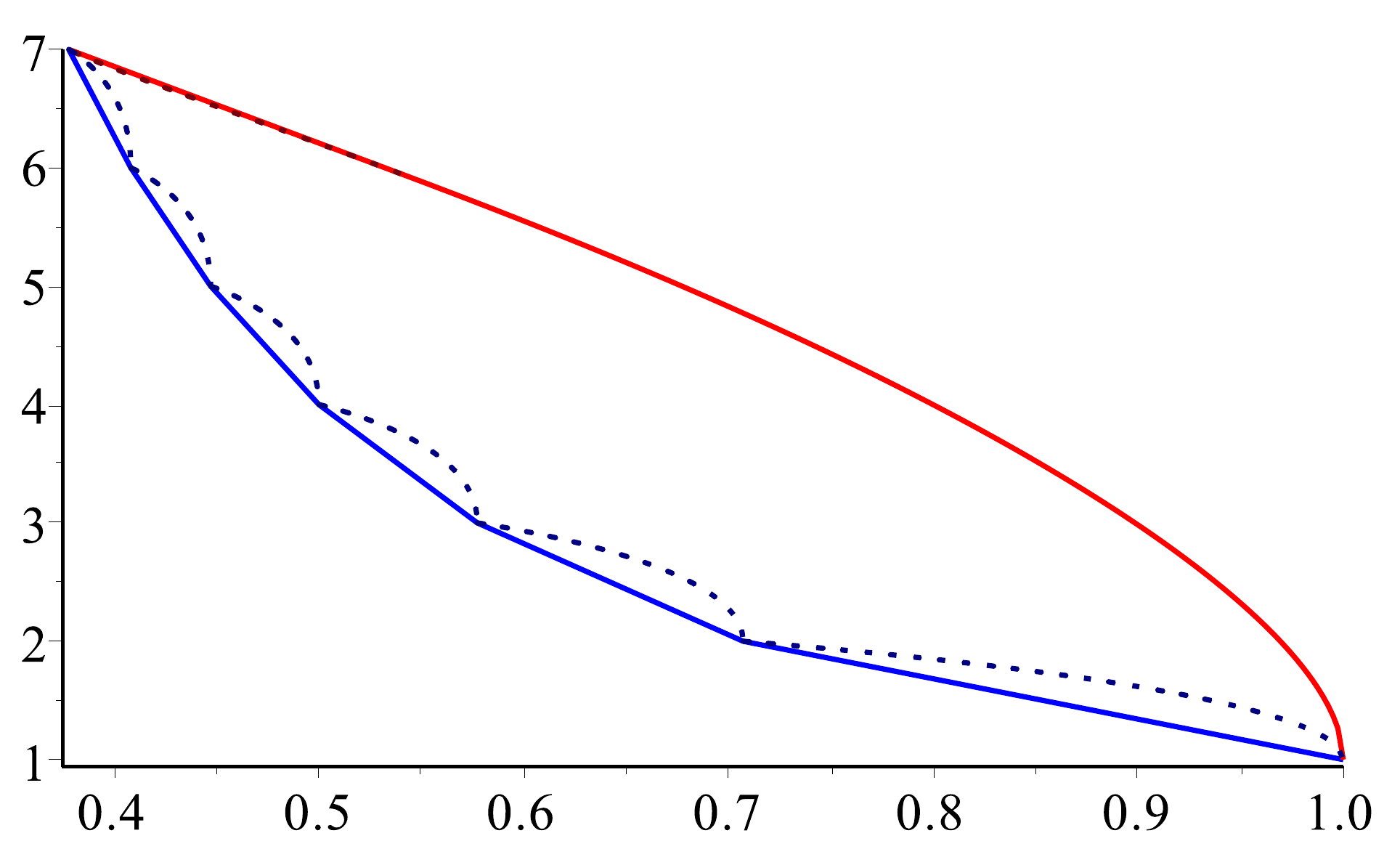}
\put(90, -2){$N_{\alpha}(X \mid Y)$}
\put(-5, 26){\rotatebox{90}{$N_{\beta}(X \mid Y)$}}
\put(18, 15){\color{blue} lower bound}
\put(60, 40){\color{red} upper bound}
\end{overpic}
}\hspace{0.05\hsize}
\subfloat[Bounds between $H_{\alpha}(X \mid Y)$ and $H_{\beta}(X \mid Y)$.]{
\label{subfig:convexhull_R}
\begin{overpic}[width = 0.45\hsize, clip]{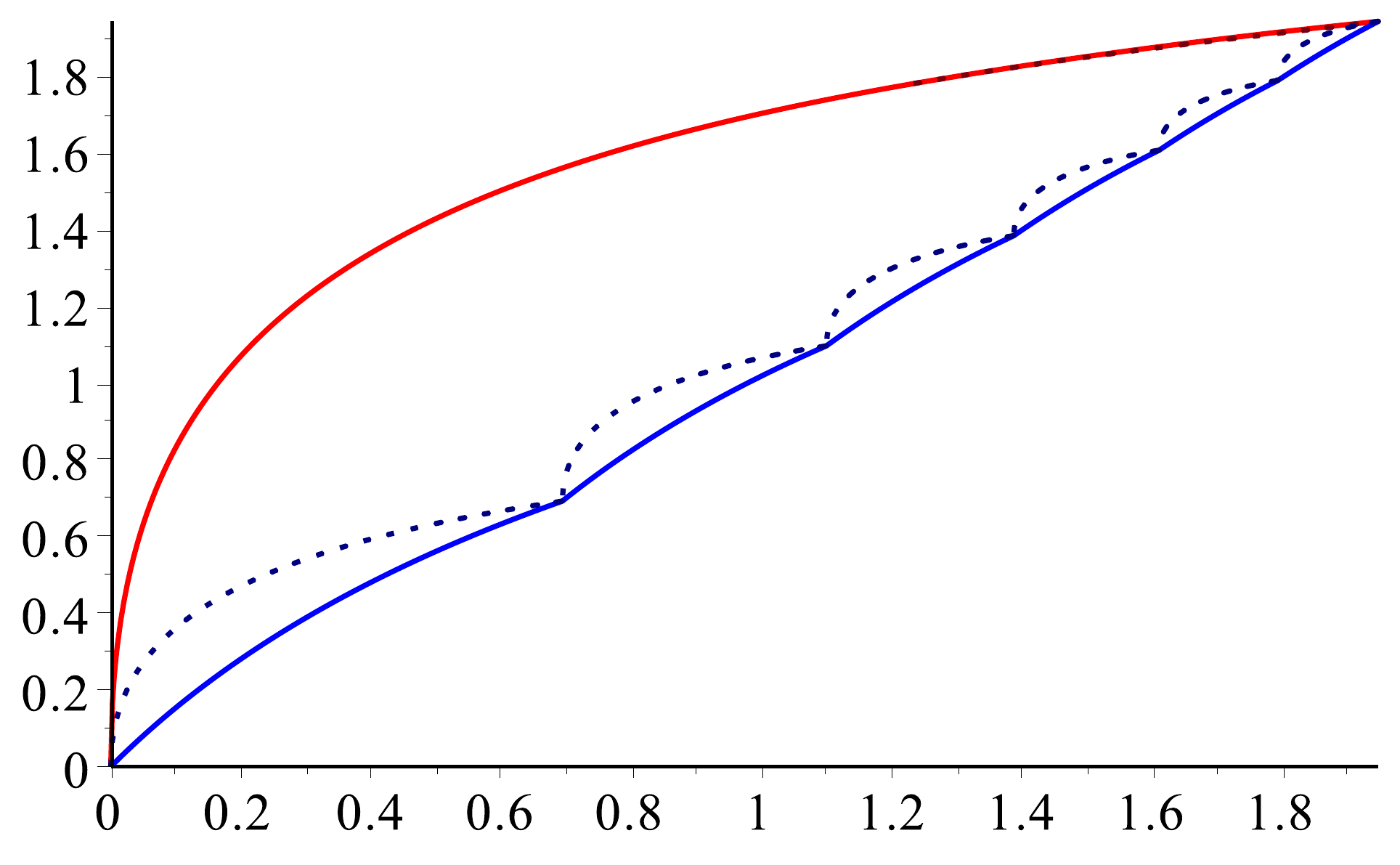}
\put(75, 10){$H_{\alpha}(X \mid Y)$}
\put(98, 2){\scriptsize [nats]}
\put(-5, 26){\rotatebox{90}{$H_{\beta}(X \mid Y)$}}
\put(-3, 60){\scriptsize [nats]}
\put(70, 35){\color{blue} lower bound}
\put(15, 54){\color{red} upper bound}
\end{overpic}
}
\caption{Boundaries of the region $\mathcal{R}_{n}^{\mathrm{cond}}( \alpha, \beta )$ with $n = 7$, $\alpha = 2$, and $\beta = 1/2$, and its application to the conditional R\'{e}nyi entropy.
The dotted lines correspond to the boundaries in \emph{un}conditional settings (cf. \figref{fig:Renyi}).}
\label{fig:convexhull}
\end{figure}

It is worth noting that $\mathcal{R}_{n}( \alpha, \beta )$ is derived from \thref{th:extremes} (cf. \eqref{eq:region_unconditional}).
Since the points $\big( N_{\alpha}(X \mid Y), N_{\beta}(X \mid Y) \big)$ and $\big( H_{\alpha}(X \mid Y), H_{\beta}(X \mid Y) \big)$ are in one-to-one correspondence (cf. \eqref{def:cond_Renyi}) for every distinct $\alpha, \beta \in (0, 1) \cup (1, \infty]$, we obtain the exact feasible regions of two conditional R\'{e}nyi entropies of distinct orders from \thref{th:convexhull}, where the exact feasible regions when the order $\alpha$ or $\beta$ is $1$ are derived from \cite[Theorem~3]{part2}.
We illustrate an application of \thref{th:convexhull} in \figref{fig:convexhull}.
In \figref{fig:convexhull}-\subref{subfig:convexhull_R}, note that the dotted lines are identical to the boundaries shown in \figref{fig:Renyi}.



\section{Uniformly focusing channels}
\label{sect:focusing}

We consider a discrete memoryless channel (DMC) as an application of the previous sections in the rest of paper.
Let finite sets $\mathcal{X}$ and $\mathcal{Y}$ be the input and output alphabets of a DMC, respectively.
Let random variables $X \in \mathcal{X}$ and $Y \in \mathcal{Y}$ be the input and output of a DMC, respectively.
The input distribution of a DMC is denoted by $P_{X}( x )$ for $x \in \mathcal{X}$.
In particular, let $U_{X}$ be the uniform input distribution on $\mathcal{X}$, i.e., $U_{X}( x ) = 1 / |\mathcal{X}|$ for all $x \in \mathcal{X}$. 
Then, the DMC consists of a transition probability distribution $\big\{ P_{Y|X}(y \mid x) \mid (x, y) \in \mathcal{X} \times \mathcal{Y} \big\}$.
We now define the following three classes of DMCs.

\begin{definition}
A DMC $P_{Y|X}$ is said to be \emph{uniformly dispersive \cite{massey}} or \emph{uniform from the input \cite{fano2}} if there exists a permutation $\pi_{x} : \mathcal{Y} \to \mathcal{Y}$ for each $x \in \mathcal{X}$ such that
\begin{align}
P_{Y|X}(\pi_{x}( y ) \mid x)
=
P_{Y|X}(\pi_{x^{\prime}}( y ) \mid x^{\prime})
\end{align}
for all $(x, x^{\prime}, y) \in \mathcal{X}^{2} \times \mathcal{Y}$.
\end{definition}

\begin{definition}
\label{def:focusing}
A DMC $P_{Y|X}$ is said to be \emph{uniformly focusing \cite{massey}} or \emph{uniform from the output \cite{fano2}} if there exists a permutation $\pi_{y} : \mathcal{X} \to \mathcal{X}$ for each $y \in \mathcal{Y}$ such that
\begin{align}
P_{Y|X}(y \mid \pi_{y}( x ))
=
P_{Y|X}(y^{\prime} \mid \pi_{y^{\prime}}( x ))
\end{align}
for all $(x, y, y^{\prime}) \in \mathcal{X} \times \mathcal{Y}^{2}$.
\end{definition}

\begin{definition}
\label{def:ssc}
A DMC is said to be \emph{strongly symmetric \cite{massey}} or \emph{doubly uniform \cite{fano2}} if it is both uniformly dispersive and uniformly focusing.
\end{definition}

If suppose that $\mathcal{X} = \mathcal{Y} = \{ 0, 1, \dots, n-1 \}$, we define the following two $n$-ary input and output strongly symmetric channels:
(i) the strongly symmetric channel $V_{Y|X} : \mathcal{X} \to \mathcal{Y}$ is defined by
\begin{align}
V_{Y|X}(y \mid x)
\coloneqq
\begin{cases}
1 - (n-1) p
& \mathrm{if} \ y = x , \\
p
& \mathrm{if} \ y \neq x
\end{cases}
\label{def:channel_v}
\end{align}
for $(x, y) \in \mathcal{X} \times \mathcal{Y}$ and some $p \in [0, 1/n]$, and
(ii) the strongly symmetric channel $W_{Y|X} : \mathcal{X} \to \mathcal{Y}$ is defined by
\begin{align}
W_{Y|X}(y \mid x)
\coloneqq
\begin{cases}
p
& \mathrm{if} \ y \equiv x + i \ (\bmod \ n) \quad \mathrm{for} \ 0 \le i < \lfloor 1/p \rfloor , \\
1 - \lfloor 1/p \rfloor \, p
& \mathrm{if} \ y \equiv x + \lfloor 1/p \rfloor \ (\bmod \ n) , \\
0
& \mathrm{otherwise}
\end{cases}
\label{def:channel_w}
\end{align}
for $(x, y) \in \mathcal{X} \times \mathcal{Y}$ and some $p \in [1/n, 1]$.
It is clear that, for all $x \in \mathcal{X}$, the decreasing orders of the conditional distributions $\{ V_{Y|X}( y \mid x ) \mid y \in \mathcal{Y} \}$ and $\{ W_{Y|X}( y \mid x ) \mid y \in \mathcal{Y} \}$ are identical to the distributions $\bvec{v}_{n}( \cdot )$ and $\bvec{w}_{n}( \cdot )$, respectively.
Note that the channel $V_{Y|X}$ is sometimes called the $n$-ary symmetric channel.

For a DMC $(X, Y) \sim P_{X} P_{Y|X}$, Arimoto \cite{arimoto2} proposed the mutual information of order $\alpha$ as
\begin{align}
I_{\alpha}( P_{X}; P_{Y|X} )
& \coloneqq
H_{\alpha}(X) - H_{\alpha}(X \mid Y)
\end{align}
for $\alpha \in (0, \infty]$, where note that $I(P_{X}; P_{Y|X}) \coloneqq I_{1}(P_{X}; P_{Y|X})$ is the (ordinary) mutual information.
Since
\begin{align}
I_{\alpha}( U_{X}; P_{Y|X} )
=
D_{\alpha}( P_{X|Y}( \cdot \mid y ) \, \| \, U_{X} )
\label{eq:mutual_Div}
\end{align}
for any uniformly focusing channel $P_{Y|X} : \mathcal{X} \to \mathcal{Y}$ and all $y \in \mathcal{Y}$ (cf. \cite[Eq.~(325)]{part1}),
we apply \corref{cor:Div_extremes} to uniformly focusing channels in \thref{th:focusing}.

\begin{theorem}
\label{th:focusing}
For any uniformly focusing channel $P_{Y|X}$ and $\alpha \in (0, \infty]$, there exist unique channels $V_{Y|X}$ and $W_{Y|X}$ such that
\begin{align}
I_{\alpha}(U_{X}; P_{Y|X})
=
I_{\alpha}(U_{X}; V_{Y|X})
=
I_{\alpha}(U_{X}; W_{Y|X}) ,
& \quad
\\
I_{\beta}(U_{X}; V_{Y|X})
\le
I_{\beta}(U_{X}; P_{Y|X})
\le
I_{\beta}(U_{X}; W_{Y|X})
& \quad \mathrm{for} \ \mathrm{all} \ \beta \in (0, \alpha) ,
\\
I_{\beta}(U_{X}; W_{Y|X})
\le
I_{\beta}(U_{X}; P_{Y|X})
\le
I_{\beta}(U_{X}; V_{Y|X})
& \quad \mathrm{for} \ \mathrm{all} \ \beta \in (\alpha, \infty] .
\end{align}
\end{theorem}

\begin{IEEEproof}[Proof of \thref{th:focusing}]
It is easy to see that
\begin{align}
\sum_{x \in \mathcal{X}} V_{Y|X}(y \mid x)
& =
\sum_{x \in \mathcal{X}} W_{Y|X}(y \mid x)
=
1
\label{eq:sum_vw_1}
\end{align}
for all $y \in \mathcal{Y}$.
For a DMC $(X, Y) \sim U_{X} V_{Y|X}$, defined in \eqref{def:channel_v}, we readily see that
\begin{align}
P_{X|Y}(\cdot \mid y)
& =
\frac{ (1 / |\mathcal{X}|) V_{Y|X}(y \mid \cdot) }{ \sum_{x^{\prime} \in \mathcal{X}} (1 / |\mathcal{X}|) V_{Y|X}( y \mid x^{\prime} ) }
\\
& =
\frac{ V_{Y|X}(y \mid \cdot) }{ \sum_{x^{\prime} \in \mathcal{X}} V_{Y|X}( y \mid x^{\prime} ) }
\\
& \overset{\eqref{eq:sum_vw_1}}{=}
V_{Y|X}(y \mid \cdot)
\\
& =
\bvec{v}_{n}( p_{v} )
\end{align}
for all $y \in \mathcal{Y}$ and some $p_{v} \in [0, 1/n]$.
Similarly, for a DMC $(X, Y) \sim U_{X} W_{Y|X}$, defined in \eqref{def:channel_w}, it also follows that
\begin{align}
P_{X|Y}(\cdot \mid y)
& =
W_{Y|X}(y \mid \cdot)
\\
& =
\bvec{w}_{n}( p_{w} )
\end{align}
for all $y \in \mathcal{Y}$ and some $p_{w} \in [1/n, 1]$.
Hence, \thref{th:focusing} directly follows from \corref{cor:Div_extremes} and \eqref{eq:mutual_Div}.
\end{IEEEproof}

In \thref{th:focusing}, we note that the input alphabets of channels $P_{Y|X}$, $V_{Y|X}$, and $W_{Y|X}$ are identical.
Therefore, as with $\bvec{v}_{n}( \cdot )$ and $\bvec{w}_{n}( \cdot )$ for \thref{th:R_extremes} and \corref{cor:Div_extremes}, the strongly symmetric channels $V_{Y|X}$ and $W_{Y|X}$ have extremal properties in the sense of the mutual information of order $\alpha$ for uniformly focusing channels.

For a DMC $P_{Y|X}$, we now consider the $E_{0}$ function
\begin{align}
E_{0}(\rho, P_{X}, P_{Y|X})
\coloneqq
- \ln \sum_{y \in \mathcal{Y}} \Bigg( \sum_{x \in \mathcal{X}} P_{X}( x ) P_{Y|X}(y \mid x)^{1/(1+\rho)} \Bigg)^{\! 1+\rho}
\notag
\end{align}
for $\rho \in (-1, \infty)$, which is defined by Gallager \cite{red}.
The $E_{0}$ function is used in the random coding exponent \cite{red}%
\footnote{In the paper, we omit the maximizing over the input distribution $P_{X}$.}
\begin{align}
E_{\mathrm{r}}( R, P_{X}, P_{Y|X} )
\coloneqq
\max_{\rho \in [0, 1]} \big\{ E_{0}(\rho, P_{X}, P_{Y|X}) - \rho R \big\}
\end{align}
for a rate $R \ge 0$, and other error exponents \cite{sphere, arimoto}.
It is known, e.g., \cite[Eq. (6)]{alsan}, that
\begin{align}
\frac{ E_{0}(\rho, U_{X}, P_{Y|X}) }{ \rho }
=
I_{1/(1+\rho)}(U_{X}; P_{Y|X})
\label{eq:E0_mutual}
\end{align}
for $\rho \in (-1, 0) \cup (0, \infty)$.
Note that the limiting value of the left-hand side of \eqref{eq:E0_mutual} as $\rho \to 0$ is the (ordinary) mutual information $I(U_{X}; P_{Y|X})$.
Namely, the $E_{0}$ function is closely related to the mutual information of order $\alpha$, and sharp bounds of two distinct $E_{0}$ functions for uniformly focusing channels can be obtained in a similar manner to \thref{th:focusing}.
Therefore, we present the sharp bounds of the error exponent with a fixed mutual information of order $\alpha$ in \thref{th:Er}.

\begin{theorem}
\label{th:Er}
For any uniformly focusing channel $P_{Y|X}$ and $\alpha \in (0, 1/2] \cup [1, \infty]$, there exist unique channels $V_{Y|X}$ and $W_{Y|X}$ such that satisfy
\begin{align}
I_{\alpha}(U_{X}; P_{Y|X})
=
I_{\alpha}(U_{X}; V_{Y|X})
=
I_{\alpha}(U_{X}; W_{Y|X})
\end{align}
and the following:
(i) if $\alpha \in (0, 1/2]$, then
\begin{align}
E_{\mathrm{r}}(R, U_{X}, W_{Y|X})
\le
E_{\mathrm{r}}(R, U_{X}, P_{Y|X})
\le
E_{\mathrm{r}}( R,U_{X}, V_{Y|X}) ,
\end{align}
for all $R \ge 0$, and
(ii) if $\alpha \in [1, \infty]$, then
\begin{align}
E_{\mathrm{r}}(R, U_{X}, V_{Y|X})
\le
E_{\mathrm{r}}(R, U_{X}, P_{Y|X})
\le
E_{\mathrm{r}}(R, U_{X}, W_{Y|X})
\end{align}
for all $R \ge 0$.
\end{theorem}

\begin{IEEEproof}[Proof of \thref{th:Er}]
Let $P_{Y|X}$ be a uniformly focusing channel.
For a fixed $\alpha \in (0, \infty]$, assume that
\begin{align}
I_{\alpha}(U_{X}; P_{Y|X})
=
I_{\alpha}(U_{X}; V_{Y|X})
=
I_{\alpha}(U_{X}; W_{Y|X}) .
\label{eq:fixed_mutual}
\end{align}
Then, it follows from \thref{th:focusing} that
\begin{align}
I_{\beta}(U_{X}; V_{Y|X}) - R
\le
I_{\beta}(U_{X}; P_{Y|X}) - R
\le
I_{\beta}(U_{X}; W_{Y|X}) - R
& \qquad \mathrm{for} \ \beta \in (0, \alpha) ,
\label{ineq:mutual_beta_1} \\
I_{\beta}(U_{X}; W_{Y|X}) - R
\le
I_{\beta}(U_{X}; P_{Y|X}) - R
\le
I_{\beta}(U_{X}; V_{Y|X}) - R
& \qquad \mathrm{for} \ \beta \in (\alpha, \infty]
\label{ineq:mutual_beta_2}
\end{align}
for all $R \ge 0$.
By change the variable as
\begin{align}
\beta
=
\frac{ 1 }{ 1 + \rho }
\iff
\rho
=
\frac{ 1 - \beta }{ \beta } ,
\end{align}
we rewrite \eqref{ineq:mutual_beta_1} and \eqref{ineq:mutual_beta_2} as
\begin{align}
I_{1/(1+\rho)}(U_{X}; V_{Y|X}) - R
\le
I_{1/(1+\rho)}(U_{X}; P_{Y|X}) - R
\le
I_{1/(1+\rho)}(U_{X}; W_{Y|X}) - R
& \qquad \mathrm{for} \ \rho \in \bigg( \frac{ 1 - \alpha }{ \alpha }, \infty \bigg) ,
\label{ineq:mutual_beta_3} \\
I_{1/(1+\rho)}(U_{X}; W_{Y|X}) - R
\le
I_{1/(1+\rho)}(U_{X}; P_{Y|X}) - R
\le
I_{1/(1+\rho)}(U_{X}; V_{Y|X}) - R
& \qquad \mathrm{for} \ \rho \in \bigg( -1, \frac{ 1 - \alpha }{ \alpha } \bigg)
\label{ineq:mutual_beta_4}
\end{align}
for all $R \ge 0$.
Then, since
\begin{align}
E_{0} (\rho, U_{X}, P_{Y|X}) - \rho R
& \overset{\eqref{eq:E0_mutual}}{=}
\rho \, I_{1/(1+\rho)}(U_{X}; P_{Y|X}) - \rho R
\\
& =
\rho \Big( I_{1/(1+\rho)}(U_{X}; P_{Y|X}) - R \Big) ,
\end{align}
it follows from \eqref{ineq:mutual_beta_3} and \eqref{ineq:mutual_beta_4} that
\begin{align}
E_{0} (\rho, U_{X}, V_{Y|X}) - \rho R
\le
E_{0} (\rho, U_{X}, P_{Y|X}) - \rho R
\le
E_{0} (\rho, U_{X}, W_{Y|X}) - \rho R
\label{ineq:mutual_beta_5}
\end{align}
for $\rho \in \big( -1, \min \big\{ 0, ( 1 - \alpha ) / \alpha \big\} \big) \cup \big( \max \big\{ 0, ( 1 - \alpha ) / \alpha \big\}, \infty \big)$ and $R \ge 0$, and
\begin{align}
E_{0} (\rho, U_{X}, W_{Y|X}) - \rho R
\le
E_{0} (\rho, U_{X}, P_{Y|X}) - \rho R
\le
E_{0} (\rho, U_{X}, V_{Y|X}) - \rho R
\label{ineq:mutual_beta_6}
\end{align}
for $\rho \in \big( \min \big\{ 0, ( 1 - \alpha ) / \alpha \big\}, \max \big\{ 0, ( 1 - \alpha ) / \alpha \big\} \big)$ and $R \ge 0$.
Note that
\begin{align}
E_{\mathrm{r}}(U_{X}, P_{Y|X}, R)
=
\max_{\rho \in [0, 1]} \Big\{ E_{0} (\rho, U_{X}, P_{Y|X}) - \rho R \Big\} .
\end{align}
By dividing the range of the order $\alpha \in (0, \infty]$, we consider \eqref{ineq:mutual_beta_5} and \eqref{ineq:mutual_beta_6} as follows:

\subsection*{Case (i): $0 < \alpha \le 1/2$}

If $\alpha \in (0, 1/2]$, then $( 1 - \alpha ) / \alpha \ge 1$;
hence, it follows from \eqref{ineq:mutual_beta_6} that
\begin{align}
E_{0} (\rho, U_{X}, W_{Y|X}) - \rho R
\le
E_{0} (\rho, U_{X}, P_{Y|X}) - \rho R
\le
E_{0} (\rho, U_{X}, V_{Y|X}) - \rho R
\end{align}
for all $\rho \in [0, 1]$ and $R \ge 0$.
Therefore, we get
\begin{align}
E_{\mathrm{r}}(U_{X}, W_{Y|X}, R)
\le
E_{\mathrm{r}}(U_{X}, P_{Y|X}, R)
\le
E_{\mathrm{r}}(U_{X}, V_{Y|X}, R)
\end{align}
for all $R \ge 0$.

\subsection*{Case (ii): $1 \le \alpha \le \infty$}

If $\alpha \in [1, \infty]$, then $\lim_{x \to \alpha} ( 1 - x ) / x \le 0$;
hence, it follows from \eqref{ineq:mutual_beta_5} that
\begin{align}
E_{0} (\rho, U_{X}, V_{Y|X}) - \rho R
\le
E_{0} (\rho, U_{X}, P_{Y|X}) - \rho R
\le
E_{0} (\rho, U_{X}, W_{Y|X}) - \rho R
\end{align}
for all $\rho \in [0, 1]$ and $R \ge 0$.
Therefore, we get
\begin{align}
E_{\mathrm{r}}(U_{X}, V_{Y|X}, R)
\le
E_{\mathrm{r}}(U_{X}, P_{Y|X}, R)
\le
E_{\mathrm{r}}(U_{X}, W_{Y|X}, R)
\end{align}
for all $R \ge 0$.
\end{IEEEproof}

\begin{figure}[!t]
\centering
\subfloat[Fixed symmetric cutoff rate $I_{1/2}( U_{X}; P_{Y|X} ) = (\ln 8) / 4$.]{
\begin{overpic}[width = 0.45\hsize, clip]{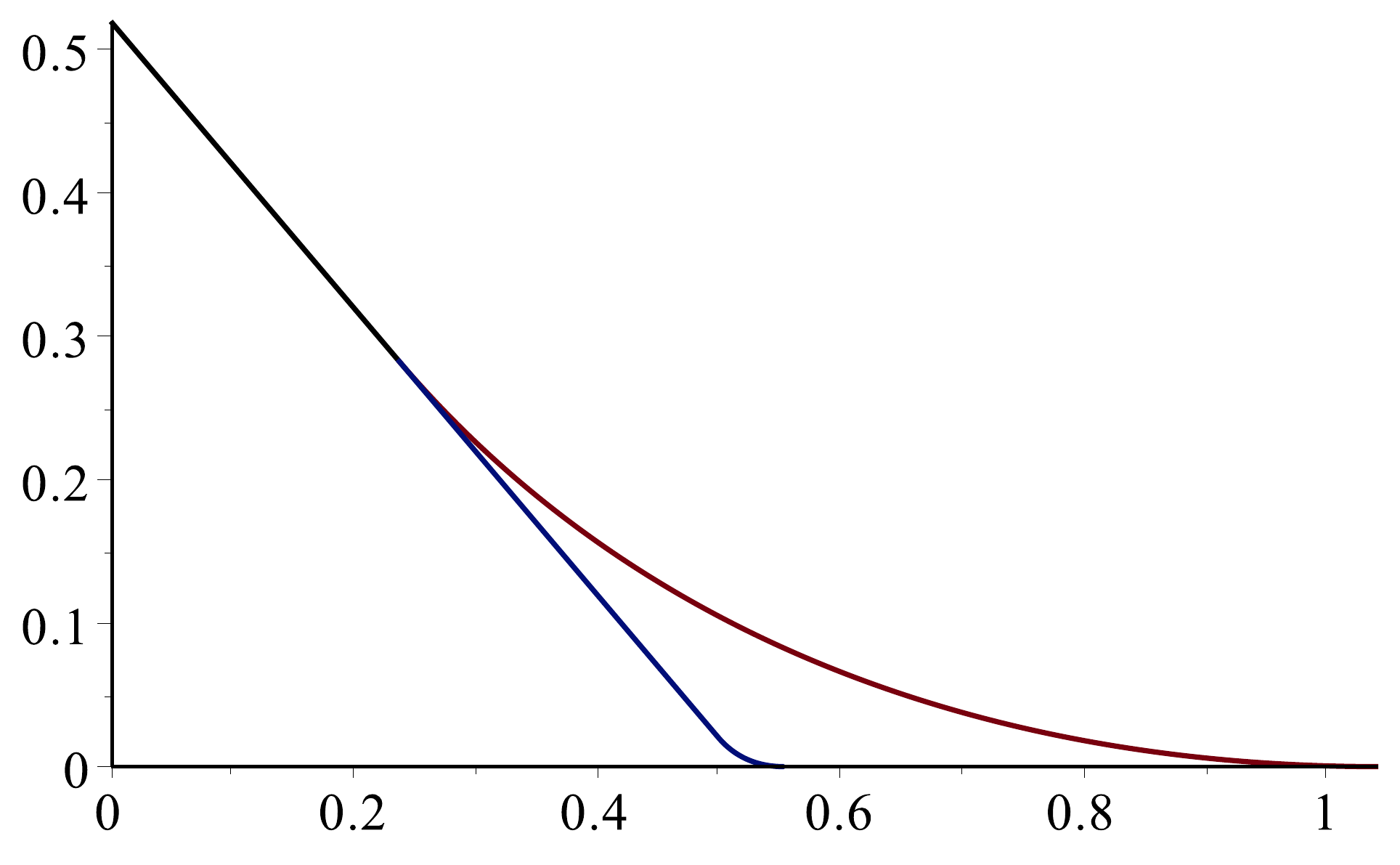}
\put(38, 10){\color{navyblue} $W_{Y|X}$}
\put(60, 17){\color{burgundy} $V_{Y|X}$}
\put(85, 0){$R$}
\put(99, 2){\scriptsize [nats]}
\put(-5, 20){\rotatebox{90}{$E_{\mathrm{r}}(R, U_{X}, P_{Y|X})$}}
\put(-2, 61){\scriptsize [nats]}
\put(30, 59){$E_{\mathrm{r}}( 0, U_{X}, P_{Y|X} ) = I_{1/2}(U_{X}; P_{Y|X})$}
\put(61, 50){$= (\ln 8) / 4$}
\put(29, 60){\vector(-1, 0){20}}
\end{overpic}
}\hspace{0.05\hsize}
\subfloat[Fixed symmetric capacity $I( U_{X}; P_{Y|X} ) = (\ln 8) / 2$.]{
\begin{overpic}[width = 0.45\hsize, clip]{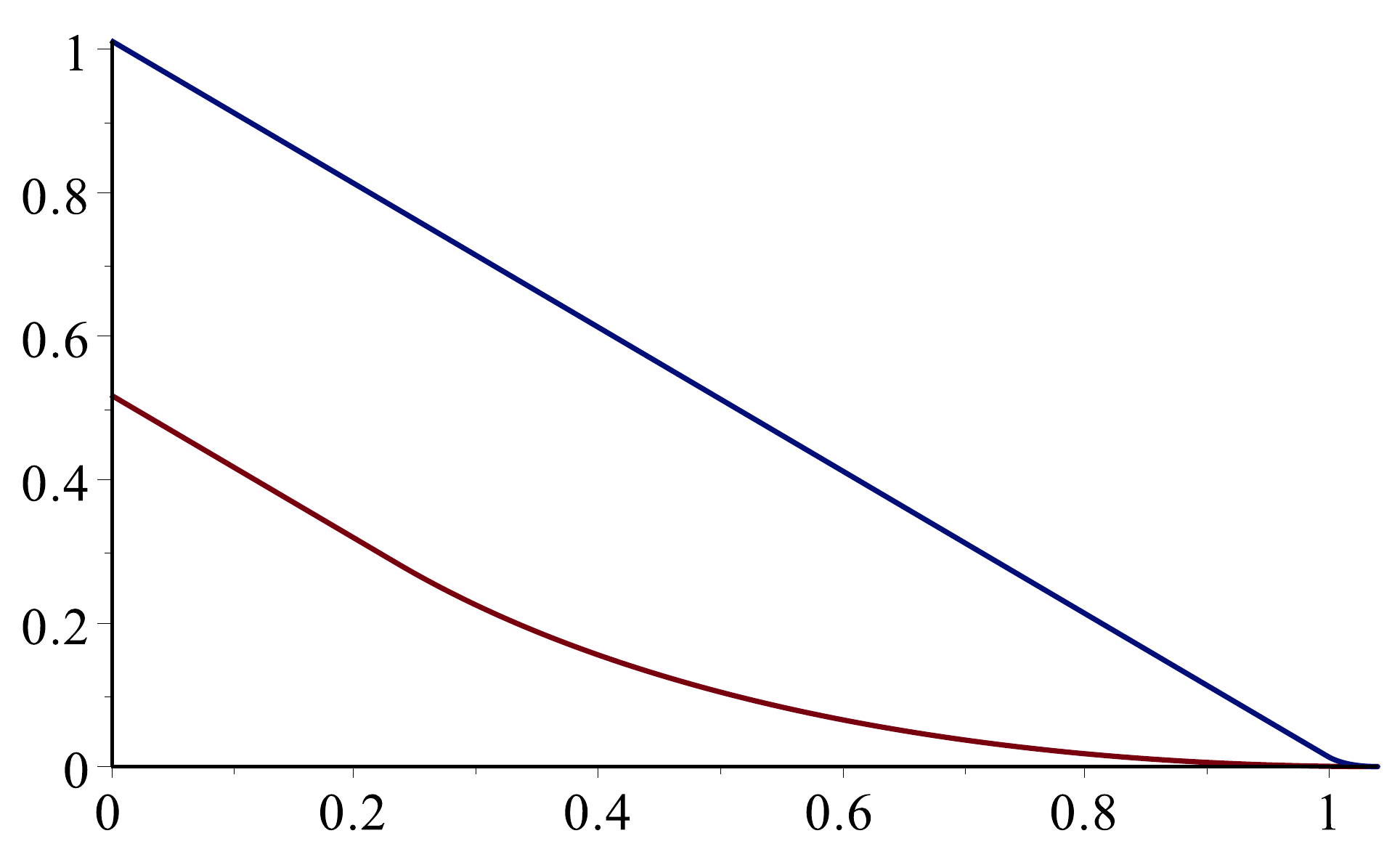}
\put(20, 17){\color{burgundy} $V_{Y|X}$}
\put(60, 31){\color{navyblue} $W_{Y|X}$}
\put(85, 0){$R$}
\put(99, 2){\scriptsize [nats]}
\put(-5, 20){\rotatebox{90}{$E_{\mathrm{r}}(R, U_{X}, P_{Y|X})$}}
\put(-2, 61){\scriptsize [nats]}
\put(60, 45){$I( U_{X}; P_{Y|X} ) = (\ln 8) / 2$}
\put(98.5, 42){\vector(0, -1){34}}
\end{overpic}
}
\caption{Upper and lower bounds of $E_{\mathrm{r}}(R, U_{X}, P_{Y|X})$ for octonary-input uniformly focusing channels $P_{Y|X}$ with fixed mutual information of order $\alpha \in \{ 1/2, 1 \}$.
The upper and lower bounds correspond to either the strongly symmetric channel $V_{Y|X}$ of \eqref{def:channel_v} or $W_{Y|X}$ of \eqref{def:channel_w}.}
\label{fig:Er}
\end{figure}
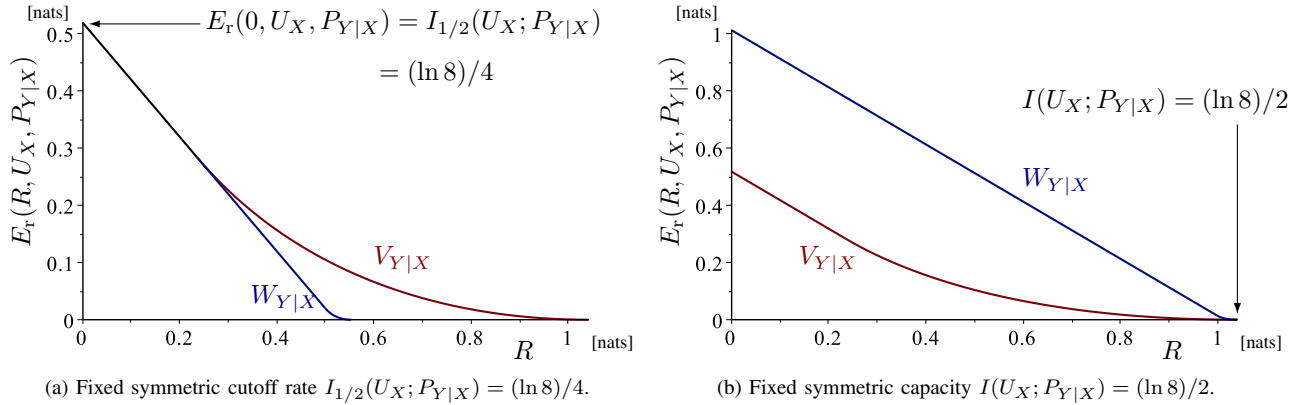

\thref{th:Er} shows that, among all $n$-ary input uniformly focusing channels with a fixed mutual information of order $\alpha \in (0, 1/2] \cup [1, \infty]$ under the uniform input distribution, the strongly symmetric channels $V_{Y|X}$ and $W_{Y|X}$ take the extremal random coding exponents.
We illustrate the sharp bounds of $E_{\mathrm{r}}(U_{X}, P_{Y|X}, R)$ for uniformly focusing channels $P_{Y|X}$ in \figref{fig:Er}.

Finally, note that the uniform input distribution maximizes the mutual information of order $\alpha$ if a channel is strongly symmetric (cf. \cite[Eq.~(20)]{arimoto2} and \cite[p.~145]{red}).

\section{Conclusion}

In this paper, we examined the sharp bounds between the $\ell_{\alpha}$-norm and the $\ell_{\beta}$-norm of $n$-dimensional probability vectors for distinct $\alpha, \beta \in (0, 1) \cup (1, \infty]$, as shown in \thref{th:extremes}.
By using the result, \thref{th:R_extremes} established the sharp bounds between two R\'{e}nyi entropies of distinct positive orders.
In \remref{remark:note}, we mentioned that sharp bounds on other axiomatic definitions of entropy \cite{tsallis, havrda, behara, daroczy, boekee} can be obtained by using \thref{th:extremes}, as with \thref{th:R_extremes}.
In \sectref{sect:conditional}, we considered to extend the above results from \emph{un}conditional settings to \emph{conditional} settings.
Then, \thref{th:convexhull} identified the exact feasible regions between two expectations of $\ell_{\alpha}$- and $\ell_{\beta}$-norms, which implies the exact feasible regions between two conditional R\'{e}nyi entropies of distinct orders (cf. \figref{fig:convexhull}).
Finally, \sectref{sect:focusing} examined the sharp bounds on channel reliability functions, such as the mutual information of order $\alpha$ and the $E_{0}$ function, for uniformly focusing channels under the uniform input distribution.
Then, \thref{th:Er} provided the sharp bounds on the random coding exponent of uniformly focusing channels under the uniform input distribution with a fixed mutual information of order $\alpha$. 
Finally, we remark that the sharp bounds of error exponents, such as the strong converse \cite{arimoto} and the sphere packing \cite{sphere} exponents, of uniformly focusing channels under the uniform input distribution with a fixed mutual information of order $\alpha$ are also obtained from \thref{th:focusing}, as with \thref{th:Er}.


\section*{Acknowledgment}

This study was partially supported by the Ministry of Education, Science, Sports and Culture, Grant-in-Aid for Scientific Research (C) 26420352.



\bibliographystyle{IEEEtran}
%

\end{document}